\documentclass[prx,graphicx,superscriptaddress,amsmath,amssymb,twocolumn,floatfix]{revtex4-1}

\usepackage{graphicx}
\usepackage{bm}
\usepackage[utf8]{inputenc}
\usepackage[english]{babel}
\usepackage[T1]{fontenc}
\usepackage{mathptmx}
\usepackage{siunitx}
\usepackage{braket}
\usepackage{amsmath,amssymb}
\usepackage{hyperref}
\hypersetup{
    colorlinks=true,
    breaklinks=true,
    urlcolor=blue,
    linkcolor=blue,
    citecolor=blue}

\usepackage{fancyhdr}
\usepackage[raggedright]{titlesec}
 
\titleformat{\subsection}[runin]{}{}{}{}[]

\newcommand{\minus}{\scalebox{0.7}[1.0]{\(-\)}}
\newcommand{\plus}{\scalebox{0.7}[1.0]{\(+\)}}
\newcommand{\medminus}{\scalebox{0.5}[0.7]{\(-\)}}

\usepackage{titlesec}
\titlespacing\section{0pt}{12pt plus 4pt minus 2pt}{2pt plus 2pt minus 2pt}
\titlespacing\subsection{0pt}{12pt plus 4pt minus 2pt}{2pt plus 2pt minus 2pt}
\titlespacing\subsubsection{0pt}{12pt plus 4pt minus 2pt}{2pt plus 2pt minus 2pt}

\begin{document}

\title{Designing spin and orbital sources of Berry curvature at oxide interfaces}

\author{Edouard~Lesne}
    \affiliation{Kavli Institute of Nanoscience, Delft University of Technology,  Lorentzweg 1, 2628CJ Delft, Netherlands.}
\author{Yildiz~G.~Sa\v{g}lam}
    \affiliation{Kavli Institute of Nanoscience, Delft University of Technology,  Lorentzweg 1, 2628CJ Delft, Netherlands.}
\author{Raffaele~Battilomo}
    \affiliation{Institute for Theoretical Physics, Center for Extreme Matter and Emergent Phenomena, Utrecht University, Princetonplein 5, 3584 CC Utrecht, Netherlands.}
 \author{Maria~Teresa~Mercaldo}
    \affiliation{Dipartimento di Fisica “E.~R.~Caianiello”, Universitá di Salerno, IT-84084 Fisciano, Italy.}
\author{Thierry~C.~van~Thiel}
    \affiliation{Kavli Institute of Nanoscience, Delft University of Technology,  Lorentzweg 1, 2628CJ Delft, Netherlands.}
\author{Ulderico~Filippozzi}
    \affiliation{Kavli Institute of Nanoscience, Delft University of Technology,  Lorentzweg 1, 2628CJ Delft, Netherlands.}
\author{Canio~Noce}
     \affiliation{Dipartimento di Fisica “E.~R.~Caianiello”, Universitá di Salerno, IT-84084 Fisciano, Italy.}
\author{Mario~Cuoco}
    \affiliation{Consiglio Nazionale delle Ricerche, CNR-SPIN, Italy}
    \affiliation{Dipartimento di Fisica “E.~R.~Caianiello”, Universitá di Salerno, IT-84084 Fisciano, Italy.}
\author{Gary~A.~Steele}
    \affiliation{Kavli Institute of Nanoscience, Delft University of Technology,  Lorentzweg 1, 2628CJ Delft, Netherlands.}
\author{Carmine~Ortix}
    \affiliation{Institute for Theoretical Physics, Center for Extreme Matter and Emergent Phenomena, Utrecht University, Princetonplein 5, 3584 CC Utrecht, Netherlands.}
    \affiliation{Dipartimento di Fisica “E.~R.~Caianiello”, Universitá di Salerno, IT-84084 Fisciano, Italy.}
\author{Andrea~D.~Caviglia}
    \affiliation{Department of Quantum Matter Physics, University of Geneva, 24 Quai Ernest Ansermet, CH-1211 Geneva, Switzerland}

\setlength{\abovedisplayskip}{3pt}
\setlength{\belowdisplayskip}{3pt}

\maketitle

\textbf{Quantum materials can display physical phenomena rooted in the geometry of electronic wavefunctions. The corresponding geometric tensor is characterized by an emergent field known as Berry curvature (BC). Large BCs typically arise when electronic states with different spin, orbital or sublattice quantum numbers hybridize at finite crystal momentum.  In all materials known to date, the BC is triggered by the hybridization of a single type of quantum number.  Here, we report the discovery of the first material system having both spin and orbital-sourced BC: LaAlO$_3$/SrTiO$_3$ interfaces grown along the [111] direction. 
We detect independently these two sources and directly probe the BC associated to the spin quantum number through measurements of an anomalous planar Hall effect. The observation of a nonlinear Hall effect with time-reversal symmetry  signals large orbital-mediated BC dipoles. The coexistence of different forms of BC enables the combination of spintronic and optoelectronic functionalities in a single material.}

When moving along closed paths, 
electrons can accumulate a geometric Berry phase related to  the flux of a field, called Berry curvature, encoding the geometric properties of the electronic wavefunctions. 
In magnetic materials  adiabatic motion of the electrons around the Fermi surface provides such a Berry phase. It is directly observable since it governs the intrinsic part of the anomalous Hall conductivity~\cite{Haldane2004,Nagaosa2010}. Anomalous Hall effect measurements therefore represent a charge transport footprint of the intrinsic geometric structure of the electronic wavefunctions. 
In non-magnetic materials the BC field is forced to vanish by symmetry when summed over the occupied electronic states. However, local concentrations of positive and negative BC in momentum space are allowed by acentric crystalline arrangements~\cite{Xiao2010}. 
This segregation of BC in different regions of momentum space appears whenever electronic states with different internal quantum numbers are coupled to each other by terms that depend linearly on the crystalline momentum $\textit{k}$. In these regions the electronic bands typically resemble the dispersion relations of relativistic Dirac or Weyl fermions. 
The spin-orbit linear in $\textit{k}$ coupling between different spin states shapes  
the Dirac cones at the surfaces of three-dimensional topological insulators~\cite{Zhang2009,He2021} and the Weyl cones of topological semimetals~\cite{arm18}. 
Couplings between different atomic orbital and sublattice states give rise instead to the (gapped) Dirac cones of transitional metal dichalcogenides and graphene. 
Conceptually speaking, 
the appearance of BC beyond this Dirac/Weyl paradigm is entirely allowed. 
The fundamental conditions for the occurrence of BC only involve the crystalline geometry of a material, with no restrictions on the specific properties of its low-energy electronic excitations. 
Achieving this challenge is of great interest. First, it 
could in principle result in the coexistence of different mechanisms of BC generation. This could be used, in turn, to endow a single material system with different Berry curvature-mediated effects, as for instance spin- and orbital-Hall effects. 
Second, searching for BCs without Dirac or Weyl cones might allow the design of materials with interplay of correlated and topological physics --  an unexplored frontier in condensed matter physics. 

Here, we reach these two milestones in the two-dimensional electron system (2DES)
confined at a (111)$-$oriented oxide interface with a high-temperature trigonal crystalline structure. This model system satisfies the crystalline symmetry properties for a non-vanishing BC. The combination of spin-orbit coupling, orbital degrees of freedom associated with the low-energy $t_{2g}$ electrons, and crystal fields leads to the coexistence of a spin-sourced and an orbital-sourced BC. The two sources are independently probed using two different charge transport diagnostic tools. 
The observation of the BC-mediated anomalous planar Hall effect~\cite{Battilomo2021,Cullen2021} grants direct access to the spin-sourced BC whereas nonlinear Hall transport measurements in time-reversal symmetric conditions~\cite{Sodemann2015,Ortix2021}  detect an orbital-mediated Berry curvature dipole -- a quantity measured so far only in gapped Dirac systems~\cite{Sodemann2015,Ortix2021,Xu2018,Ma2019,Ho2021,Battilomo2019,You2018,Son2019,Zhang2018a,Du2018,Kang2019} and three-dimensional topological semimetals~\cite{Kumar2021,Matsyshyn2019,Singh2020,Facio2018,Zhang2018b,Wawrzik2021}. 
We identify (111)$-$LaAlO$_3$/SrTiO$_3$ heterointerfaces as an ideal material system because its two-dimensional electron system features many-body correlations and a two-dimensional superconducting ground state~\cite{Ohtomo2004,Reyren2007,Rout2017,Monteiro2017,Monteiro2019}. 

\begin{figure*}
\centering
\includegraphics[width=18cm,scale=1]{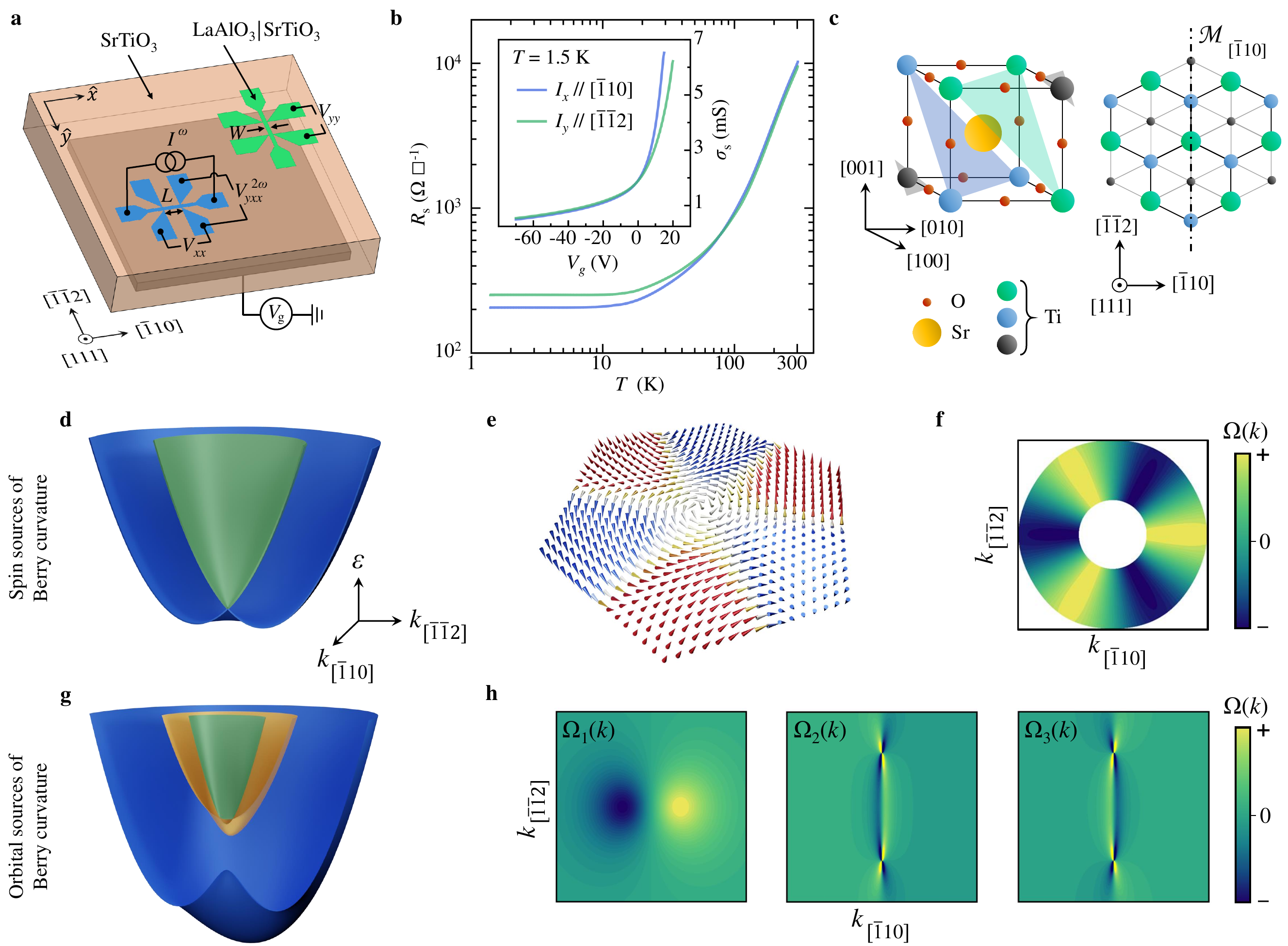}
    \caption{\textbf{Crystal and model band structures of the (111)$-$LaAlO$_{3}$/SrTiO$_{3}$ 2DES, and basic electrical characterization.}  \textbf{a}, Schematic layout of the electrical measurement configurations of two 
    Hall bars, hosting a 2DES, and oriented along the $[\bar{1}10]$ and $[\bar{1}\bar{1}2]$ crystallographic axis. 
    $W$ is the width of the channel and $L$ is the distance between the longitudinal voltage probes. $V_\textrm{g}$ stands for the high-voltage source used to tune the 2DES band occupation (Fermi energy) in a back-gate geometry.
    \textbf{b}, Sheet resistance, $R_\textrm{s}$, versus temperature, $T$, of the 2DES for the $[\bar{1}10]$ and $[\bar{1}\bar{1}2]$ Hall bar devices, showing a nearly isotropic metallic character. Inset: Sheet conductance, $\smash{\sigma_\textrm{s}=R_\textrm{s}^{\minus 1}}$, as a function of back-gate voltage, $V_\textrm{g}$, showing gate-tunability of the 2DES at 1.5~K. \textbf{c}, Left: schematic representation of an SrTiO$_{3}$ perovskite cubic unit cell displaying the nonequivalent (111) titanium planes (shaded areas). Right: top view along the [111] crystallographic direction, of the same unit cell, showing only the Ti atoms. The dash-dotted line indicates the mirror line $\mathcal{M}_{[\bar{1} 1 0]}$.
    \textbf{d}, Schematics of a single pair of spin-split bands  
    forming a Kramers pair at the $\Gamma$ point up to the Fermi level. 
    \textbf{e}, Each spin band is characterised by a non-trivial spin texture  with an out-of-plane spin components induced by the effect of trigonal warping. 
    \textbf{f}, Exclusion plot of the Berry curvature $\Omega_{k}$ over the Fermi surfaces of the two spin sub-bands. 
    \textbf{g}, Schematics  band structure of spin-orbit free orbital bands corresponding to $t_{2g}$ electrons subject to a ${\mathcal C}_s$ crystal field. At the center of the BZ all levels are split. The orbital Rashba coupling $\propto \alpha_m$ leads to mirror-symmetry protected crossings. 
    \textbf{h}, Band-resolved Berry curvature displaying dipolar hot-spots (left panel) in the lowest energy band and singular pinch points in the highest energy bands.}
\label{fig1}
\end{figure*}

We synthesise (111)$-$LaAlO$_{3}$/SrTiO$_{3}$ heterostructures by pulsed laser deposition, as detailed in the Methods section. 
The samples are lithographically patterned into Hall bars oriented along the two orthogonal principal in-plane crystallographic directions: the $[\bar{1}10]$ and $[\bar{1}\bar{1}2]$ axis (see Fig.\,\hyperref[fig1]{1a}). 
The sheet conductance and carrier density of the 2DES are controlled by electrostatic field effects in a back-gate geometry (see Fig.\,\hyperref[fig1]{1b}). 
We source an oscillating current ($I^{\omega}$) with frequency $\omega/2\pi$ along each Hall bar, and concomitantly measure the longitudinal response as well as the first or second harmonic transverse voltages in a conventional lock-in detection scheme (see Fig.\,\hyperref[fig1]{1a}). 

The non-trivial geometric properties of the electronic waves in the 2DES derive entirely from the triangular arrangement of the titanium atoms at the (111)$-$LaAlO$_3$/SrTiO$_3$ interface (see Fig.\,\hyperref[fig1]{1c}). Together with the ${\mathcal M}_{{\bar 1} 1 0}$ mirror line symmetry, this yields a ${\mathcal C}_{3v}$ crystallographic point group symmetry. 
As a result of this trigonal crystal field and the concomitant presence of 
spin-orbit coupling, the entire $d$-orbital manifold of the Ti atoms located at the center of the surface Brillouin zone (BZ) is split into five distinct Kramers pairs (Supplementary Note I). 
The energy bands of the pairs are shifted in momentum due to spin-orbit coupling. In their simplest form, they acquire a parabolic dispersion reminiscent of a Rashba 2DES (see Fig.\,\hyperref[fig1]{1d}). 
However, the trigonal crystal field brings about a specific hexagonal warping~\cite{roe14,bar14} that has
a twofold effect. First, for each time-reversal related pair of bands, the Fermi lines acquire an hexagonal ``snowflake" shape~\cite{Fu2009}. 
Second, and most important, the spin texture in momentum space acquires a characteristic out-of-plane component~\cite{He2018,tra22}, with alternating meron and antimeron wedges respecting the symmetry properties of the crystal (see Fig.\,\hyperref[fig1]{1e}). 
This unique spin-momentum locking enables a non-vanishing local BC entirely generated by spin-orbit coupling (Supplementary Note I).
The local BC of the spin-split bands of each pair cancel each other at the same crystal momentum. However, there is a region of crystal momenta populated by a single spin band. In this region -- the annulus between the two Fermi lines of the system --  alternating positive and negative regions of non-vanishing BC are present (see Fig.\,\hyperref[fig1]{1f}). 

Apart from the spin channel, an inherently different source of BC exists. 
In systems with orbital degrees of freedom, lack of crystal centrosymmetry yields coupling that are linear in $\textit{k}$, 
and mix different atomic orbital states. These orbital Rashba couplings~\cite{Kim13} are
independent of the presence 
of spin-orbit coupling. 
Precisely as its spin counterpart, 
the orbital Rashba coupling can generate a finite BC~\cite{Mercaldo2022}, but only when all rotational symmetries are broken (see Methods section and Supplementary Note I). With a reduced ${\mathcal C}_s$ symmetry, low-lying $t_{2g}$ orbitals are split into three non-degenerate levels. 
The corresponding orbital bands then realize a gapped Rashba-like spectrum with protected crossings along the mirror-symmetric lines of the two-dimensional Brillouin zone (BZ) (see Fig.\,\hyperref[fig1]{1g}). These characteristics result in the appearance of dipolar BC hot-spots and singular pinch points (see Fig.\,\hyperref[fig1]{1h}).
Such orbital sources of Berry curvature are fully active at (111) oxide interfaces thanks to reduced low-temperature symmetries. 
The cubic-to-tetragonal structural phase transition~\cite{Rimai1962,Fleury1968} occurring at 110~K breaks the  threefold rotational symmetry along the [111] direction.
In addition, the 
tetragonal to locally triclinic structural distortions at temperatures below $\simeq$~70~K together with the ferroelectric instability~\cite{ros2013} 
below 50~K is expected to strongly enhance the orbital Rashba strength. 

The orbital-sourced BC is expected to be very stiff in response to externally applied in-plane magnetic fields due to the absence of symmetry-protected orbital degeneracies. In contrast, the spin-sourced BC is significantly more susceptible to planar magnetic fields.  
As shown in Fig.\,\hyperref[fig1]{2a,b}, 
an  
in-plane magnetic field is capable of generating a BC hot-spot 
within the Fermi surface annulus. 
This BC hot-spot corresponds to a field-induced avoided level crossing between the two spin-split bands 
that occurs
whenever the applied magnetic field breaks the residual crystalline mirror symmetry. 
The momentum-integrated net Berry curvature is then non-zero (Supplementary Note II), and yields a transverse Hall conductance satisfying the antisymmetric property $\sigma_\textrm{xy} \rho_\textrm{yx}=\minus1$, even in the absence of any Lorentz force. 
This effect, theoretically predicted in Refs.~\citenum{Battilomo2021,Cullen2021} and known as the anomalous planar Hall effect (APHE), is different in nature with respect to the conventional planar Hall effect, which is instead related to the anisotropy in the longitudinal magnetoresistance and thus characterised by a symmetric response $\sigma_\textrm{xy}(B)=\sigma_\textrm{xy}(\minus B)$. 

\begin{figure*}
\centering
\includegraphics[width=18cm,scale=1]{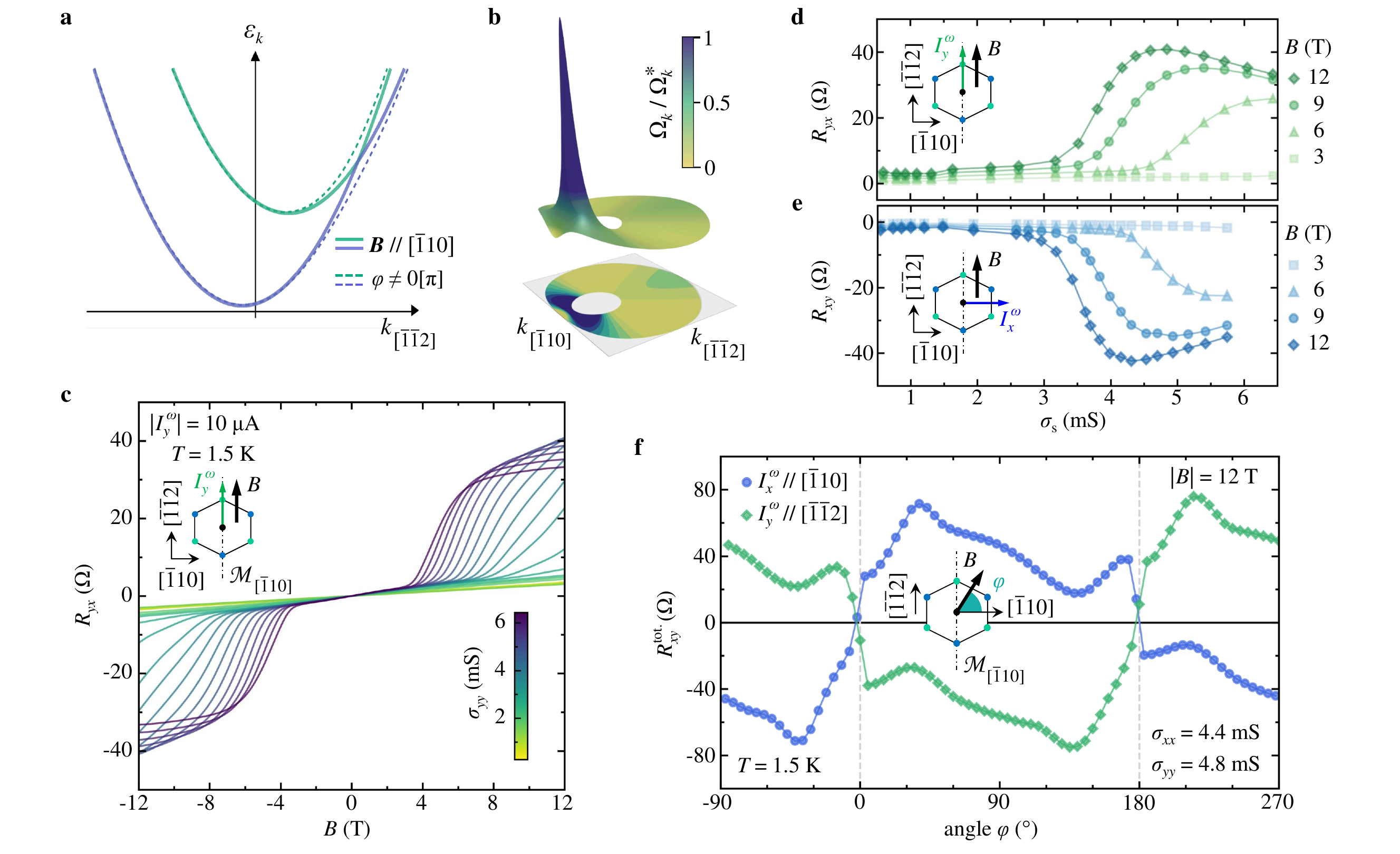}
\label{fig2}
    \caption{\textbf{Anomalous planar Hall response of the 2DES induced by the spin-sourced Berry curvature.} \textbf{a}, Schematic energy dispersion of the spin-split bands along the mirror line of the Brillouin zone $k_{[\bar{1} 1 0]}=0$ in the presence of a planar magnetic field. When the latter is oriented along the $[\bar{1} 1 0]$ direction there is a mirror-symmetry protected crossing of the spin-split bands that evolves into an anticrossing for other directions of the magnetic field. The angle $\varphi$ is defined by the orientation of the magnetic field  w.r.t. the $[\bar{1} 1 0]$ crystallographic direction (\textit{c.f.} schematic inset in panel \textbf{f}).
    \textbf{b}, Sketch of the spin-sourced Berry curvature normalised magnitude $\Omega_{k}/\Omega_{k}^{\star}$ when the magnetic field is directed along the $[\bar{1} \bar{1} 2]$ direction. 
    When the anticrossing point enters the Fermi surface annulus, the integral of the curvature is strongly enhanced and the anomalous planar Hall response reaches its maximum. \textbf{c}, Experimentally measured field-antisymmetric planar Hall resistance $R_\textrm{xy}=V_\textrm{xy}^{\omega}/I_\textrm{x}^{\omega}$ at $T=1.5$~K, with $I_\textrm{y}^{\omega}$ along $[\bar{1}\bar{1}2] \parallel B$ (see inset schematic), for different sheet conductance values $\sigma_\textrm{yy}$ indicated by the inset colored scale bar. \textbf{d}, Corresponding dependence of $R_\textrm{yx}$ versus $\sigma_\textrm{yy}$ showing a non-monotonic behavior for fixed values of $B>3$~T, and an onset above a threshold value of $\sigma_\textrm{yy}$. \textbf{e}, Dependence of the field-antisymmetric contribution $R_\textrm{xy}$ versus $\sigma_\textrm{xx}$ for $I_\textrm{x}^{\omega}$ along $[\bar{1}10] \perp B$ (see Extended Data Fig.\,\hyperref[extfig1]{1}). \textbf{f}, In-plane angular dependence of the raw total transverse resistance response $R_\textrm{xy}^\textrm{tot.}$, \textit{i.e.} not field-(anti)symmetrised, for the two Hall bar devices at $B=12$~T. The planar Hall response obeys Onsager relation $R_\textrm{xy}(B)=R_\textrm{yx}(-B)$, as seen by the near identical angular dependence upon imposing a $\pm\pi$ translation to either curve.  Remarkably, $R_\textrm{xy}^\textrm{tot.}$ goes to zero at $\varphi=0^{\circ}$ and $\varphi=180^{\circ}$, \textit{i.e.} when the mirror symmetry is preserved even in the presence of an external magnetic field.}
\end{figure*}

Figure~\hyperref[fig2]{2c} shows the transverse (Hall) resistance measured with a current applied along the $[\bar{1}\bar{1} 2]$ crystal direction and with collinear current and magnetic field. This ensures a vanishing symmetric planar Hall effect \cite{Battilomo2021}. 
At fields well-below 4 T, a small signal increasing linearly with the field strength is detected. This feature can be
attributed to an out-of-plane misalignment of the magnetic field smaller than $1.5^{\circ}$ (Supplementary Note III). 
Above a magnetic field threshold instead, a large transverse Hall signal sharply emerges (see also Extended Data Fig.\,\hyperref[extfig3]{3}). 
At even larger fields this response saturates. Electrostatic gating is found to decrease the magnetic field threshold and promotes a non-monotonic evolution of the response amplitude (Fig.\,\hyperref[fig2]{2d,e}).
The experimental features of this Hall response can be captured by 
considering
a single pair of spin-split bands
coupled to the external field by the Zeeman interaction. In this picture, the sudden onset of the transverse response is associated to the 
appearance 
of the BC hot-spot inside the Fermi surface annulus occurring at a critical magnetic field strength (Supplementary Note II). 
Magnetoconductance measurements in the weak antilocalization regime (Extended Data Figs.\,\hyperref[extfig4]{4},\hyperref[extfig5]{5}) show that the onset of the transverse Hall signal coincides precisely with the appearance of the spin-sourced BC hot-spot. 
The non-monotonic behavior of the transverse response as a function of electrostatic gating and magnetic field strength can be also ascribed to the BC origin of the Hall response. 
The angular dependence of the transverse resistance as shown in Fig.\,\hyperref[fig2]{2f} indicates a vanishing transverse linear conductivity when the planar magnetic field is along the $[\bar{1} 1 0]$ direction, due to the mirror symmetry ${\mathcal M}_{[\bar{1} 1 0]}$. This is independent on whether the driving current is along the $[\bar{1}10]$ or the $[\bar{1}\bar{1}2]$ direction. Note that the two angular dependencies are related to each other by a 180$^{\circ}$ shift in agreement with the Onsager reciprocity relations~\cite{Onsager1931a}.

\begin{figure*}
\centering
\includegraphics[width=18cm,scale=1]{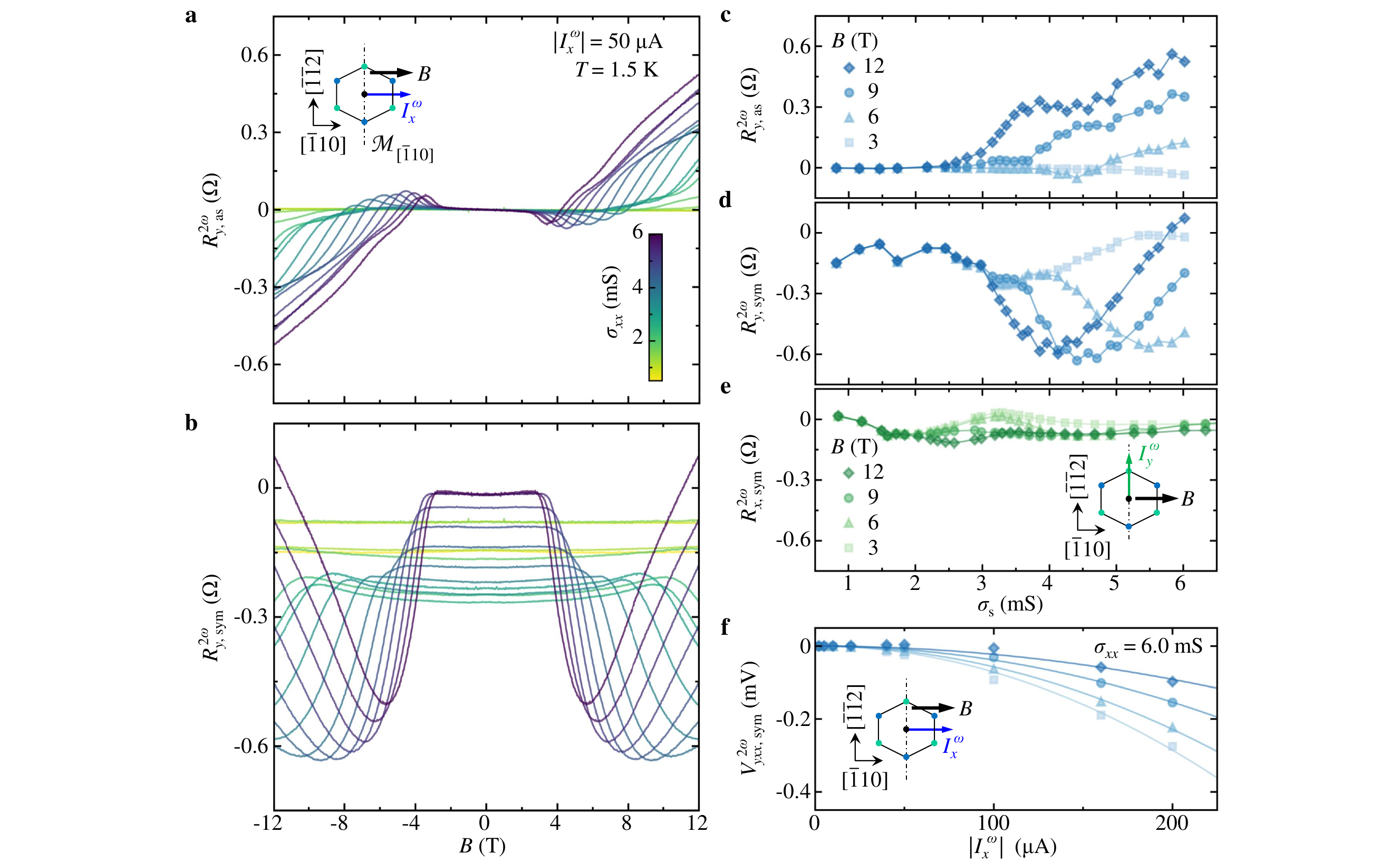}
\label{fig3}
    \caption{\textbf{Nonlinear Hall response of the 2DES in a planar magnetic field.} \textbf{a},\textbf{b}, Field-antisymmetric, $R_\textrm{y,as}^{2\omega}$ (\textbf{a}), and field-symmetric $R_\textrm{y,sym}^{2\omega}$ (\textbf{b}) second harmonic transverse resistance responses, for $I_\textrm{x}^{\omega}$ along $[\bar{1}10] \parallel B$, and for different values of the sheet conductance $\sigma_\textrm{xx}$. \textbf{c}, \textbf{d}, Corresponding dependence of $R_\textrm{y,as}^{2\omega}$ and $R_\textrm{y,sym}^{2\omega}$, respectively, versus $\sigma_\textrm{xx}$ for different values of the in-plane magnetic field $B$. The field-symmetric nonlinear transverse resistance displays a strong non-monotonic response, attributed to a Zeeman-induced Berry phase contribution. \textbf{e}, Field-symmetric second harmonic transverse resistance response, $R_\textrm{x,sym}^{2\omega}$, for $I_\textrm{y}^{\omega}$ along $[\bar{1}\bar{1}2] \perp B$ versus sheet conductance $\sigma_\textrm{yy}$ at fixed values of the planar magnetic field (see also Extended Data Fig.\,\hyperref[extfig2]{2}). The full scale ordinate axis is chosen to be the same as for panel \textbf{d}, for better comparison. \textbf{f}, Field-symmetric nonlinear transverse second harmonic voltage $V_\textrm{yxx,sym}^{2\omega}$\ versus the a.c. current amplitude $I_\textrm{x}^{\omega}$, at fixed values of $B$ (see Extended Data Fig.\,\hyperref[extfig6]{6}). Solid lines are quadratic fits.
    }
\end{figure*}

The absence of linear conductivity makes this configuration the ideal regime to investigate the presence of nonlinear transverse responses, which are symmetry-allowed when the driving current is collinear with the magnetic field (Supplementary Note II). We have therefore performed systematic measurements of the second harmonic, \textit{i.e.} at $2 \omega$, transverse responses (see Fig.\,\hyperref[fig3]{3a,b}) by sourcing the a.c. current along the $[\bar{1}10]$ direction. We have subsequently disentangled 
the field-antisymmetric $R_\textrm{y,as}^{2\omega}=\left[ R_\textrm{y}^{2\omega}(B) \, \medminus \, R_\textrm{y}^{2\omega}(\minus B) \right]$/2, and the field-symmetric contributions $R_\textrm{y,sym}^{2\omega}=\left[ R_\textrm{y}^{2\omega}(B) \, \plus \, R_\textrm{y}^{2\omega}(\minus B) \right]$/2 since they originate from distinct physical effects. 
In particular, the antisymmetric part contains a
semiclassical contribution that only depends upon the conventional 
group velocity of the carriers at the Fermi level (Supplementary Note II). Conversely, the symmetric part originates from the anomalous velocity term of the carriers. It is a purely quantum contribution and can be expressed in terms of a Berry curvature dipole (BCD). 
We observe the following features 
in Fig.\,\hyperref[fig3]{3a,b}. 
The semiclassical antisymmetric contribution has a sudden onset above a characteristic magnetic field (see Fig.\,\hyperref[fig3]{3a}) that is sensitive to gating (see Fig.\,\hyperref[fig3]{3c}). The gate dependence displays a monotonic growth consistent with its physical origin. On the contrary, the symmetric contribution displays the typical non-monotonous gate and field amplitude dependence (see Fig.\,\hyperref[fig3]{3d}) of BC-mediated effects. 
The gate dependence of the nonlinear symmetric contribution obtained by sourcing the current along the $[\bar{1}\bar{1}2]$ direction is instead strongly suppressed and featureless (see Fig.\,\hyperref[fig3]{3e}). This is consistent with a $[\bar{1}10]-$oriented BCD, which gives a vanishing response in this configuration. 
We note
that the symmetric nonlinear transverse resistance has a 
characteristic quadratic current-voltage ($I^\omega-V^{2\omega}$), which, combined with the response at double the driving frequency, establishes its  second-order nature (see Fig.\,\hyperref[fig3]{3f}). 

\begin{figure*}
\centering
\includegraphics[width=18cm,scale=1]{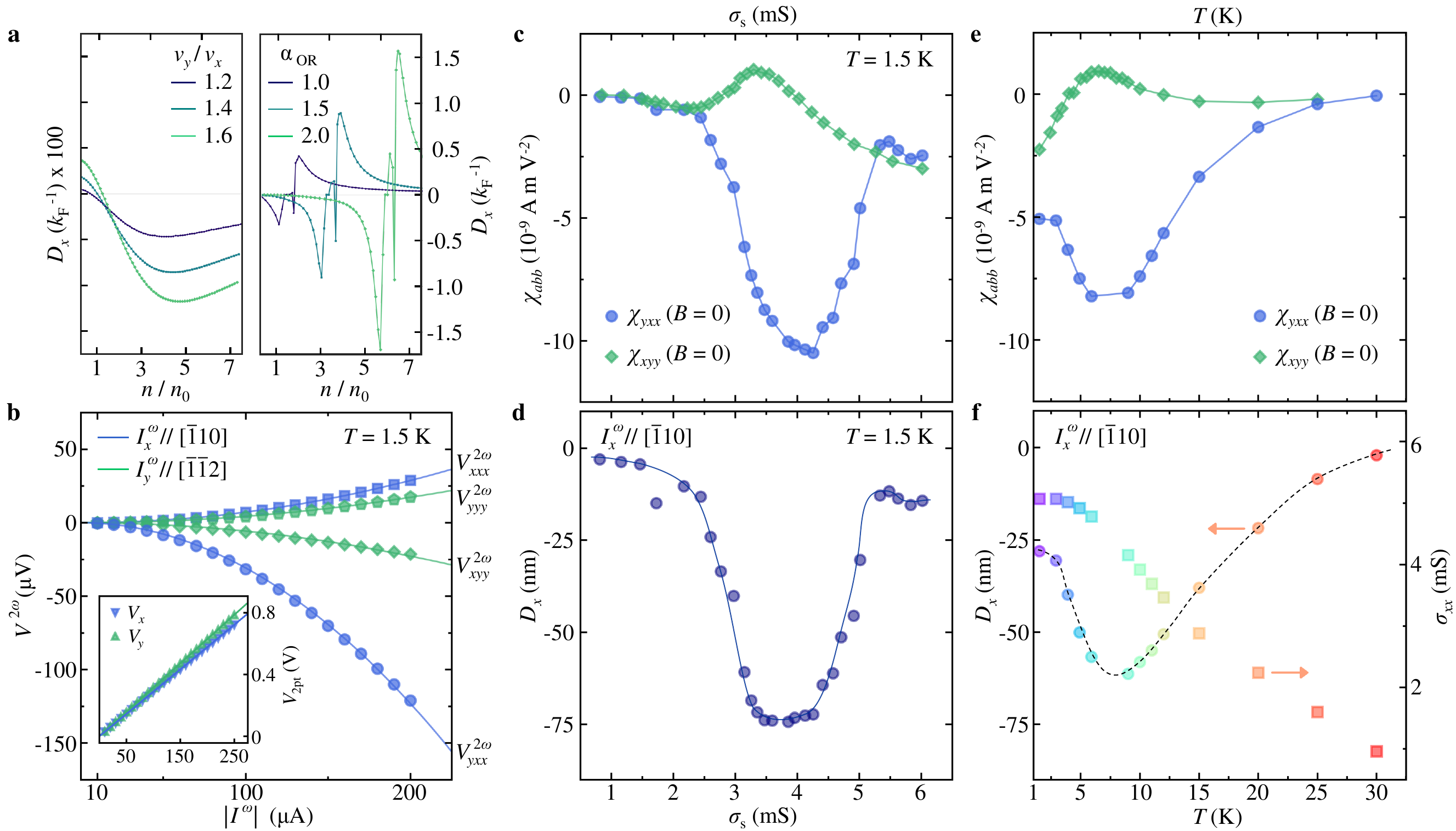}
\label{fig4}
    \caption{\textbf{Berry curvature dipole under time-reversal symmetric conditions} \textbf{a}, Calculated spin-sourced and orbital-sourced Berry curvature dipole as a function of the sheet carrier density. The spin-sourced  dipole (left panel) has been evaluated for different strength of the rotational symmetry breaking distortion $\propto v_\textrm{y}/v_\textrm{x}$ (see Methods section), whereas the orbital-sourced dipole (right panel) has been computed varying the strength of the orbital Rashba coupling $\alpha_\textrm{OR}$.
    In both cases, the dipole has a strongly non-monotonic behavior, goes to zero for large densities, and is directed along the $[\bar{1}10]$ direction. The orbital-sourced dipole is two order of magnitude larger in all the density range. \textbf{b}, Measured $I^\omega-V^{2\omega}$ characteristics at zero magnetic field. Longitudinal $V_\textrm{xxx(yyy)}^{2\omega}$ and transverse $V_\textrm{yxx(xyy)}^{2\omega}$ voltage drops versus the a.c. excitation bias $|I^{\omega}|$ for $I_\textrm{x(y)}^{\omega}$ along $[\bar{1}10]$ ($[\bar{1}\bar{1}2]$, respectively) at $\sigma_\textrm{xx} \approx \sigma_\textrm{yy} \approx 4.5-4.6$~mS. 
    Solid lines are quadratic fits. Inset: Linear two-terminal $I-V$ characteristics highlighting the Ohmic behaviour of the electrical contacts to the 2DES. Solid lines are linear fits. \textbf{c}, Sheet conductance dependence of the measured nonlinear transverse conductivity tensor elements $\chi_\textrm{yxx}$ and $\chi_\textrm{xyy}$ for $I^{\omega}$ sourced along the two orthogonal in-plane principal crystallographic directions. \textbf{d}, Berry curvature dipole's magnitude, $D_\textrm{x}$, under time-reversal symmetric conditions ($B = 0$). The BCD estimated after Eq.\,\hyperref[eq1]{(1)} is found to peak strongly at intermediate doping levels, where it reaches a maximum value of nearly -75~nm. \textbf{e}, Temperature dependence of the nonlinear transverse conductivities $\chi_\textrm{yxx}$ and $\chi_\textrm{xyy}$. The two quantities go to zero as the temperature increases and strontium titanate recovers a higher (non-polar tetragonal) crystal symmetry. Concomitantly, the BCD is forced to vanish by symmetry. \textbf{f}, Temperature dependence of the BCD, $D_\textrm{x}$ (left axis), and corresponding change in sheet conductance $\sigma_\textrm{xx}(T)$ of the 2DES (right axis). Solid and dashed lines in panels \textbf{d} and \textbf{f} are guides to the eye.}
\end{figure*}

The fact that only the symmetric contribution persists even in the zero-field limit (see Fig.\,\hyperref[fig3]{3a,b}) indicates the presence of a finite BCD in the absence of externally applied magnetic fields, and thus of a 
nonlinear Hall effect in time-reversal symmetric conditions. 
To support the existence of a finite BCD with time-reversal symmetry, we have evaluated individually the dipole originating from the spin-sourced BC and the dipole related to the orbital-sourced BC (see the Methods section). Figure \hyperref[fig4]{4a} shows that in all the parameter space of our low-energy theory model, the spin-sourced BCD is two order of magnitudes smaller than the orbital-sourced BCD. The latter exceeds the inverse characteristic Fermi wavenumber $k_\textrm{F}\,^{-1} \sim 0.5$~nm.  
Beside the intrinsic contribution to the BCD, the nonlinear Hall response with time-reversal symmetry also contains disorder-induced contributions~\cite{Du2019,Ortix2021} due to nonlinear skew and side-jumps scattering. 
We experimentally access such contributions by measuring the longitudinal signal $V_\textrm{yyy}^{2\omega}$ that is symmetry-allowed but does not possess any intrinsic BCD contribution. As displayed in Fig.\,\hyperref[fig4]{4b}, the strong difference in amplitude between the longitudinal signal and the transverse $V_\textrm{yxx}^{2\omega}$ signal over a large driving current range proves both the absence of threefold rotation symmetry and a nonlinear Hall effect completely dominated by the intrinsic BCD. 
The anisotropy between longitudinal and transverse nonlinear signals also allows us to exclude a leading role played by thermoelectric effects due to Joule heating (see also inset of Fig.\,\hyperref[fig4]{4b}). 
We further observe that both the longitudinal $V_\textrm{xxx}^{2\omega}$ and transverse $V_\textrm{xyy}^{2\omega}$ responses have an amplitude comparable with the longitudinal signal $V_\textrm{yyy}^{2\omega}$, thus suggesting their disorder-induced nature. We point out that the finite amplitudes of $V_\textrm{xxx}^{2\omega}$ and  $V_\textrm{xyy}^{2\omega}$ 
imply ${\mathcal M}_{{\bar 1} 1 0}$ symmetry breaking (Supplementary Note II). This can be 
related to the mirror breaking arrangements of the oxygen atoms caused by the antiferrodistortive octahedron rotations. It might be also due to the presence of structural domain patterns appearing at the cubic-to-tetragonal structural transition. 

We have systematically verified the occurrence of a sizeable nonlinear transverse response over the full range of sheet conductances, 
and concomitantly observed a large difference between the two nonlinear transverse conductivity tensor component $\chi_\textrm{yxx}$ and $\chi_\textrm{xyy}$, as seen in Fig.\,\hyperref[fig4]{4c}. This further proves a main intrinsic BCD contribution to the nonlinear Hall response. 
By further evaluating the momentum relaxation time, $\tau$ (Supplementary Note II), we can estimate the size of the Berry curvature dipole using (see Methods section): 
\begin{equation}
    D_\mathrm{x} = \frac{ 2 \hbar^2}{e^3 \tau} \,\chi_\textrm{yxx}\,. 
\label{eq1}
\end{equation}
The resulting BCD (see Fig.\,\hyperref[fig4]{4d}) is two order of magnitudes larger than the dipole observed in systems with massive Dirac fermions, such as bilayer WTe$_2$~\cite{Ma2019,Xu2018} and -- over a finite density range -- a factor two larger than the dipole observed in corrugated bilayer graphene~\cite{Ho2021}. We attribute the large magnitude of the effect to the fact 
that the orbital-sourced BC is naturally equipped with a large dipolar denisty due to the presence of singular pinch points and hot-spots with dipolar arrangements. 
We also monitored the temperature dependence of the transverse conductivity tensor components $\chi_\textrm{yxx}$ and $\chi_\textrm{xyy}$ (see Fig.\,\hyperref[fig4]{4e}) and the corresponding behavior of the BCD $D_{\textrm{x}}$ (see Fig.\,\hyperref[fig4]{4f}). All these quantities rapidly drop approaching 30~K, \textit{i.e.} the temperature above which the strong polar quantum fluctuations of SrTiO$_3$ vanish. This further establishes the orbital Rashba coupling
as the physical mechanism behind the orbital-sourced BC. 

The pure orbital-based mechanism of BCD featured here paves the way to the atomic scale design of quantum sources of nonlinear electrodynamics persisting up to room-temperature. Oxide-based 2DES could be for instance combined with a room-temperature polar ferroelectric layer, triggering symmetry lowering and 
thus inducing
orbital Rashba coupling, by interfacial design. This and other alternative platforms combining a low symmetric crystal with orbital degrees of freedom and polar modes, including room-temperature polar metals~\cite{Kim2016} and conducting ferroelectric domain walls, are candidate oxide architectures to perform operations such as rectification~\cite{Zhang2021} and frequency mixing. Moreover, multiple sources of Berry curvature can be implemented for combined optoelectronic and spintronic functionalities in a single material system: photogalvanic currents due to the orbital-sourced BC can be employed to create spin-Hall voltages exploiting the spin-sourced BC. 
Our study also establishes a general approach to generate topological charge distributions in strongly correlated materials, opening a vast space for exploration at the intersection between topology and correlations.

\section*{Methods}
\subsection*{\textbf{Sample growth.}}

The 9 unit cells (u.c.) thick LaAlO$_3$ crystalline layer is grown on the TiO-rich surface of a (111)$-$oriented SrTiO$_3$ substrate, from the ablation of a high purity (> 99.9\%)  LaAlO$_3$ sintered target by pulsed laser deposition (PLD), using a KrF excimer laser (wavelength 248~nm). We perform real-time monitoring of the growth by following the intensity oscillations, in a layer-by-layer growth mode, of the first diffraction spot using reflection high-energy electron diffraction (RHEED), as shown in Extended Data Fig.\,\hyperref[extfig7]{7a}. This allows us to stop the growth at precisely the critical thickness of 9~unit cells of LaAlO$_3$~\cite{Herranz2012} necessary for the (111)$-$LaAlO$_3$/SrTiO$_3$ 2DES to form. The SrTiO$_3$(111) substrate was first heated to 700$^{\circ}$C in an oxygen partial pressure of 6\,x\,10$^{\medminus5}$~mbar. The LaAlO$_3$ layer was grown in those conditions at a laser fluence of 1.2~J\,cm$^{\medminus2}$ and at a laser repetition rate of 1~Hz. Following the growth of the LaAlO$_3$ layer, the temperature is ramped down to 500°C before performing a one hour long \textit{in situ} annealing in a static background pressure of 300~mbar of pure oxygen, in order to recover the oxygen stoichiometry of the reduced heterostructure. Finally the sample is cooled down at -20$^{\circ}$C min$^{\medminus1}$, and kept in the same oxygen environment at zero heating power for at least 45 minutes.

\subsection*{\textbf{Devices fabrication.}}

The (111)$-$LaAlO$_{3}$/SrTiO$_{3}$ blanket films were lithographically patterned into two Hall bars (with dimensions: $W=40$~$\mu$m, $L=180$~$\mu$m), oriented along the two orthogonal crystal axis directions $[\bar{1}10]$ and $[\bar{1}\bar{1}2]$. The Hall bars are defined by electron beam lithography into a PMMA resist, which is used as a hard mask for Argon ion milling (see see Extended Data Fig.\,\hyperref[extfig7]{7c}). The dry etching duration is calibrated and timed to be stopped precisely when the LaAlO$_{3}$ layer is fully removed, in order to avoid the creation of an oxygen-deficient conducting SrTiO$_{3-\delta}$ surface. This leaves an insulating SrTiO$_3$ matrix surrounding the protected LaAlO$_{3}$/SrTiO$_{3}$ areas, which 
host a geometrically confined 2DES.

\subsection*{\textbf{Electrical transport measurements.}} 

The Hall bars are connected to a chip carrier by ultrasonic wedge-bonding technique whereby the Aluminum wires form ohmic contacts to the 2DES through the LaAlO$_{3}$ overlayer. The sample is anchored to the chip carrier by homogeneously coating the back-side of the SrTiO$_3$ substrate with silver paint. A d.c. voltage $V_\textrm{g}$ is sourced between the silver back-electrode and the desired Hall bar device to enable electrostatic field-effect gating of the 2DES, leveraging the large dielectric permittivity of strontium titanate at low-$T$ ($\approx2$\,x\,10$^{4}$ below 10~K) ~\cite{Muller1979,Thiel2006}. Non-hysteretic dependence of $\sigma_\textrm{xx}$ ($\sigma_\textrm{yy}$) on $V_\textrm{g}$ is achieved following an initial gate-forming procedure~\cite{Biscaras2014}.

Standard four-terminal electrical (magneto-)transport measurements were performed at 1.5~K in a liquid Helium-4 flow cryostat, equipped with a superconducting magnet (maximum magnetic field $B=\pm$12~T). An a.c. excitation current $I^{\omega} \propto  |I^{\omega}|\cdot \sin(\omega t)$, of frequency $\omega/(2\pi)=17.77$~Hz, is sourced along the desired crystallographic direction. The sheet resistance, $R_\textrm{s}=\sigma_\textrm{xx}^{\minus 1}$, of a Hall bar device is related to the first harmonic longitudinal voltage drop $V_\textrm{xx}$ according to:  $R_\textrm{s}=\left( V_\textrm{xx}/I_\textrm{x} \right) \left(W/L \right)$. When the a.c. current is sourced along $\hat{\textbf{\textit{x}}} \parallel [\bar{1}10]$  ($\hat{\textbf{\textit{y}}} \parallel [\bar{1}\bar{1}2]$, respectively), we make use of a standard lock-in detection technique to concomitantly measure the first harmonic longitudinal response $V_\textrm{xx}$ ($V_\textrm{yy}$), and either the in-phase first-harmonic $V_\textrm{xy}^\omega$ ($V_\textrm{yx}^\omega$) or out-of-phase second harmonic $V^{2\omega}_\textrm{yxx}$ ($V^{2\omega}_\textrm{xyy}$) transverse voltages (see Fig.\,\hyperref[fig1]{1a}). We define the first and second harmonic transverse resistances as $R_\textrm{xy}^\omega = V_\textrm{xy}^\omega/|I^\omega_\textrm{x}|$ and $R_\textrm{y}^{2\omega} = V_\textrm{yxx}^{2\omega}/|I^\omega_\textrm{x}|^2$, respectively. First and second harmonic measurements are performed at 10~$\mu$A and 50~$\mu$A respectively. 

We systematically decompose both the first and second harmonic magneto-responses into their field-symmetric, $\smash{R^{(2)\omega}_\textrm{sym}}$, and field-antisymmetric, $\smash{R^{(2)\omega}_\textrm{as}}$, contributions according to:
\begin{subequations}
\begin{alignat}{2}
R^{(2)\omega}_{\textrm{sym}} = \left[R^{(2)\omega}(B) + R^{(2)\omega}(\minus B) \right] /2 \,,\\
R^{(2)\omega}_{\textrm{as}} = \left[R^{(2)\omega}(B) - R^{(2)\omega}(\minus B) \right] /2 \,.
\end{alignat}
\end{subequations}
In particular, the first harmonic transverse resistance is purely field-antisymmetric, hence we chose the simplified notation $R_\textrm{xy} \equiv R^\omega_\textrm{xy,as}$. 

\subsection*{\textbf{Estimation of the Rashba spin-orbit energy from magnetoconductance measurements in the weak antilocalization regime.}}

In a 2DES, in the presence of a spin relaxation mechanism induced by an additional spin-orbit interaction, 
the conductance is subject at low temperature to weak localization corrections. Extended Data Fig.\,\hyperref[extfig4]{4\textbf{a}} shows the gate-modulated magnetoconductance curves of the 2DES, which exhibit a characteristic low field weak antilocalization (WAL) behaviour. The magnetoconductance curves, normalized to the quantum of conductance $\textrm{G}_\textrm{Q} = e^2/\left(\pi \hbar \right)$, are fitted using a Hikami-Larkin-Nagaoka (HLN) model that expresses the change of conductivity $\Delta \sigma\left(B_{\perp}\right)=\sigma\left(B_\perp\right)-\sigma(0)$ of the 2DES under an external out-of-plane magnetic field $B_\perp$, in the diffusive regime (with negligible Zeeman splitting), as~\cite{HLN1980,MF1981}:
 
\begin{align}
\begin{split}
\frac{\Delta \sigma(B_{\perp})}{G_\textrm{Q}}&{} = -\frac{1}{2}\Psi\left( \frac{1}{2}+\frac{B_\textrm{i}}{B_\perp} \right)+\frac{1}{2} \ln\left(\frac{B_\textrm{i}}{B} \right)\\
&+\Psi\left(\frac{1}{2}+\frac{B_\textrm{i}+B_\textrm{so}}{B_\perp} \right)- \ln\left(\frac{B_\textrm{i}+B_\textrm{so}}{B_\perp}\right)\\
&+\frac{1}{2}\Psi\left(\frac{1}{2}+\frac{B_\textrm{i}+2B_\textrm{so}}{B_\perp} \right)-\frac{1}{2} \ln\left(\frac{B_\textrm{i}+2B_\textrm{so}}{B_\perp} \right)\\
&-A_\textrm{k} \frac{\sigma(0)}{G_\textrm{Q}} B_\perp^2\,.
\end{split}
\label{eq:HLN}
\end{align}
where $\Psi$ is the digamma function, $B_\textrm{i,so}=\hbar/\left(4eD\tau_\textrm{i,so}\right)$ are the effective fields related to the inelastic and spin-orbit relaxation times ($\tau_\textrm{i}$ and $\tau_\textrm{so}$ respectively), with $D=\pi \hbar^2\sigma(0)/(e^2 m^*)$ the diffusion constant. The last term in Eq.\,\hyperref[eq:HLN]{(3)}, proportional to $B_\perp^2$, contains $A_\textrm{k}$, the so-called Kohler coefficient, which accounts for orbital magnetoconductance.

Hence, from the fit to the WAL magnetoconductance curves, the effective Rashba spin-orbit coupling $\alpha_\textrm{R}$ can be calculated as:
\begin{equation}
    \alpha_\textrm{R}=\hbar^2/\left[2m^*\sqrt{\left(D\tau_\textrm{so}\right)}\right]\,,
\end{equation}
based on a D'yakonov-Perel' spin relaxation mechanism~\cite{MF1981}. A summary of the dependence of the extracted parameters on the 2DES' sheet conductance is plotted in Extended Data Fig.\,\hyperref[extfig5]{5b}.
The spin-orbit energy $\Delta_\textrm{so}$ can then be estimated according to:
\begin{equation}
    \Delta_\textrm{so}=2\alpha_\textrm{R}k_\textrm{F}\,,
\end{equation}
where, in 2 dimensions, the Fermi wavevector is given by $k_\textrm{F}=\sqrt{2\pi n_\textrm{2D}}$, assuming a circular Fermi surface. The sheet carrier density $n_\textrm{2D}$ is experimentally obtained for each doping value from the (ordinary) Hall effect (Supplementary Note III), measured concomitantly with the magnetoconductance traces.

\subsection*{\textbf{Spin-sourced and orbital-sourced Berry curvature dipole calculations.}}
We first estimate the Berry curvature dipole due to spin-sources in time-reversal symmetric condition as a function of carrier density considering the low-energy Hamiltonian for a single Kramers' related pair of bands (Supplementary Note I): 
\begin{equation}
\mathcal{H}=\frac{\textbf{\textit{k}}^2}{2m(\textbf{{\textit{k}})}} - \alpha_\textrm{R}\, \pmb{\sigma}\cdot \textbf{\textit{k}}\times\hat{\mathbf{z}} + \frac{\lambda}{2}(k_{+}^3+k_{-}^3)\sigma_\textrm{z} \,.
\label{eq:fullH}
\end{equation}
where the momentum dependent mass can be negative close to $\Gamma$ point (Supplementary Note I).
Although this model Hamiltonian is equipped with a finite Berry curvature, its dipole is forced to vanish by the threefold rotation symmetry (Supplementary Note I). 
We capture the rotation symmetry breaking of the low-temperature structure at the leading order by assuming inequivalent coefficients for the spin-orbit coupling terms linear in momentum. In other words we make the substitution $\alpha_\textrm{R} \left( \sigma_\textrm{x} k_\textrm{y} - \sigma_\textrm{y} k_\textrm{x} \right) \rightarrow v_\textrm{y} k_\textrm{y} \sigma_\textrm{x} - v_\textrm{x} k_\textrm{x} \sigma_\textrm{y}$. 
Since the dipole is a pseudo-vector, the residual mirror symmetry $\mathcal{M}_{x}$ forces it to be directed along the $\hat{\textbf{\textit{x}}}$ direction. In the relaxation time approximation it is given by
\begin{equation}
D_{x} = \int_\textbf{\textit{k}} \partial_{k_{x}}\Omega_\mathrm{z}(\textbf{\textit{k}})
\end{equation}
with $\Omega_\textrm{z}$ the Berry curvature of our two-band model which we write in dimensionless form by measuring energies in units of $k_\textrm{F}^2/2m(k_\textrm{F})$, lengths in units of $1/k_\textrm{F}$ and densities in units of $n_0=k_\textrm{F}^2/2\pi$. Here, $k_F$ is a reference Fermi wavevector.
For simplicity, we have considered a positive momentum-independent effective mass.
For the Berry curvature dipole shown in Fig.\,\hyperref[fig4]{4a} the remaining parameters have been chosen as $v_\textrm{x}=0.4$, $v_\textrm{y}=(1.2,1.4,1.6) \times v_\textrm{x}$ and $\lambda=0.1$. Moreover, we account for the orbital degeneracy by tripling the dipole of a single Kramers' pair. This gives an upper bound for the spin-sourced BCD. 

We have also evaluated the BCD due to orbital sources considering the low-energy Hamiltonian for spin-orbit free $t_{2g}$ electrons derived from symmetry principles (Supplementary Note I) and reading
\begin{eqnarray}
\label{eq:Ham}
{\mathcal H}({\bf k})&=& \dfrac{\hbar^2 {\bf k}^2}{2 m} \Lambda_0 + \Delta \left(\Lambda_3 + \dfrac{1}{\sqrt{3}}  \Lambda_8 \right) + \Delta_\textrm{m} \left(\dfrac{1}{2} \Lambda_3 -\dfrac{\sqrt{3}}{2} \Lambda_8 \right) \nonumber \\ & & - \alpha_\textrm{OR} \left[k_\textrm{x} \Lambda_5 + k_\textrm{y} \Lambda_2 \right] - \alpha_\textrm{m} k_\textrm{x} \Lambda_7,
\end{eqnarray}
where we introduced the Gell-Mann matrices
\begin{eqnarray*}
 \Lambda_{2}=\begin{pmatrix}0 & -i & 0\\
i & 0 & 0\\
0 & 0 & 0
\end{pmatrix}& \hspace{.2cm}, \hspace{.2cm}  &
\Lambda_{3}=\begin{pmatrix}1 & 0 & 0\\
0 & -1 & 0\\
0 & 0 & 0
\end{pmatrix} \\
\Lambda_{5}=\begin{pmatrix}0 & 0 & -i\\
0 & 0 & 0\\
i & 0 & 0
\end{pmatrix} & \hspace{.2cm}, \hspace{.2cm}  & \Lambda_{7}=\begin{pmatrix}0 & 0 & 0\\
0 & 0 & -i\\
0 & i & 0
\end{pmatrix} \\
\quad\Lambda_{8}=\begin{pmatrix}\tfrac{1}{\sqrt{3}} & 0 & 0\\
0 & \tfrac{1}{\sqrt{3}} & 0\\
0 & 0 & \tfrac{-2}{\sqrt{3}}
\end{pmatrix}. & & 
\end{eqnarray*}
and $\Lambda_0$ the identity matrix.
In the Hamiltonian above, $\Delta$ is the splitting between the $a_{1g}$ singlet and the $e_g^{\prime}$ doublet resulting from the $t_{2g}$ orbitals in a trigonal crystal field. $\Delta_\textrm{m}$ is the additional splitting between the doublet caused by the rotational symmetry breaking. Finally, $\alpha_\textrm{OR}$ and $\alpha_\textrm{m}$ are the strength of the orbital Rashba coupling. Note that in the presence of a threefold rotation symmetry $\alpha_\textrm{m} \equiv 0$, in which case the Berry curvature is forced to vanish. 
For simplicity, we have evaluated the Berry curvature for the ${\mathcal C}_s$ point group-symmetric case assuming $\alpha_\textrm{m} \equiv \alpha_\textrm{OR}$. In our continuum $SU(3)$ model, the BC can be computed using the method outlined in Ref.~\cite{bar12}. We have subsequently computed the corresponding dipole measuring, as before, energies in units of $k_\textrm{F}^2/2m$, lengths in units of $1/k_\textrm{F}$ and densities in units of
$n_0=k_\textrm{F}^2/2\pi$. 
The dimensionless orbital Rashba coupling has been varied between $\alpha_\textrm{OR}=1$ and $\alpha_\textrm{OR}=2$ whereas we have fixed $\Delta=-0.1$ and $\Delta_\textrm{m}=0.005$. 
The value of the crystal field splitting $\Delta$ is consistent with the amplitude determined by x-ray absorption spectrosopy~\cite{del18} 
of the order 8 meV, and therefore almost one order of magnitude smaller than our energy unit $\simeq 40$meV for a reference $k_F^{-1} \simeq 0.5$~nm and an effective mass $m \simeq 3 m_e$ (see Supplementary Note III).
The calculated dipole displayed in Fig.\,\hyperref[fig4]{4a} has been finally multiplied by two to account for spin degeneracy. 
We remark that, as shown in the Supplementary Information (Supplementary Note I), both the model Hamiltonian for the spin sources of Berry curvature Eq.~\hyperref[eq:fullH]{(6)} and the model Hamiltonian for the orbital sources Eq.~\hyperref[eq:Ham]{(8)} derive from a single six-band model where orbital and spin degree of freedom are treated on an equal footing.

\subsection*{\textbf{Estimation of the Berry curvature dipole magnitude from nonlinear Hall measurements.}}

The nonlinear current density is mathematically given by $j_\alpha^{2\omega}=\chi_{\alpha\beta\gamma}\,E_\beta\, E_\gamma$, where $\chi_{\alpha\beta\gamma}$ is the nonlinear transverse conductivity tensor. When an a.c. current density $I^\omega_\textrm{x}/W = \sigma_\textrm{xx} E^\omega_\textrm{x}$ is sourced along $\hat{\textbf{\textit{x}}}$, the transverse second harmonic current density developing along $\hat{\textbf{\textit{y}}}$ is related to the BC dipole $\textbf{\textit{D}}$ according to~\cite{Sodemann2015}:
\begin{equation}
\textbf{\textit{j}}_\mathrm{y}^{2\omega}~=~\frac{e^3\tau}{2\hbar^2(1+i\omega\tau)}\left(\hat{\mathbf{z}} \times \textbf{\textit{E}}_\mathrm{x}^\omega \right)\left(\textbf{\textit{D}} \cdot \textbf{\textit{E}}_\mathrm{x}^\omega \right) \,,
\label{eq:j2omega}
\end{equation}
where $\tau$ is the momentum relaxation time, and $e$ the elementary charge. 
Due to the mirror symmetry $\mathcal{M}_\textrm{x} \equiv \mathcal{M}_{[\bar{1} 1 0]}$, the dipole is found to point along $\hat{\textbf{\textit{x}}}$, and in the quasi-d.c. limit, \textit{i.e.} $(\omega \tau) \ll 1$, the BC dipole expression reduces to:
\begin{equation}
    D_\mathrm{x} = \frac{2\hbar^2}{e^3\tau} \frac{j^{2\omega}_\mathrm{y}}{\left(E^{\omega}_\mathrm{x}\right)^2}= \frac{2\hbar^2}{e^3 \tau}\frac{V^{2\omega}_\textrm{yxx}\,\sigma_{xx}^3\,W}{|I_\textrm{x}^\omega|^2},
\label{eq:BCDexplicit}
\end{equation}
which is the explicit expression for Eq.\,(\hyperref[eq1]{1}), in terms of experimentally measurable quantities only, and where:
\begin{subequations}
\begin{align}
    \chi_\textrm{yxx}=\frac{j^{2\omega}_\textrm{y}}{\left(E_\textrm{x}^\omega\right)^2}, \\
    \chi_\textrm{xyy}=\frac{j^{2\omega}_\textrm{x}}{\left(E_\textrm{y}^\omega\right)^2}
\end{align}
\label{eq:chi_abb}
\end{subequations}
are the measured nonlinear transverse conductivity tensor elements shown in Fig.\,\hyperref[fig4]{4c,e}.

\section*{Acknowledgements}

This work was supported by the Swiss State Secretariat for Education, Research and Innovation (SERI) under contract number MB22.00071, the Gordon and Betty Moore Foundation (Grant No. 332 GBMF10451 to A. D. C.), the European Research Council (ERC, Grants No. 677458), by the project Quantox of QuantERA ERA-NET Cofund in Quantum Technologies, and by the Netherlands Organisation for Scientific Research (NWO/OCW) as part of the VIDI [project 680-47-543 to C.O., project 016.Vidi.189.061 to A.D.C.] and Frontiers of Nanoscience program (NanoFront). E.L. acknowledges funding from the EU Horizon 2020 research and innovation programme under the Marie Sk\l{}odowska-Curie grant agreement no. 707404. The authors acknowledge A.~M.~Monteiro, L.~Hendl, J.~R.~Hortensius, P.~Bruneel, and M.~Gabay for valuable discussions.

\section*{Author contributions}
\noindent A.D.C. proposed and supervised the experiments. C.O. proposed the theory models and supervised their analysis. E.L. grew the crystalline LaAlO$_3$ thin films by PLD and performed the structural characterizations. E.L. and Y.G.S. lithographically patterned the samples, performed the magnetotransport experiments, and analysed the experimental data, with help from T.C.v.T. and U.F. R.B. and M.T.M. performed the Berry curvature and semiclassical transport calculations with help from M.C., C.N. and C.O. 
C.O., E.L., R.B. and A.D.C. wrote the manuscript, with input from all authors.
\newline

\section*{Competing interests}
\noindent The authors declare no competing interests.
\newline

\section*{Corresponding authors} 
\noindent edouard.lesne@gmail.com \\
cortix@unisa.it \\
andrea.caviglia@unige.ch

\clearpage

\renewcommand{\figurename}{Extended Data Fig.}

\renewcommand{\thefigure}{1}

\begin{figure*}
\centering
\includegraphics[width=8cm,scale=1]{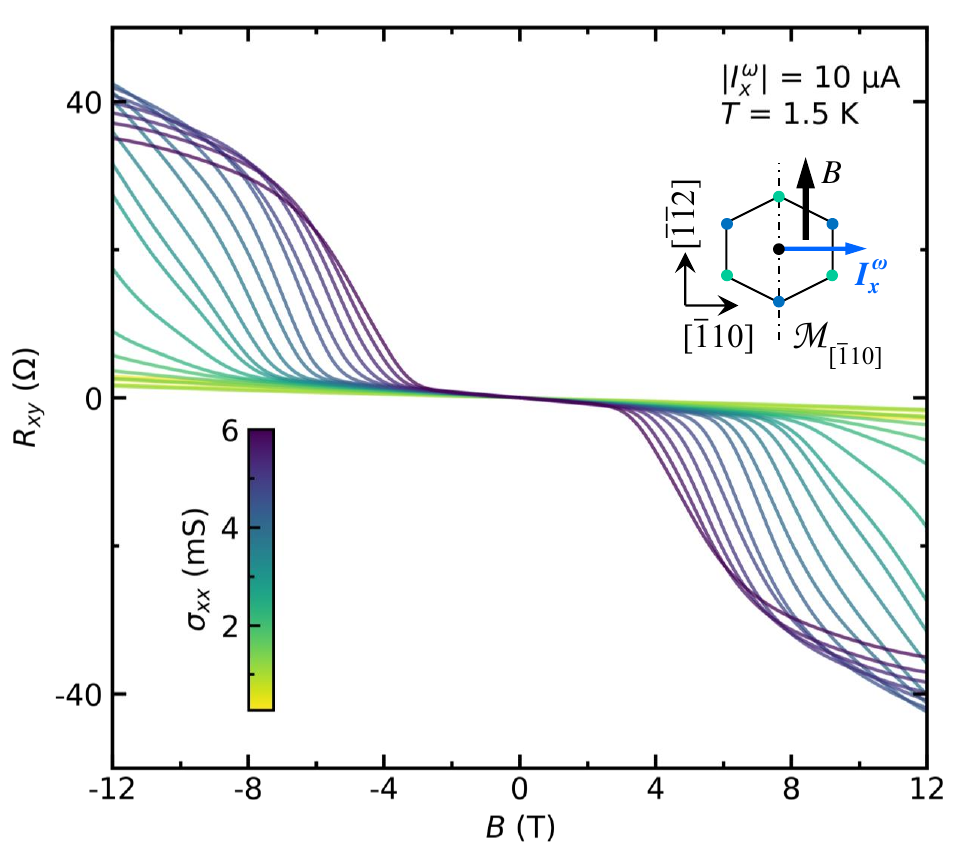}
    \caption{\textbf{Hall effect in a planar magnetic field.} Experimentally measured field-antisymmetric transverse magnetoresistance $R_\textrm{yx}=V_\textrm{yx}^{\omega}/I_\textrm{y}^{\omega}$ at $T=1.5$~K, with $I_\textrm{x}^{\omega}$ along $[\bar{1}10] \perp B$ (see inset schematic), for varying sheet conductance values $\sigma_\textrm{xx}$ (indicated by the inset colored scale bar), and tuned via electrostatic field effect in a back-gate geometry.
    }
\label{extfig1}
\end{figure*}

\renewcommand{\thefigure}{2}

\begin{figure*}
\centering
\includegraphics[width=8cm,scale=1]{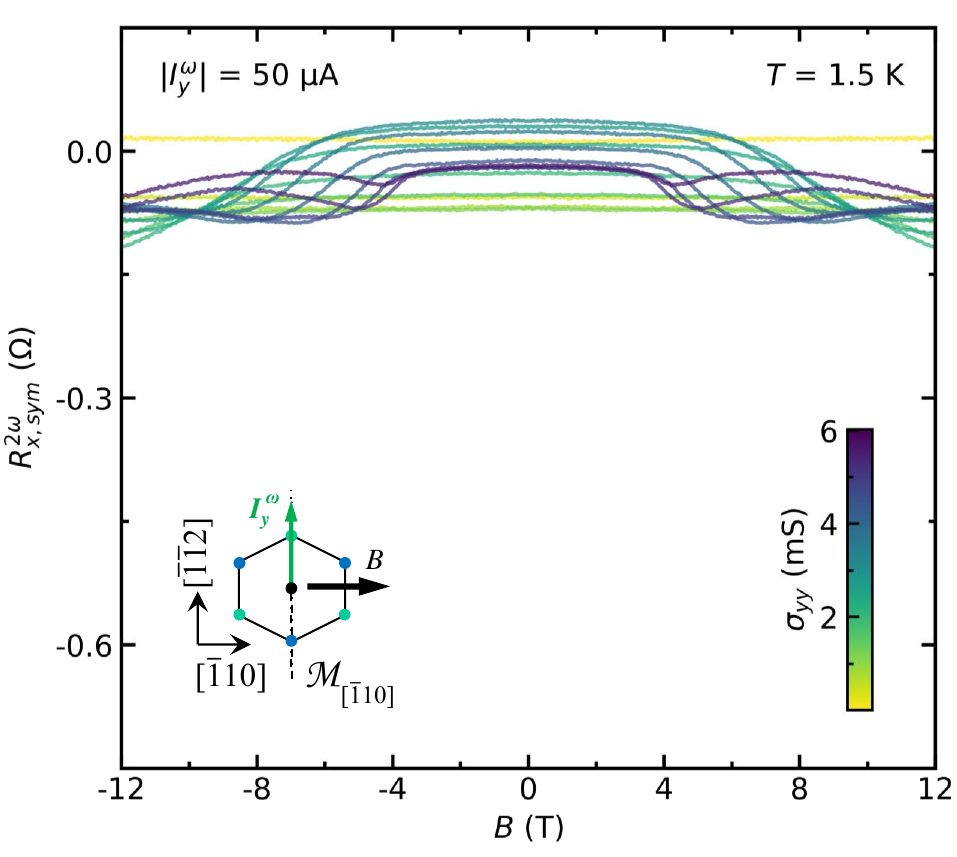}
    \caption{\textbf{Nonlinear Hall response in a planar magnetic field.} Experimentally measured field-symmetric nonlinear transverse magnetoresponse $R_\textrm{x}^{2\omega}$ at $T=1.5$~K with $I_\textrm{y}^{\omega}$ along $[\bar{1}\bar{1}2] \perp B$ (see inset schematic), for varying sheet conductance values $\sigma_\textrm{yy}$ (indicated by the inset colored scale bar). The full scale ordinate axis is chosen to be the same as for panel \textbf{b} of Fig.\,\hyperref[fig3]{3}, for better comparison, and highlights the comparatively small nonlinear Hall response when the current is sourced along the ${\mathcal M}_{{\bar 1} 1 0}$ mirror line corresponding to a symmetry demanded zero BCD. 
    }
\label{extfig2}
\end{figure*}

\renewcommand{\thefigure}{3}

\begin{figure*}
\centering
\includegraphics[width=15cm,scale=1]{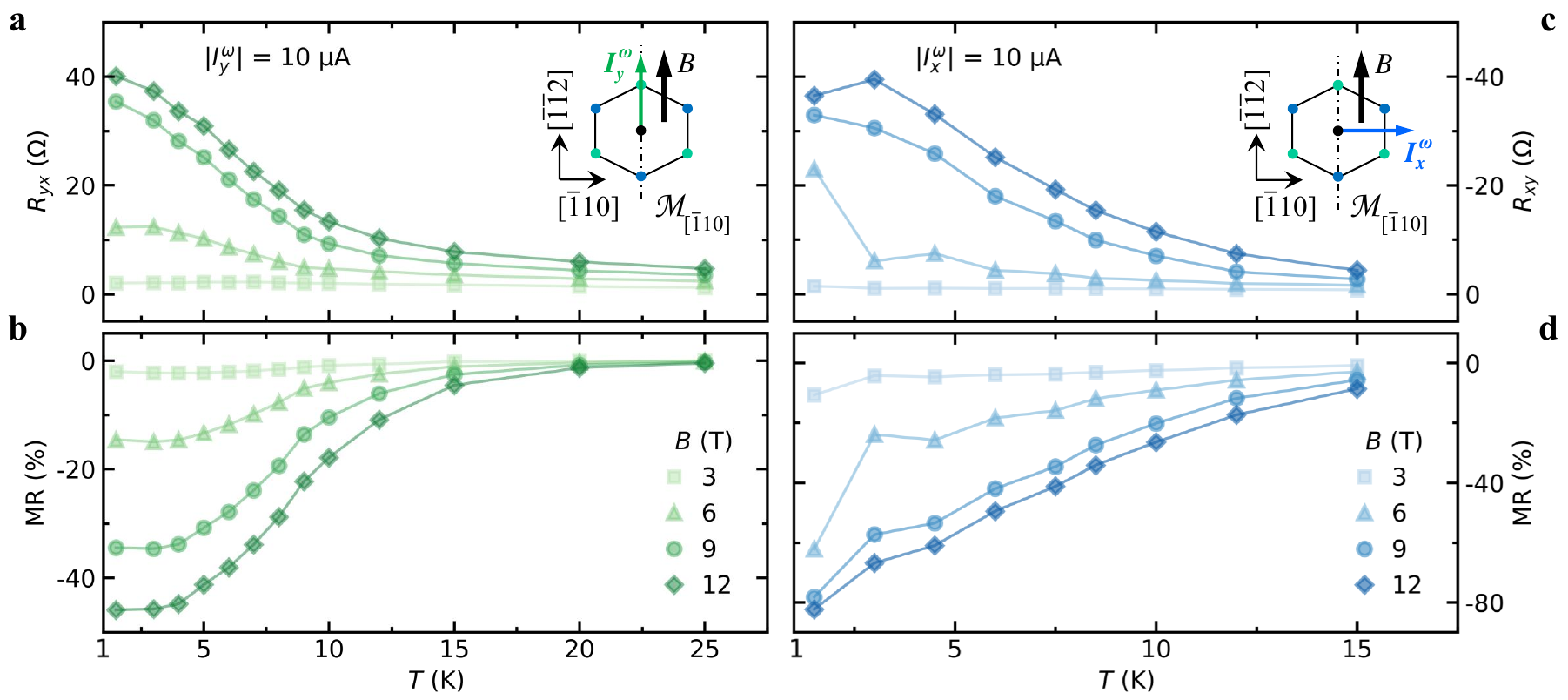}
    \caption{\textbf{Temperature dependence of the Hall and longitudinal magnetoresistances in a planar magnetic field.} Temperature dependent planar Hall resistance  (\textbf{a},\textbf{c}) and longitudinal MR (\textbf{b},\textbf{d}) in the linear response regime for various strength of the in-plane magnetic field (and at fixed gate-voltage), with $I_\textrm{y}^{\omega}$ along $[\bar{1}\bar{1}2] \parallel B$ (\textbf{a-b}) and $I_\textrm{x}^{\omega}$ along $[\bar{1}10] \perp B$ (\textbf{c-d}). Both $R_\textrm{yx}$ and $R_\textrm{xy}$, as well as the corresponding longitudinal MR, asymptotically go to zero as the temperature increases toward strontium titanate's nonpolar tetragonal phase (above $\approx 30$~K).
    }
\label{extfig3}
\end{figure*}

\renewcommand{\thefigure}{4}

\begin{figure*}
\centering
\includegraphics[width=16cm,scale=1]{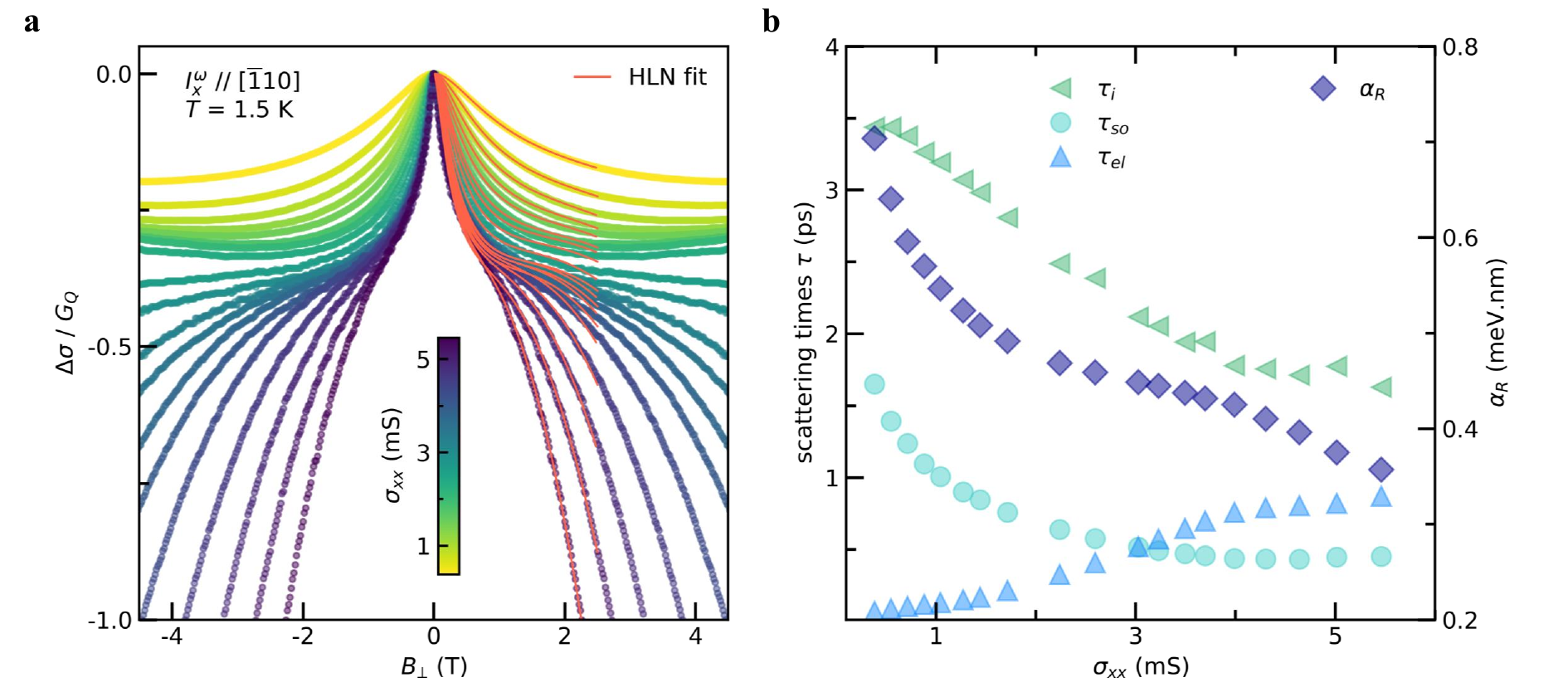}
    \caption{\textbf{Weak antilocalization regime and Rashba spin-orbit coupling.} \textbf{a}, Gate-modulated magnetoconductance curves (normalized to the quantum of conductance $\textrm{G}_\textrm{Q}$) for $I^\omega_\textrm{x}$ along $[\bar{1}10]$ (see also Supplementary Note III for WAL measurements along $[\bar{1}\bar{1}2]$). Hikami-Larkin-Nagaoka fits (solid red lines) are performed following Eq.\,(\hyperref[eq:HLN]{3}). $B_\perp$, the out-of-plane magnetic field. \textbf{(c)} Left axis: Experimentally estimated momentum, inelastic and spin-orbit relaxation times, $\tau_\textrm{el}$ (named $\tau$ throughout the manuscript), $\tau_\textrm{i}$ and $\tau_\textrm{so}$, respectively. Right axis: Strength of the Rashba spin-orbit coupling $\alpha_\textrm{R}$ versus sheet conductance $\sigma_\textrm{xx}$.
    }
\label{extfig4}
\end{figure*}

\renewcommand{\thefigure}{5}

\begin{figure*}
\centering
\includegraphics[width=9cm,scale=1]{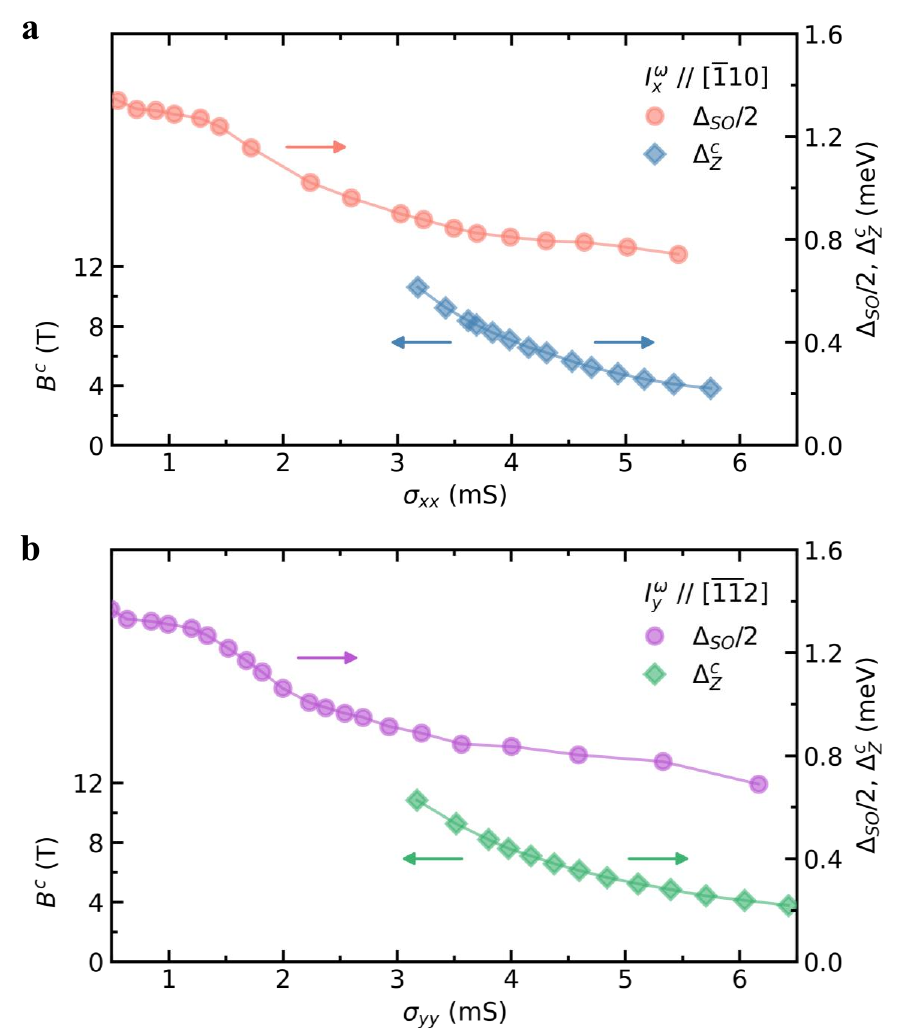}
    \caption{\textbf{Effective Zeeman energy and Rashba spin-orbit energy of the 2DES.} \textbf{a}-\textbf{b}, Right axis: Sheet conductance dependence of the Rashba spin-orbit energy $\Delta_\textrm{so}/2=\alpha_\textrm{R}k_\textrm{F}$ and effective Zeeman energy $\Delta_\textrm{Z}^\textrm{c}=g\mu_\textrm{B}B^\textrm{c}/2$ at the critical in-plane field $B^\textrm{c}$ (left axis), for $I_\textrm{x}^{\omega}$ along $[\bar{1}10]$ (panel \textbf{a}) and for $I_\textrm{y}^{\omega}$ along $[\bar{1}\bar{1}2]$ (panel \textbf{b}). See Supplementary Note III for details regarding the determination of $B^\textrm{c}$.}
\label{extfig5}
\end{figure*}

\renewcommand{\thefigure}{6}

\begin{figure*}
\centering
\includegraphics[width=16cm,scale=1]{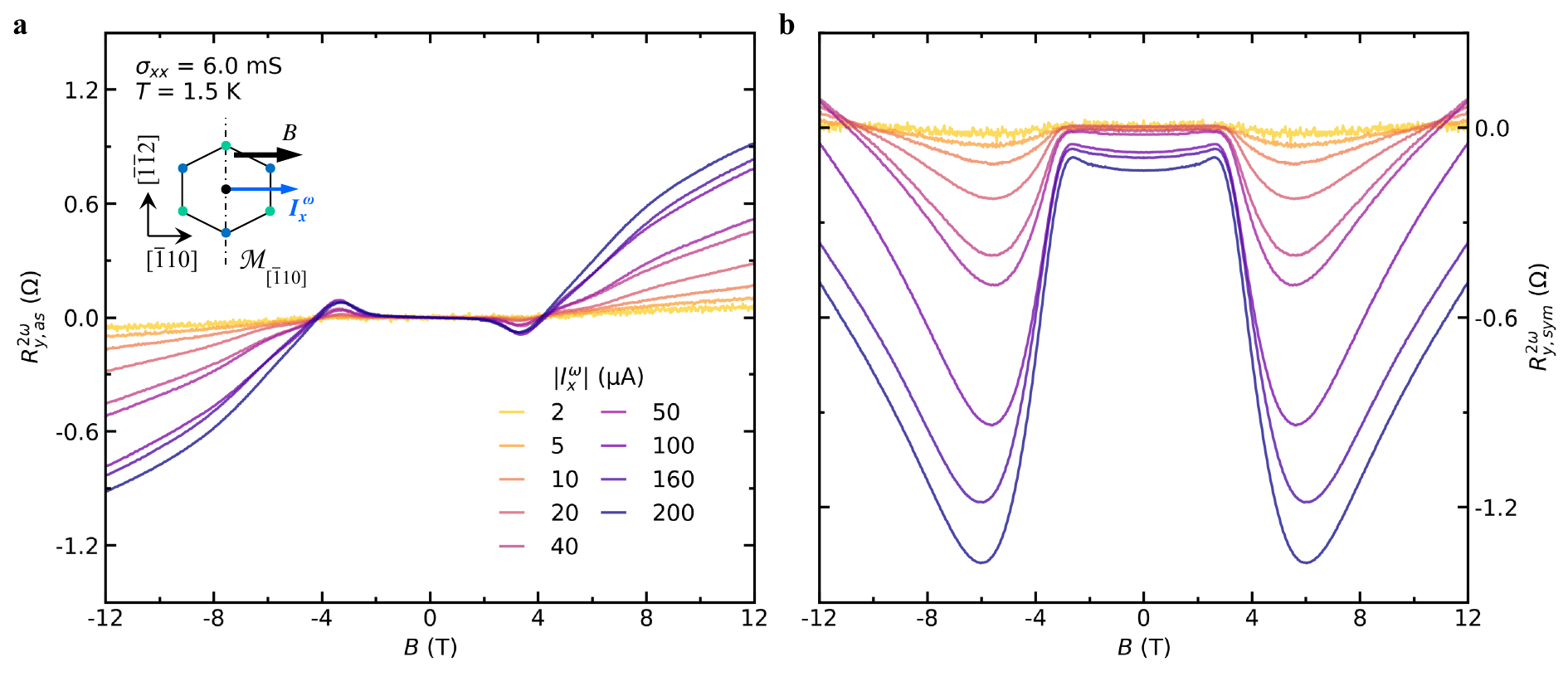}
    \caption{\textbf{Current bias dependent nonlinear transverse signal in a planar magnetic field.} \textbf{a},\textbf{b}, Field-antisymmetric, $R_\textrm{y,as}^{2\omega}$, and field-symmetric $R_\textrm{y,sym}^{2\omega}$ second harmonic transverse resistance responses, for various magnitudes of the excitation a.c. current along $[\bar{1}10] (\parallel B$, see inset schematic).
    }
\label{extfig6}
\end{figure*}

\renewcommand{\thefigure}{7}

\begin{figure*}
\centering
\includegraphics[width=14cm,scale=1]{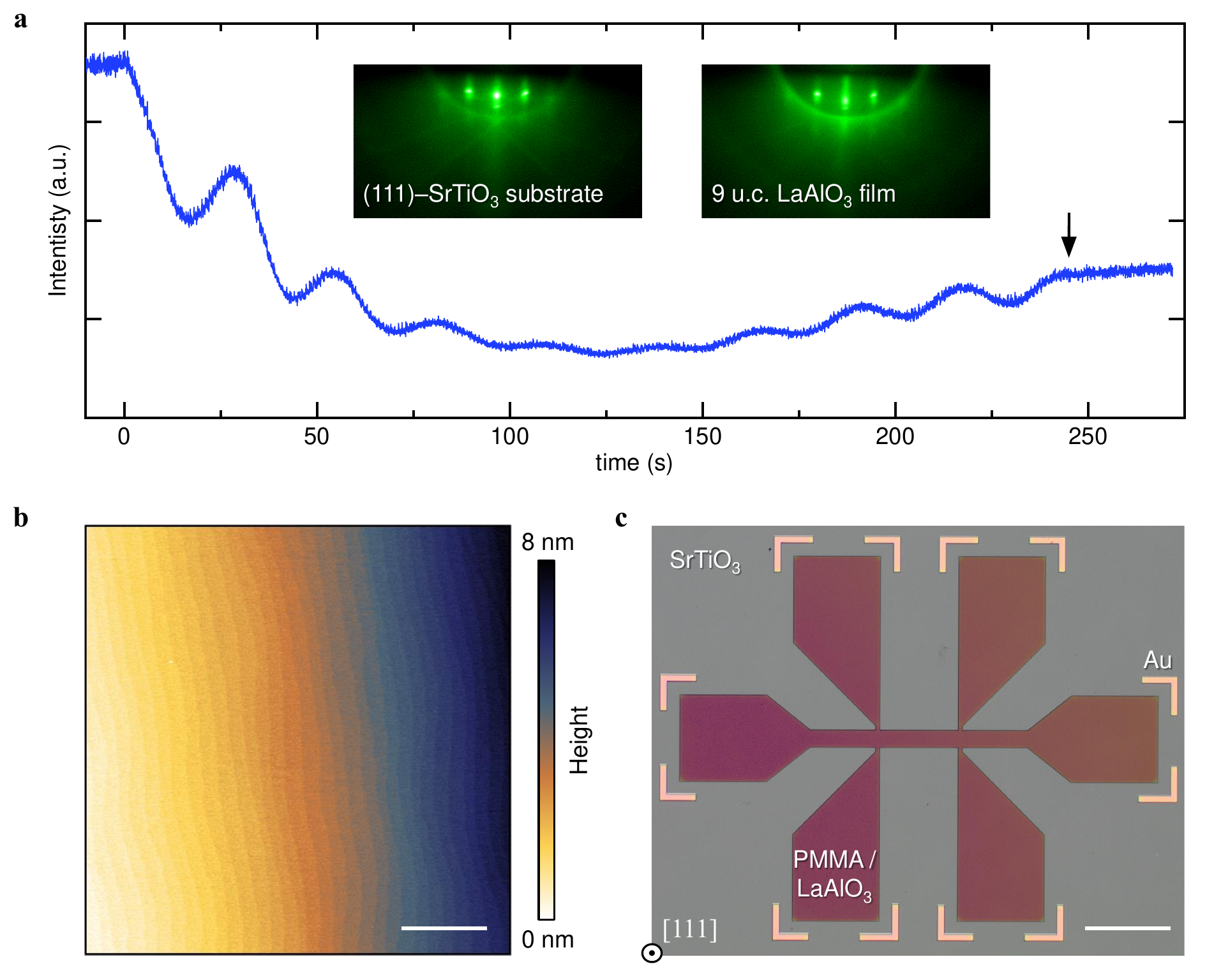}
    \caption{\textbf{Growth, structural characterization and device fabrication of (111)$-$LaAlO$_3$/SrTiO$_3$ 2DES.} \textbf{a}, \textit{In situ} real-time RHEED monitoring of the layer-by-layer PLD growth of a 9 u.c. LaAlO$_3$ film on a (111)$-$oriented SrTiO$_3$ substrate. Insets: RHEED patterns acquired prior to (left) and following the film growth (right), highlighting the high crystalline quality of the epitaxial LaAlO$_3$ film. The vertical arrow marks the end of the growth (at 245~s). \textbf{b}, Atomic force microscopy image of a 9~u.c. thick LaAlO$_3$ film on SrTiO$_3$(111). The film's topography reproduces the characteristic atomically sharp steps-and-terraces structure of the substrate's vicinal surface. Horizontal scale bar: 1~$\mu$m. \textbf{c}, Optical micrograph of a patterned PPMA resist hard mask on a 9 u.c. blanket film of LaAlO$_3$/SrTiO$_3$ (prior to the Ar ion milling step) which ultimately defines the Hall bar device area where the 2DES subsists. Pre-patterned gold markers are used to signify the position of the Hall bars' contact pads, as the optical contrast between the etched area (bare substrate) and the crystalline LaAlO$_3$/SrTiO$_3$ devices is extremely low. Scale bar: 200~$\mu$m.
    }
\label{extfig7}
\end{figure*}

\end{document}


\renewcommand{\figurename}{Fig. S}

\title{Supplementary Information: \\ Designing spin and orbital sources of Berry curvature at oxide interfaces}

\author{Edouard~Lesne}
    \affiliation{Kavli Institute of Nanoscience, Delft University of Technology,  Lorentzweg 1, 2628CJ Delft, Netherlands.}
\author{Yildiz~G.~Sa\v{g}lam}
    \affiliation{Kavli Institute of Nanoscience, Delft University of Technology,  Lorentzweg 1, 2628CJ Delft, Netherlands.}
\author{Raffaele~Battilomo}
    \affiliation{Institute for Theoretical Physics, Center for Extreme Matter and Emergent Phenomena, Utrecht University, Princetonplein 5, 3584 CC Utrecht, Netherlands.}
 \author{Maria~Teresa~Mercaldo}
    \affiliation{Dipartimento di Fisica “E.~R.~Caianiello”, Universitá di Salerno, IT-84084 Fisciano, Italy.}
\author{Thierry~C.~van~Thiel}
    \affiliation{Kavli Institute of Nanoscience, Delft University of Technology,  Lorentzweg 1, 2628CJ Delft, Netherlands.}
\author{Ulderico~Filippozzi}
    \affiliation{Kavli Institute of Nanoscience, Delft University of Technology,  Lorentzweg 1, 2628CJ Delft, Netherlands.}
\author{Canio~Noce}
     \affiliation{Dipartimento di Fisica “E.~R.~Caianiello”, Universitá di Salerno, IT-84084 Fisciano, Italy.}
\author{Mario~Cuoco}
    \affiliation{Consiglio Nazionale delle Ricerche, CNR-SPIN, Italy}
    \affiliation{Dipartimento di Fisica “E.~R.~Caianiello”, Universitá di Salerno, IT-84084 Fisciano, Italy.}
\author{Gary~A.~Steele}
    \affiliation{Kavli Institute of Nanoscience, Delft University of Technology,  Lorentzweg 1, 2628CJ Delft, Netherlands.}
\author{Carmine~Ortix}
    \affiliation{Institute for Theoretical Physics, Center for Extreme Matter and Emergent Phenomena, Utrecht University, Princetonplein 5, 3584 CC Utrecht, Netherlands.}
    \affiliation{Dipartimento di Fisica “E.~R.~Caianiello”, Universitá di Salerno, IT-84084 Fisciano, Italy.}
\author{Andrea~D.~Caviglia}
    \affiliation{Department of Quantum Matter Physics, University of Geneva, 24 Quai Ernest Ansermet, CH-1211 Geneva, Switzerland}

\setlength{\abovedisplayskip}{3pt}
\setlength{\belowdisplayskip}{3pt}

\maketitle

\section{Supplementary Note I: Berry curvature at trigonal oxide interfaces}
\label{Snote:SI}

To derive the electronic properties of the two-dimensional system realized at the [111]$-$LaAlO$_3$/SrTiO$_3$ interface in the high-temperature trigonal phase, we first discuss the orbital and symmetry character of the electronic levels at the center of the Brillouin zone. Let us first neglect spin-orbit coupling. Due to the trigonal coordination, the three $t_{2g}$ Ti orbitals are split into an $a_{1g}$ orbital representing a real one-dimensional irreducible representation of the ${\mathcal C}_{3v}$ point group, and an $e_g^{\prime}$ doublet forming a real two-dimensional irreducible representation of the group. The $e_{g}$ Ti orbitals are not affected by the trigonal crystal field as they also form a doublet. Let us now include the effect of spin-orbit coupling breaking the $SU(2)$ spin rotation symmetry. 
The two spin-orbit coupled states originating from the $a_{1g}$ orbital have a symmetry-protected degeneracy since they form the $\Gamma_4$ two-dimensional irreducible representation of the ${\mathcal C}_{3v}$ double-point group.
Spin-orbit coupling instead splits the quartet of states originating from the $e_g^{\prime}$ doublets. Specifically, two spin-orbit coupled orbitals are singly degenerate and form the $\Gamma_5$ and the $\Gamma_6$ one-dimensional irreducible representations of the double point group. The remaining two states are instead degenerate and form a $\Gamma_4$ representation. Since the $\Gamma_5$ and $\Gamma_6$ representations are complex, time-reversal symmetry implies that these states must stick together thus forming a Kramers' doublet. The irreducible two-dimensional $\Gamma_4$ representation is instead quaternionic and therefore is already equipped with time-reversal invariance. 
 As a result, we have that all levels at the BZ center correspond to an effective spin-$\frac{1}{2}$ Kramers' doublet. 
 
The minimal model Hamiltonian close to each of these Kramers' doublets can be derived in a ${\bf k \cdot p}$ expansion accounting for all symmetry-allowed terms. To do so, we note that in the surface Kramers' doublet basis the time-reversal symmetry can be represented as $\mathcal{T}=i\sigma_y \mathcal{K}$ with ${\mathcal K}$ the complex conjugation. The mirror symmetry is instead represented by $\mathcal{M}=i\sigma_x$. Note that from here onwards $\hat{x}$ will indicate the $\left[\bar{1} 1 0 \right]$ direction. In the basis $\ket{\psi^{\uparrow\downarrow}}$ the threefold rotation operator takes the form $\mathcal{C}_3=e^{-i\sigma_z\pi/3}$. Under the operation of $\mathcal{C}_3$ and $\mathcal{M}$, momentum and spin transform as follows, 
\begin{align*}
&\mathcal{C}_3: & k_\pm &\rightarrow e^{\pm i 2\pi/3}k_\pm,& \sigma_\pm &\rightarrow e^{\pm i 2\pi/3}\sigma_\pm, & \sigma_z&\rightarrow\sigma_z\\
&\mathcal{M}: & k_+ &\rightarrow -k_-& \sigma_x &\rightarrow \sigma_x, & \sigma_{y,z}&\rightarrow-\sigma_{y,z}\\
\end{align*}
where $k_\pm=k_x\pm i k_y$ and $\sigma_\pm=\sigma_x\pm i \sigma_y$. The Hamiltonian must also be invariant under time reversal which adds the constraint $\mathcal{H}(\mathbf{k})=\mathcal{T}\mathcal{H}(\mathbf{-k})\mathcal{T}^{-1}=\sigma_y\mathcal{H}^*(\mathbf{-k})\sigma_y$.

At linear order in the momentum $\mathbf{k}$, and including an effective regularizing quadratic term, the minimal two-band Hamiltonian for a Kramers' related pair of bands reads:
\begin{equation}
\mathcal{H}_\textrm{R}(\mathbf{k})=\frac{\mathbf{ k}^2}{2m} \sigma_0-\alpha_\textrm{R}\, \pmb{\sigma}\cdot \mathbf{k}\times\hat{\mathbf{z}}\,,
\label{eq:Hrashba}
\end{equation}

\noindent where $\pmb{\sigma}$ is a vector of Pauli matrices, $\sigma_0$ is the identity matrix,$\alpha_\textrm{R}$ is the ``Rashba" spin-orbit coupling strength, while $m$ is the effective electron mass. 
Note that nothing prevents the effective electron mass to be negative. In this case, electron-like transport can be ensured by adding a term quartic in momentum ${\mathbf k}^4 / (2 m_1) ~ \sigma_0$ with $m_1>0$. Overall, this amounts to consider a momentum dependent effective electron mass.
The Hamiltonian in Eq.~(\ref{eq:Hrashba}) does not capture crystalline anisotropy effects. In addition, the Berry curvature associated to this minimal model is zero, since there is no term proportional to $\sigma_z$. However, higher order momentum terms change this situation. 
The first symmetry allowed term accounting for crystalline anisotropy is third order in momentum and takes the form,

\begin{equation}
\mathcal{H}_w(\mathbf{k})=\frac{\lambda}{2}(k_+^3+k_-^3)\sigma_z \,.
\label{eq:Hwarping}
\end{equation}
This warping Hamiltonian is proportional to the Pauli matrix $\sigma_z$, which is crucial to obtain a non-zero Berry curvature and leads to out-of-plane spin textures. Note that since the full Hamiltonian is invariant under the mirror symmetry $\mathcal{M}$, $\mathcal{H}_w(\mathbf{k})$ is forced to vanish along the mirror line.

To show the presence of a finite Berry curvature induced by warping, we 
recall that in a two band model the Berry curvature can be calculated by rewriting the full Hamiltonian as 
$\mathcal{H}(\mathbf{k})=\mathbf{d}(\mathbf{k})\cdot\boldsymbol{\sigma}+ 
 \mathbf{ k}^2 \sigma_0 / [2 m(\mathbf{ k})]$, 
where ${\mathbf d}$ is a momentum dependent vector, which for our specific model has components ${\mathbf d}=\left\{-k_y, k_x, \lambda \left(k_{+}^3 + k_{-}^3 \right)/2 \right\}$.
The $\mathbf{d}$ vector is independent of terms $\propto \sigma_0$, and thus of the momentum dependent effective electron mass.
The expression for the Berry curvature is then given by $\Omega^\pm_z(\mathbf{k})=\pm\hat{\mathbf{d}}\cdot (\partial_{k_x}\hat{\mathbf{d}}\times\partial_{k_y}\hat{\mathbf{d} })/2$ with $\hat{\mathbf{d} } = \mathbf{d} / \left|\mathbf{d} \right|$. For our minimal model Hamiltonian in the presence of trigonal symmetry, we have 
\begin{equation}
\Omega_\pm^{z}(k,\theta)=\pm\frac{2 \sqrt{2} \lambda   \alpha_\textrm{R}^2 k^3 \cos (3 \theta )}{\left[2\alpha_\textrm{R}^2 k^2+\lambda ^2 k^6 \cos (6 \theta )+\lambda ^2 k^6\right]^{3/2}}\,, 
\label{eq:OmegaB}
\end{equation}
where $\theta$ is the polar angle in momentum space. 

The Berry curvature is well defined in each point except the origin where the bands are degenerate. 
Note that the constraints set by time reversal symmetry and the three-fold rotational symmetry are satisfied as can be verified upon a closer inspection of Eq.~(\ref{eq:OmegaB}). Moreover $\Omega_\pm^z(k,\theta)$ vanishes along the mirror lines, in accordance with Eq.~(\ref{eq:Hwarping}). 

\begin{figure}[h!]
\centering
\includegraphics[width=\textwidth,scale=1]{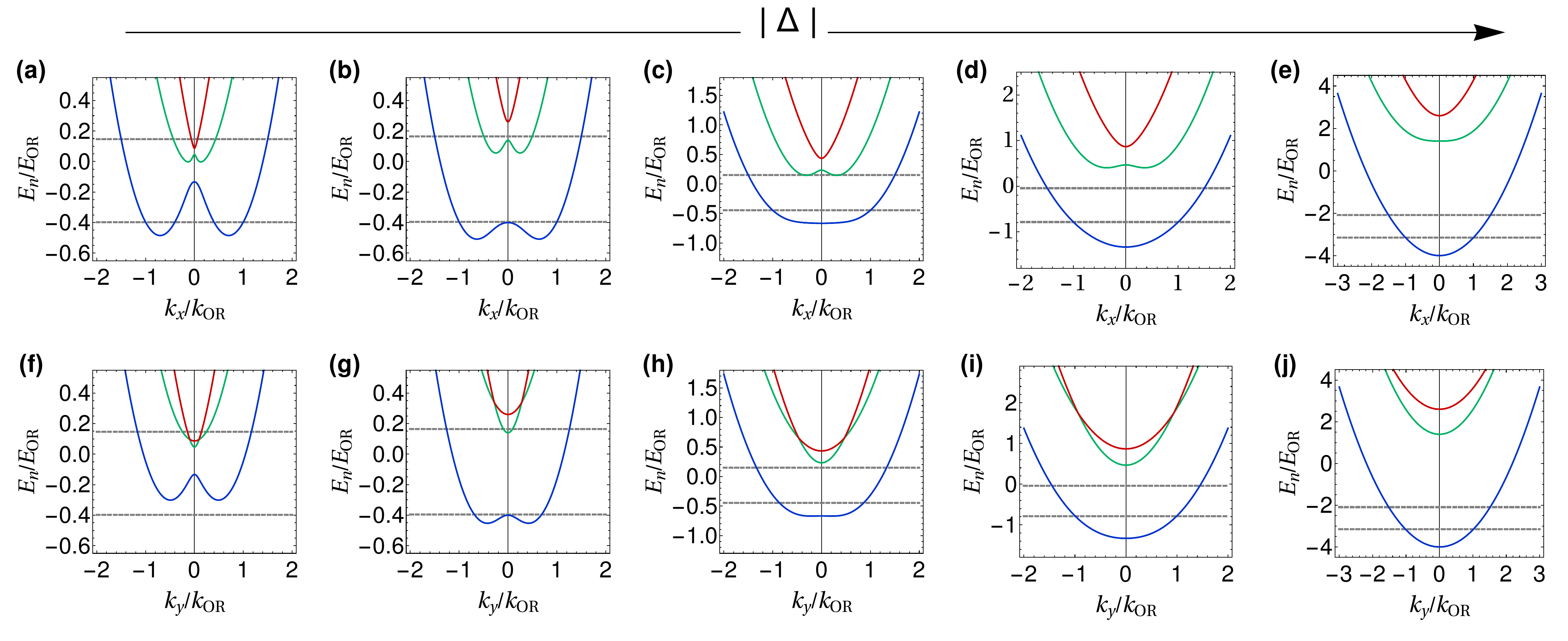}
\caption{{\bf Dependence of the orbital-sourced model on crystal field splittings.} Evolution of the energy bands on the $k_y=0$ line (top panels) and the $k_x=0$ line (bottom panels) for the spin-orbit-free orbital-sourced model by increasing the ratio between the crystal field splitting $\Delta$ and the characteristic orbital Rashba energy $E_\textrm{OR}=2 m \alpha_\textrm{OR}^2$. The panels correspond to a) $\Delta / E_\textrm{OR}=-0.1$, (b) $\Delta / E_\textrm{OR}=-0.3$, (c) $\Delta / E_\textrm{OR}=-0.5  $, (d) $\Delta / E_\textrm{OR}=-1.0 $, and (e) $\Delta / E_\textrm{OR} =-3.0 $. Same values are used for the sequence (f)-(j). The crystal field splitting $\Delta_m=0.2 \left|\Delta \right|$ and we have chosen the orbital Rashba parameter $\alpha_m=\alpha_\textrm{OR}$. All energies are measured in units of $E_\textrm{OR}$ whereas momenta are measured in units of the orbital Rashba momentum $k_\textrm{OR}=2 m \alpha_\textrm{OR} $. The gray lines are  upper and lower bounds for the Fermi energy.}
\label{fig:R1}
\end{figure}

The orbital sources of Berry curvature can be instead derived by explicitly considering interorbital mixing terms. To this end, it is convenient to neglect spin-orbit coupling. The effective Hamiltonian for the 
$a_{1g}$ and $e_g^{\prime}$ doublet in the trigonal crystal field can be derived using symmetry principles. 
Specifically, any generic $3 \times 3$ Hamiltonian can be expanded in terms of the nine Gell-Mann matrices $\Lambda_i$  as 
\begin{equation}
{\mathcal H}({\bf k})=\sum_{i=0}^8  b_i({\bf k}) \Lambda_i\,.
\end{equation}
The invariance of the Hamiltonian requires that the components of the Hamiltonian vector ${\bf b}({\bf k})$ should have the same behavior as the corresponding Gell-Mann matrices $\Lambda_i$. This means that they should belong to the same representation of the crystal point group. The representation of the Gell-Mann matrices $\Lambda_i$ and those of the polynomials of ${\bf k}$ can be found using that the generators of the point group for spinless electrons can be written as
\begin{equation*}
{\mathcal M}_x=\left( \begin{array}{ccc} 1 & 0 & 0 \\ 0 & 1 & 0 \\ 0 & 0 & -1 \end{array} \right) ; \hspace{.3cm} {\mathcal C}_3 = \left( \begin{array}{ccc} 1 & 0 & 0 \\  0 & \cos{\frac{2\pi}{3}} & \sin{\frac{2 \pi}{3}} \\ 0 & -\sin{\frac{2 \pi}{3}} & \cos{\frac{2 \pi}{3}} \end{array}
\right). 
\end{equation*}
With this, the effective Hamiltonian up to linear order in momentum reads 
\begin{equation}
\label{eq:H0}
{\mathcal H}({\bf k})=\dfrac{{\bf k}^2}{2 m} \Lambda_0 + \Delta \left(\Lambda_3 + \dfrac{1}{\sqrt{3}} \Lambda_8 \right) - \alpha_\textrm{OR} \left[k_x \Lambda_5 + k_y \Lambda_2 \right], 
\end{equation}
where $\Lambda_0$ is the identity matrix and the other Gell-Mann matrices are reported in the main text. The last term in the equation above corresponds to the orbital Rashba coupling with strength $\alpha_\textrm{OR}$. The second term gives the crystal field splitting of size $\Delta$ between the $a_{1g}$ singlet and the $e_g^{\prime}$ doublet.
As discussed in the main text, a direct computation of the Berry curvature using the method outlined in Ref.~\citenum{bar12si} yields a vanishing Berry curvature at all momenta. 
This changes by considering a crystalline symmetry lowering to ${\mathcal C}_s$. From the representation of Gell-Mann matrices and polynomials of momentum, the effective Hamiltonian now takes the form 
\begin{equation}
\label{eq:Ham}
{\mathcal H}_\textrm{OR}({\bf k})=\dfrac{{\bf k}^2}{2 m} \Lambda_0  + \Delta \left(\Lambda_3 + \dfrac{1}{\sqrt{3}}  \Lambda_8 \right) + \Delta_m \left(\dfrac{1}{2} \Lambda_3 -\dfrac{\sqrt{3}}{2} \Lambda_8 \right) \nonumber  - \alpha_\textrm{OR} \left[k_x \Lambda_5 + k_y \Lambda_2 \right] - \alpha_m k_x \Lambda_7 \,.
\end{equation}
In the equation above, the crystal field splitting $\Delta_m$ yields an energy separation between the $e_g^\prime$ states at the center of the BZ. 
This Hamiltonian is characterized by Berry curvature distributions with dipolar hot spots and singular pinch points. Furthermore, it is naturally equipped with a large non-vanishing Berry curvature dipole. 
The dispersion of the bands is strongly dependent on the ratio between the characteristic orbital Rashba energy $E_\textrm{OR}= 2 m \alpha_\textrm{OR}^2$ and the crystal field splittings. In particular, the effective mass of the lowest band is negative close to the $\Gamma$ point of the BZ whenever the orbital Rashba energy is comparable to the crystal-field splitting $\Delta$. In this regime, an intraband Lifshitz transition occurs when increasing the chemical potential. On the contrary, in the opposite regime of large crystal field splittings, all bands possess an effective electron mass that is positive at all momenta, and consequently intraband Lifshitz transitions do not take place. This is explicitly shown in Fig.~\hyperref[fig:R1]{S1} where we report the evolution of the bands by continuously increasing the $\Delta / E_\textrm{OR}$ ratio. Note that similar features are also found within the remaining doublet of bands. In this case the momentum dependence of the effective mass is controlled by the ratio between the crystal field splitting $\Delta_m$ and the effective orbital Rashba energy $E_\textrm{OR}= 2 m \alpha_m^2 $, and can acquire a non-trivial dependence along the $k_y=0$ line.

In Fig.~\hyperref[fig:R1]{S1} we also indicate the relevant range of energies in the band structure, which we evaluate using the following argument. 
To obtain a Berry curvature dipole of order $k_F$, the orbital Rashba parameter has to be comparable to $k_F/(2 m)$. This is explicitly shown in Fig.~3a of our main manuscript --  the Berry curvature dipole has a sweet spot and can exceed 
$1/k_F$ when  $2 m \alpha_{OR} \simeq k_F$. 
Hence, the orbital Rashba momentum $k_{OR} \simeq k_F$. 
Considering the finite range of the Berry curvature dipole sweet spot and the additional tuning of the carrier density by gating (see Supplementary Note III), we estimate 
that the Fermi wavevector $k_{OR}<k_F<1.5 k_{OR}$. This consequently gives an upper and lower bounds for the Fermi energy shown as gray lines in Fig.~\hyperref[fig:R1]{S1}. We note that the possibility of occupying more than one band is in line with the observations presented in Ref.~\onlinecite{Monteiro2019si}. Finally we wish to point out that since the orbital Rashba energy corresponds to the characteristic kinetic energy $k_F^2/(2 m)$, which lies in the tens of meV range, all values of the crystal field splitting $\Delta$  considered in Fig.~\hyperref[fig:R1]{S1} could be realized in practice --  x-ray absorption spectroscopy~\cite{del18si} indicates a crystal field splitting of the order $ 8\,$meV. In other words, the material could be on the verge of a change in the sign of the effective mass close to zero momentum.

We next show that both the Hamiltonian characterizing the orbital degrees of freedom and the Hamiltonian in Eq.~(\ref{eq:Hrashba}) for the spin degree of freedom can be derived from a single model that treats orbital and spin degrees of freedom on an equal footing. 
In order to make a link between the spin and orbital degrees of freedom one needs to include the atomic spin-orbit coupling.

Then, the starting point for the coupled spin and orbital degrees of freedom is provided by the following model Hamiltonian:
\begin{eqnarray}
\label{eq:1}
{\mathcal H}_{tot}({\bf k})&=& {\mathcal H}_\textrm{OR}({\bf k}) \otimes \sigma_0 + \lambda_\textrm{so} (L_x \otimes \sigma_x+ L_y \otimes \sigma_y +L_z \otimes \sigma_z),
\end{eqnarray}
\noindent where we have included the atomic spin-orbit coupling with the strength $\lambda_\textrm{so}$ and $\sigma_i$ indicate the spin Pauli matrices with $i=x,y,z$. We recall that the orbital angular momentum projected on the effective subspace with the three selected orbital configurations has the following representation in terms of the Gell-Mann matrices: $L_x=\Lambda_2$, $L_y=\Lambda_5$, and $L_z=\Lambda_7$.

For simplicity, we next consider a  ${\mathcal C}_{3v}$ symmetric (i.e. $\Delta_m=\alpha_m=0$) crystalline environment. 
At the $\Gamma$ point of the BZ, and explicitly including atomic spin-orbit coupling, 
the Hamiltonian reduces to:

\begin{eqnarray}
\label{eq:3}
{\mathcal H}_{tot}^\Gamma= {\mathcal H}_\textrm{OR}^\Gamma \otimes \sigma_0 + \lambda_\textrm{so} (L_x \otimes \sigma_x+ L_y \otimes \sigma_y +L_z \otimes \sigma_z),
\end{eqnarray}
with 
\begin{eqnarray}
\label{eq:4}
{\mathcal H}_\textrm{OR}^\Gamma =\Delta \left(\Lambda_3 + \dfrac{1}{\sqrt{3}} \Lambda_8\right). 
\end{eqnarray}

${\mathcal H}_{tot}^\Gamma$
can be diagonalized and yields the following double degenerate (Kramer pairs) eigenvalues $E_A={\tilde{E_A}}+\frac{4}{3} \Delta$,$E_B={\tilde{E_B}}+\frac{4}{3} \Delta$, $E_C={\tilde{E_C}}+\frac{4}{3} \Delta$ with:
\begin{eqnarray}
\label{eq:5}
{\tilde{E}}_C&=& \frac{1}{2} \left( -2 \Delta -\lambda_\textrm{so} + \sqrt{4 \Delta^2 +4 \Delta \lambda_\textrm{so}+ 9 \lambda_\textrm{so}^2 }\right) \\
{\tilde{E}}_B&=&- 2 \Delta + \lambda_\textrm{so} \\
{\tilde{E}}_A&=& \frac{1}{2} \left( -2 \Delta -\lambda_\textrm{so} - \sqrt{4 \Delta^2 +4 \Delta \lambda_\textrm{so}+ 9 \lambda_\textrm{so}^2 }\right) .
\end{eqnarray}

According to the character of the trigonal splitting and the Hund's rule  (spin-orbit coupling favors the highest total angular momentum at low filling) $\lambda_\textrm{so}$ and $\Delta$ have negative values. Thus, the following energy relation holds $E_A<E_B<E_C$. $E_A$ is the lowest energy in the multiplet.

The eigenvectors can be also written in a compact vectorial form as:
\begin{eqnarray}
\label{eq:6}
|v_{C,1}\rangle&=&\frac{1}{\sqrt{2+c^2}} \left[-c, 0, 0, -i, 0, 1 \right] \\
|v_{C,2}\rangle&=&\frac{1}{\sqrt{2+c^2}} \left[0, c, i, 0, 1, 0 \right] \\
|v_{B,1}\rangle&=&\frac{1}{\sqrt{2}} \left[0, 0, 0, i, 0, 1  \right] \\
|v_{B,2}\rangle&=&\frac{1}{\sqrt{2}} \left[0, 0, -i, 0, 1, 0 \right] \\
|v_{A,1}\rangle&=&\frac{1}{\sqrt{2+a^2}} \left[-a, 0, 0, -i, 0, 1 \right] \\
|v_{A,2}\rangle&=&\frac{1}{\sqrt{2+a^2}} \left[0, a, i, 0, 1, 0 \right],
\end{eqnarray}

with

\begin{eqnarray}
\label{eq:7}
a&=&-\frac{{\tilde{E}}_C}{\lambda_\textrm{so}} \\
c&=&-\frac{{\tilde{E}}_A}{\lambda_\textrm{so}} .
\end{eqnarray}

In the new spin-orbit coupled basis the Hamiltonian ${\mathcal {{H}}}_{tot}$ has the following matrix form:
\begin{eqnarray}
\label{eq:8}
{\mathcal {{H}}}_{tot}=\begin{pmatrix} H_{CC} & H_{CB} & H_{CA}\\
H^{\dagger}_{CB} & H_{BB} & H_{BA} \\
H^{\dagger}_{CA} & H^{\dagger}_{BA} & H_{AA}
\end{pmatrix}
\end{eqnarray}

\noindent with $H_{MN}$ being 2x2 matrices.

Let us now focus on the matrix elements of the block Hamiltonian  $H_{AA}$ associated with the lowest energy doublet assuming that the energy splitting among the levels is larger than the strength occurring in the inter-doublet terms (i.e. $H_{CB}$,$H_{BA}$, $H_{CA}$). 
We emphasize that since the energy splitting between the spin-orbit Kramers' doublets is controlled (at fixed values of the crystal fields) by the spin-orbit coupling strength, the intradoublet hybridization channels $H_{BA},H_{CA},H_{CB}$ can be safely neglected in the large spin-orbit coupling regime $\lambda_\textrm{so} \gg E_\textrm{OR}$. In this case, one can directly determine the effective spin Hamiltonian
by evaluating the following expectation values:
\begin{eqnarray}
\label{eq:9}
H_{AA}(i,j)=\langle v_{A,i}|{\mathcal {{H}}}_{tot}|v_{A,j}\rangle .
\end{eqnarray}

Taking into account the expressions of the eigenvectors $v_{A,i}$ and of the k-dependent terms of the total Hamiltonian, the matrix of the block $H_{AA}$ can be written as:

\begin{eqnarray}
H_{AA}= E_A + \dfrac{ {\bf k}^2}{2 m} + \begin{pmatrix}  \frac{\lambda}{2}  (k_{+}^3+k_{-}^3) & -\alpha_\textrm{R} \left(i k_x -ky \right) \\
-\alpha_\textrm{R} \left(-i k_x -ky \right) & - \frac{\lambda}{2} (k_{+}^3+k_{-}^3) 
\end{pmatrix}
\label{eq:Hspinderived}
\end{eqnarray}

\noindent with the effective intra-doublet Rashba interaction given by 
\begin{eqnarray}
\alpha_\textrm{R}=\frac{2 \alpha_\textrm{OR} \lambda_\textrm{so}}{\sqrt{4 \Delta^2 +4 \Delta \lambda_\textrm{so}+ 9 \lambda_\textrm{so}^2 }}\,.
\end{eqnarray}
Note that in the Hamiltonian above the trigonal warping in the spin sector is found by equipping the orbital model Hamiltonian with a symmetry-allowed cubic term $\propto \gamma (k_{+}^3+k_{-}^3) \Lambda_7 $. Then 
the warping coupling $\lambda$ has the following expression
\begin{eqnarray}
\lambda=\gamma \frac{16 \lambda_\textrm{so}^2}{8 \lambda_\textrm{so}^2+{\tilde{E}}_A^2}\,.
\end{eqnarray}

\begin{figure}
\centering
\includegraphics[width=\textwidth,scale=1]{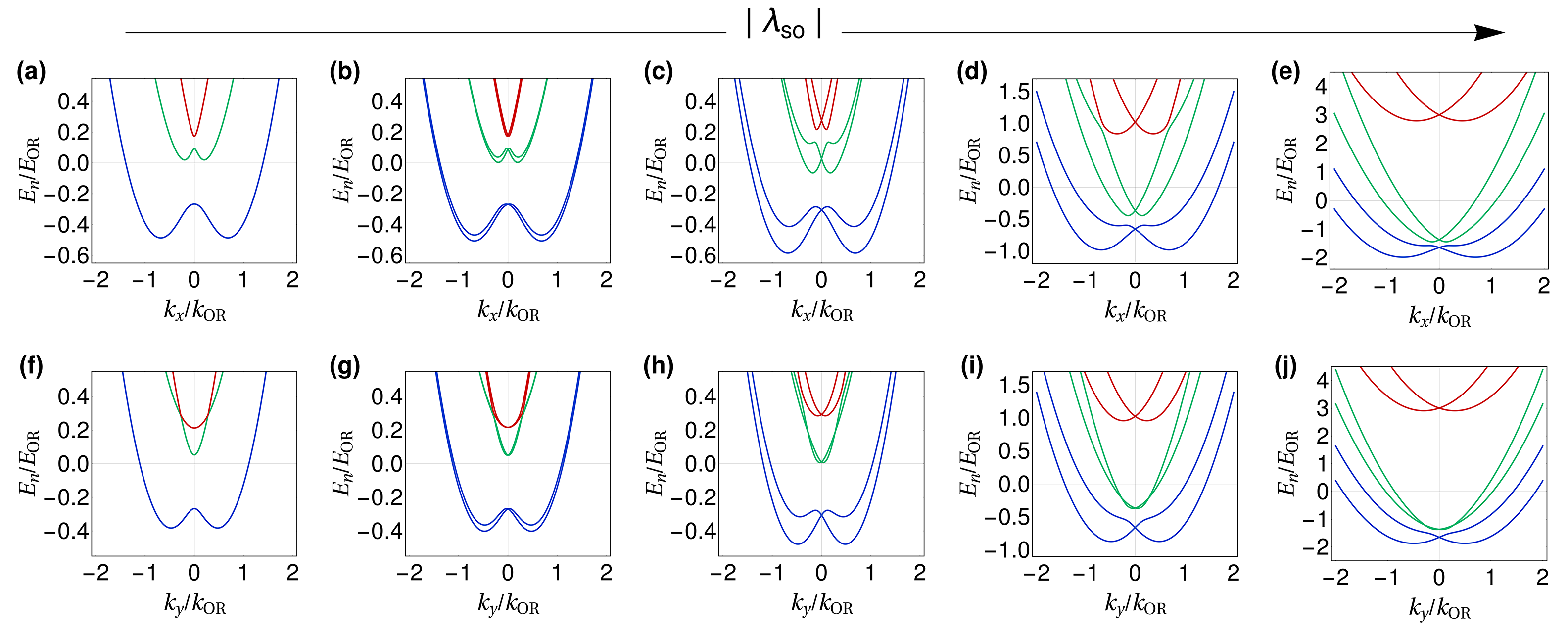}
\caption{{\bf Dependence of the energy bands on spin-orbit coupling.} Evolution of the energy bands on the $k_y=0$ line (top panels) and the $k_x=0$ line (bottom panels) for the full model Eq.~(\ref{eq:1}) by increasing the ratio between the spin-orbit coupling strength $\lambda_\textrm{so}$
and the characteristic orbital Rashba energy $E_\textrm{OR}=2 m \alpha_\textrm{OR}^2 $. The panels correspond to (a) $\lambda_\textrm{so}=0$, (b) $\lambda_\textrm{so} / E_\textrm{OR}=-0.02$, (c) $\lambda_\textrm{so} / E_\textrm{OR}=-0.1 $, (d) $\lambda_\textrm{so} / E_\textrm{OR}=-0.5 $, and (e) $\lambda_\textrm{so} / E_\textrm{OR}=-1.5 $. Same values are used for the sequence (f)-(j). The crystal field splittings have been chosen as $\Delta=-0.2  E_\textrm{OR}$ and $\Delta_m=0.2|\Delta|$. Finally the orbital Rashba parameter $\alpha_m=\alpha_\textrm{OR}$. All energies are measured in units of $E_\textrm{OR}$ whereas momenta are measured in units of the orbital Rashba momentum $k_\textrm{OR}=2 m \alpha_\textrm{OR}$. }
\label{fig:R2}
\end{figure}

Therefore, the effective low-energy spin model can be directly linked to the orbital model with its sources of Berry curvature dipole. Two remarks are in order here. First, using a L\"owdin procedure one can verify that there exist higher-order corrections, {\it e.g.} momentum dependent renormalization of the Rashba coupling $\alpha_\textrm{R} \rightarrow \alpha_\textrm{R} (1 + \delta {\bf k}^2)$. Second, the bare effective electron mass appearing in Eq.~(\ref{eq:Hspinderived}) is corrected by intraorbital mixing terms due to the finiteness of the spin-orbit coupling strength. This generally results in a momentum dependent mass that can be negative close to zero momentum in agreement with the features of the spin-orbit free orbital bands discussed above.  This is explicitly proven in Fig.~\hyperref[fig:R2]{S2} where we show the evolution of the spin-orbit coupled bands by increasing $\lambda_\textrm{so}$. Spin-orbit coupling generally weakens the momentum dependence of the effective mass eventually leading to a positive effective mass at all momenta. Importantly, and as mentioned above, the symmetry-based spin model can be applied without restrictions even in the presence of a strongly momentum dependent mass. Therefore, the spin sources of Berry curvature are correctly captured even for intermediate spin-orbit coupling strenghts.

In this case, the spin-sourced model can be applied up to momenta for which the corresponding energy of the spin bands is comparable to the splitting $\Delta E_{\Gamma}$ between two different Kramers doublets at the center of the BZ. This gives an upper bound for the Fermi wavevector $k_\textrm{F}=\sqrt{k_\textrm{R}^2/4 + m \Delta E_\Gamma } - k_\textrm{R}/2$ with $k_\textrm{R}=2 m \alpha_\textrm{R}$.

Our experimental observations indicate the occurrence of a strong anomalous planar Hall effect at the same carrier densities at which a sizable non-linear Hall effect at zero magnetic field is measured. The characteristic features of the anomalous planar Hall effect can be explained using the spin-sourced model -- the orbital-sourced Berry curvature is stiff in response to externally applied in-plane magnetic fields.  Instead, the non-linear Hall effect at zero magnetic field cannot be quantitatively captured within this picture, and the orbital sources of Berry curvature are needed. 
Putting these observations together, we can therefore conclude that both orbital Rashba and spin-orbit coupling are relevant ingredients 
to describe the physical properties of the material structure. 
Our experimental tools independently probe the Berry curvature associated with orbital and spin degrees of freedom.

\section{Supplementary Note II: Magnetotransport theory}
\label{Snote:SII}
\addcontentsline{toc}{subsection}{Anomalous planar Hall effect}
\subsection*{Anomalous planar Hall effect}

In this section we discuss the Hall effect induced by the spin-sourced Berry curvature in the presence of a planar magnetic field using a semiclassical Boltzmann framework. 

We first note that the presence of a planar magnetic field generates a Zeeman coupling term in the low-energy Hamiltonian $\mathcal{H}_B=B(\sigma_x\cos\phi+\sigma_y\sin\phi)$, that breaks $\mathcal{C}_3$ when present, $\mathcal{T}$ and the remaining mirror $\mathcal{M}_x$ except when the direction of the magnetic field with respect to the $\hat{x}$ axis $\phi=2\pi m /6$ with $m\in \mathbb{N}$. In this case one mirror symmetry is preserved.  The energy dispersion of the total Hamiltonian is then given by:

\begin{eqnarray}
\varepsilon_{\pm}(\mathbf{k})&=&\frac{\mathbf{k}^2}{2m}+\left[(B \sin (\phi )+k_x \alpha_\textrm{R})^2+(k_y \alpha_\textrm{R}-B \cos (\phi ))^2 \right. \nonumber \\ & & \left. +\lambda ^2 \left(k_x^3-3k_x k_y^2\right)^2 \right]^{1/2}.
\end{eqnarray}

The planar magnetic field breaks the Kramer's degeneracy at the center of the BZ and completely splits the two bands. However, when a mirror is preserved there is a mirror-symmetry protected Dirac point along the mirror line at $(k,\theta)=(|B|/\alpha_\textrm{R},\phi+sgn(B)\pi/2)$.  
The condition $k_F= |B| / \alpha_\textrm{R}$ defines the critical magnetic field strength introduced in the main text. 
Moreover, a Lifshitz transition (a change in topology in the Fermi surface, in this case going from one closed Fermi line to two) appears since the Zeeman coupling shifts the minima of the two bands. The energy for which the upper band is completely depleted is given by $\varepsilon_L=B^2/(2m\alpha_\textrm{R}^2)$ if $B<m\alpha_\textrm{R}^2$ otherwise $\varepsilon_L=B-m\alpha_\textrm{R}^2/2$ if $B>m\alpha_\textrm{R}^2$.

To derive the transport characteristic associated to this low-energy model we assume  to apply an electric field $\mathbf{E}= E \mathbf{\hat{x}}$ along the $\hat{x}$ axis and rotate the magnetic field in the $x-y$ plane. 
Thus we write the magnetic field as $\mathbf{B}=B(\cos\phi \mathbf{\hat{x}}+\sin\phi \mathbf{\hat{y}})$. Generally speaking, the semiclassical equations of motions in the presence of a non-zero Berry curvature read
\begin{align}
&\mathbf{\dot{r}}=D(\mathbf{B},\Omega_k)[\mathbf{v_k}+e(\mathbf{E}\times\Omega_k)+e(\mathbf{v_k}\cdot \Omega_k) \mathbf{B}] \label{eqr}\\
&\mathbf{\dot{k}}=D(\mathbf{B},\Omega_k)[e\mathbf{E}+e(\mathbf{v_k}\times\mathbf{B})+e^2(\mathbf{E}\cdot\mathbf{B}){\Omega_k}]\,, \label{eqk}
\end{align}
where $D(\mathbf{B},\Omega_k)=[1+(e)(\mathbf{B\cdot} \Omega_k)]^{-1}$ , $\Omega_k$ is the Berry curvature and $\mathbf{v_k}$ is the group velocity. Solving the Boltzmann equation for the 
electron distribution function $f(k)$ within the relaxation time approximation allows to compute the charge current $\mathbf{J}=e\int (d^d k/(2\pi)^d)\mathbf{\dot{r}} f(k) D^{-1} $ that accounts for the modified phase space factor $D$. 
In linear response theory the charge current obeys the relation $J_a=\sigma_{ab}E_b$, where $\sigma_{ab}$ are the components of the conductivity tensor and $E_b$ the external electric field. 
Following Ref.~\citenum{Battilomo2021si}, the planar Hall conductivity, after discarding higher order contributions, is given by

\begin{eqnarray}
\sigma_{yx}&=&e^2\int\frac{d^d k}{(2\pi)^d}D\tau(-\frac{\partial f_{eq}}{\partial\epsilon})\mathcal{f}[\mathit{v_y}+eB\sin\theta(\mathbf{v_k \cdot} \Omega_k)] \nonumber \\  & & [ \mathit{v_x} +eB\cos\theta (\mathbf{v_k\cdot}\Omega_k]\mathcal{g}+ \dfrac{e^2}{h} \int \frac{d^2k}{(2\pi^2)}\Omega^z_k f_{eq},
\label{eq1}
\end{eqnarray}
where $f_{eq}$ is the equilibrium Fermi-Dirac distribution. 
Eq.~(\ref{eq1}) contains all transverse linear responses in the presence of coplanar electric and magnetic fields in both two- and three-dimensional systems. 
In three-dimensional topological Dirac and type-I Weyl semimetals, the Berry-curvature induced planar Hall effect stems from the term $\propto B^2 \sin{\theta} \cos{\theta} (\mathbf{v_k\cdot}\Omega_k)^2$. The last term instead represents the usual anomalous Hall contribution 
for magnetic materials. 
In two-dimensional systems, $\mathbf{v_k} \perp \Omega_k$. However, in trigonal crystals a non-vanishing contribution 
to the last term of Eq.~(\ref{eq1}) appears once the Zeeman interaction is explicitly taken into account, and exists provided all mirror symmetries are broken, even though the material is non-magnetic. 
This also implies that, contrary to three-dimensional topological semimetals, the Berry-curvature related planar Hall effect of two-dimensional systems does not explicitly depend on the relative angle between the electric and magnetic field but only on the angle between the planar magnetic field and the principal crystallographic directions. 
This, in turns, allows to directly observe a purely antisymmetric planar Hall effect when the electric and magnetic fields are perfectly aligned since in this configuration the classical contribution of Eq.~(\ref{eq1}), containing the $v_xv_y$ term vanishes. 

We have computed the behavior of the anomalous planar Hall conductivity assuming a magnetic field direction $\phi=\pi/2$. In this configuration all mirror symmetries are broken. Moreover the integral of the Berry curvature is maximum. 
The contribution to the linear transverse response $\sigma_{xy}$ is given by the integral of the Berry curvature over the Fermi surfaces of the two energy bands. Since the two bands contribute with opposite curvatures it is sufficient to integrate over the exclusion region of the two bands. Hence the anomalous contribution to the transverse conductivity, at zero temperature is given by

\begin{equation}
\sigma_{xy} \propto \int_{\mathcal{S}_+}\Omega^+_z(\mathbf{k}) + \int_{\mathcal{S}_-}\Omega^-_z(\mathbf{k}) =  \int_{\mathcal{S}_+\cap\mathcal{S}_-}\Omega^-_z(\mathbf{k})\,,
\end{equation}

\noindent where $\mathcal{S}_\pm$ are the Fermi surfaces of the two bands and the last integral contains $\Omega^-$ since it is the curvature of the outermost band. The magnetic field dependence at constant density can be calculated numerically by varying the Fermi energy as the magnetic field is changed. Indeed, by keeping the area of the surface $\mathcal{S}_+\cap\mathcal{S}_-$  fixed the number of electronic carriers stays constant.
To obtain the resistivity $\rho_{xy}$ it is necessary to compute the two longitudinal conductivities $\sigma_{xx}$ and $\sigma_{yy}$. In the relaxation time approximation these are given by,

\begin{equation}
\sigma_{\alpha\alpha}=e^2 \tau\sum_{\gamma=\pm}\int_{\mathbf{k}}(\partial_{\alpha} \varepsilon_{\gamma})^2( -\frac{ \partial f_0}{\partial \varepsilon_\gamma})\,,
\end{equation}

\noindent where $\alpha=(x,y)$, $\varepsilon_\gamma$ is the energy dispersion of the band $\gamma$ and $f_0$ is the equilibrium Fermi-Dirac distribution. 
The resistivity $\rho_{xy}$ is then obtained by inverting the conductivity tensor,

\begin{equation}
\rho_{xy}=\frac{\sigma_{xy}}{\sigma_{xx}\sigma_{yy}+\sigma_{xy}^2}\,.
\label{eq:rhoxy}
\end{equation}
Here we have used that the transverse conductivity is purely antisymmetric, {\it i.e.} $\sigma_{xy}(B)=-\sigma_{yx}(B)$, since the semiclassical contributions are vanishing.

\begin{figure}[h!]
\centering
\includegraphics[width=8.5cm,scale=1]{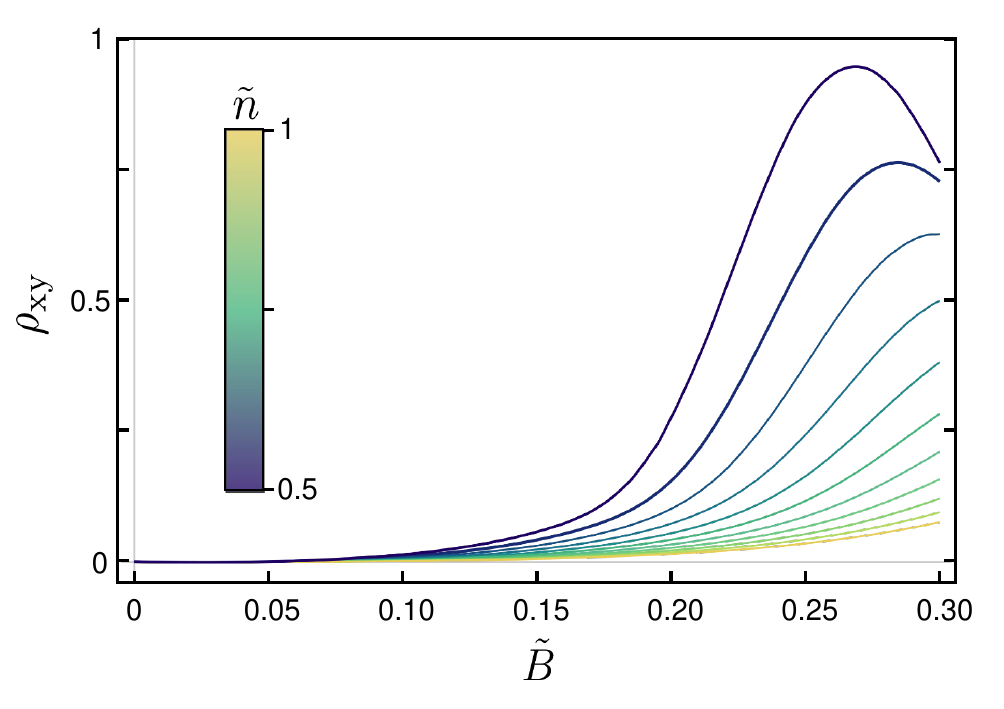}
    \caption{\textbf{Calculated planar Hall resistivity versus band filling.} Calculated magnetic field dependence of the anomalous planar Hall resistivity (measured in arbitrary units) $\rho_{xy}$, for different carrier densities. The magnetic field $\Tilde{B}$ is normalized in units of $k_\textrm{F}^2/2m$ and the densities $\Tilde{n}$ are in units of $k_\textrm{F}^2/2\pi$. The anomalous planar Hall contribution is maximum for values of the magnetic field that shift the anti-crossing point close to the Fermi energy, and goes back to zero for stronger magnetic fields.}
\label{fig:PHEcalc}
\end{figure}

Figure~\hyperref[fig:PHEcalc]{S3} shows the corresponding behavior of the anomalous planar Hall resistivity as a function of the magnetic field strength  for different values of the carrier density. The plot has been obtained by writing the low-energy Hamiltonian in dimensionless form measuring energies units of $ k_\textrm{F}^2/2m$, lengths in units of $1/k_\textrm{F}$ and density in units of $n_0=k_\textrm{F}^2/2\pi$. The remaining dimensionless parameters have been fixed to $\alpha_\textrm{R} = 0.4$, and $\lambda= 0.1$ 
Furthermore, we have used the approximate expression for the transverse resistivity $\rho_{xy} \simeq \sigma_{xy} / (\sigma_{xx} \sigma_{yy})$, which is accurate since the transverse conductivity is much smaller than the linear in $\tau$ longitudinal resistivity. 
We obtain that the anomalous planar Hall resistivity increases non-monotonically and gets enhanced by decreasing the carrier density. Note that when considering the observed decrease of the carrier density as the sheet conductivity increases (see Supplementary Note~\hyperref[Snote:SVI]{VI}), this trend is in agreement with the behavior of the antisymmetric Hall resistance reported in the main text. 

\addcontentsline{toc}{subsection}{Nonlinear transverse response with planar magnetic fields}
\subsection*{Nonlinear transverse response with planar magnetic fields}

Nonlinear transverse currents have two intrinsic contributions: the first is a semiclassical term that depends on the integral of the electronic velocities, whereas the other has a purely quantum nature stemming from the Berry curvature dipole:
 
\begin{equation}
\mathrm{\sigma^{sc}_{\alpha\alpha\beta}}\propto\int_k\,\partial^2_\alpha f_0\partial_\beta\varepsilon_k
\label{Eq:sigmasc}
\end{equation}

\begin{equation}
\mathrm{\sigma^{dip}_{\alpha\alpha\beta}}\propto\epsilon_{\alpha\beta}\int_k\,(\partial_{\beta}\Omega) f_0
\label{Eq:sigmadipc} \,,
\end{equation}

\noindent where $\epsilon_{\alpha\beta}$ is the Levi-Civita antisymmetric tensor, $\partial_\alpha = \partial_{k_\alpha}$, $f_0$ is the equilibrium Fermi Dirac distribution and $\varepsilon_k$ is the energy dispersion.
The two conductivities are the proportionality factors between the applied AC electric field $(E^{\omega}_\alpha)^2$ and the second harmonic response $j^{2\omega}_\beta$. Typically the full response is measured, and it is necessary to decouple the two contributions in order to extract the magnitude of the Berry curvature dipole. This can be done by considering how the two conductivities behave when switching the sign of the magnetic field. The $\mathrm{\sigma^{sc}_{\alpha\alpha\beta}}$ is odd in $B$: this can be seen by sending $B\rightarrow - B$ and applying the coordinate change $\mathbf{k}\rightarrow - \mathbf{k}$. While the integration measure $\partial k_x \partial k_y$ and the $f_0$ remain unchanged $\left[\varepsilon(\mathbf{k},B)= \varepsilon(-\mathbf{k},-B)\right]$ the three derivatives bring an overall minus sign. On the other hand the contribution $\mathrm{\sigma^{dip}_{\alpha\alpha\beta}}$ is even with respect to the sign of the magnetic field. Sending $B\rightarrow - B$ and $\mathbf{k}\rightarrow - \mathbf{k}$ produces a sign change in the Berry curvature $\left[\Omega(\mathbf{k},B)=-\Omega(-\mathbf{k},-B)\right]$ which is compensated by the minus sign originating from the coordinate change in the partial derivative. 

\addcontentsline{toc}{subsection}{Symmetry constraints on the linear and nonlinear resistivity tensor}
\subsection*{Symmetry constraints on the linear and nonlinear resistivity tensor}

We recall that the linear conductivity tensor is defined by the relation
\begin{equation} 
j_{\alpha}=\sigma_{\alpha \beta} E_{\beta} \,.
\end{equation}
We can derive the transformation rule of the conductivity tensor under a generic point group symmetry represented by an orthogonal matrix 
${\mathcal O}$ by simply noticing that both the current $j$ and the driving electric field $E$ transform as vectors under a generic coordinate change. Therefore, the conductivity tensor transforms as $\mathcal{O}^T \sigma \mathcal{O}$. For the point group ${\mathcal C}_{s}$ the single mirror symmetry ${\mathcal M}_x$ implies that the transverse conductivity $\sigma_{xy} \equiv \sigma_{y x} \equiv 0$. 
In crystals with ${\mathcal C}_{3v}$ point group symmetry instead, the additional threefold rotation symmetry implies that the two longitudinal conductivities along the principal crystallographic directions $\sigma_{xx}\equiv \sigma_{yy}$. 
Crystalline symmetries also pose constraints on the nonlinear conductivity tensor defined by 
\begin{equation}
j_{\alpha}=\chi_{\alpha \beta \gamma}~E_{\beta}~E_{\gamma}\,.
\end{equation}
The transformation rule of the nonlinear conductivity tensor imply that in the presence of a ${\mathcal M}_x$  mirror symmetry, we have $\chi_{x x x}=\chi_{x y y }=\chi_{y x y}=\chi_{y y x}=0$. 
The additional threefold rotation symmetry in the ${\mathcal C}_{3v}$ point group symmetry implies that the non-zero component of the nonlinear conductivity tensor satisfy the relation $\chi_{x x y} = \chi_{x y x} = \chi_{y x x} = - \chi_{y y y}$. 
A violation of this relation implies that the trigonal symmetry is broken and only a mirror symmetry is present in the system. 

\addcontentsline{toc}{subsection}{Planar magnetoresistance computation}
\subsection*{Planar magnetoresistance computation} 

\begin{figure}[h!]
\centering
\includegraphics[width=18cm,scale=1]{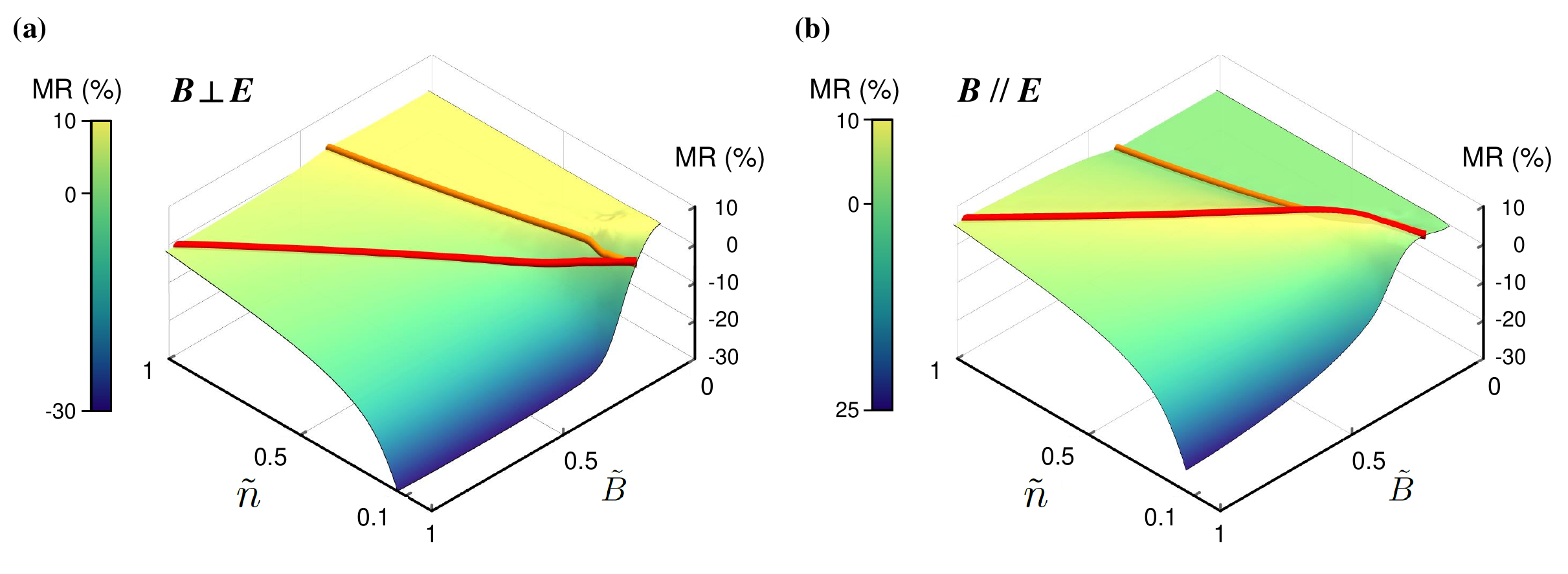}
    \caption{\textbf{Calculated planar magnetoresistance.} Planar magnetoresistance as obtained by considering a planar magnetic field in the $[\bar{1} 1 0]$ direction and an orthogonal \textbf{(a)}, or collinear \textbf{(b)} driving electric field. Since the magnetic field is mirror-symmetry preserving $\rho_{xx,yy}=1/\sigma_{xx,yy}$. The magnetoresistance has been obtained using the same parameter set as in Fig. S1, and is shown also as a function of the carrier density. The magnetic field $\Tilde{B}$ is normalized in units of $ k_\textrm{F}^2/2m$ and the densities $\Tilde{n}$ are in units of $k_\textrm{F}^2/2\pi$.  }
\label{fig:MRcalc}
\end{figure}

\begin{figure}[h!]
\centering
\includegraphics[width=16cm,scale=1]{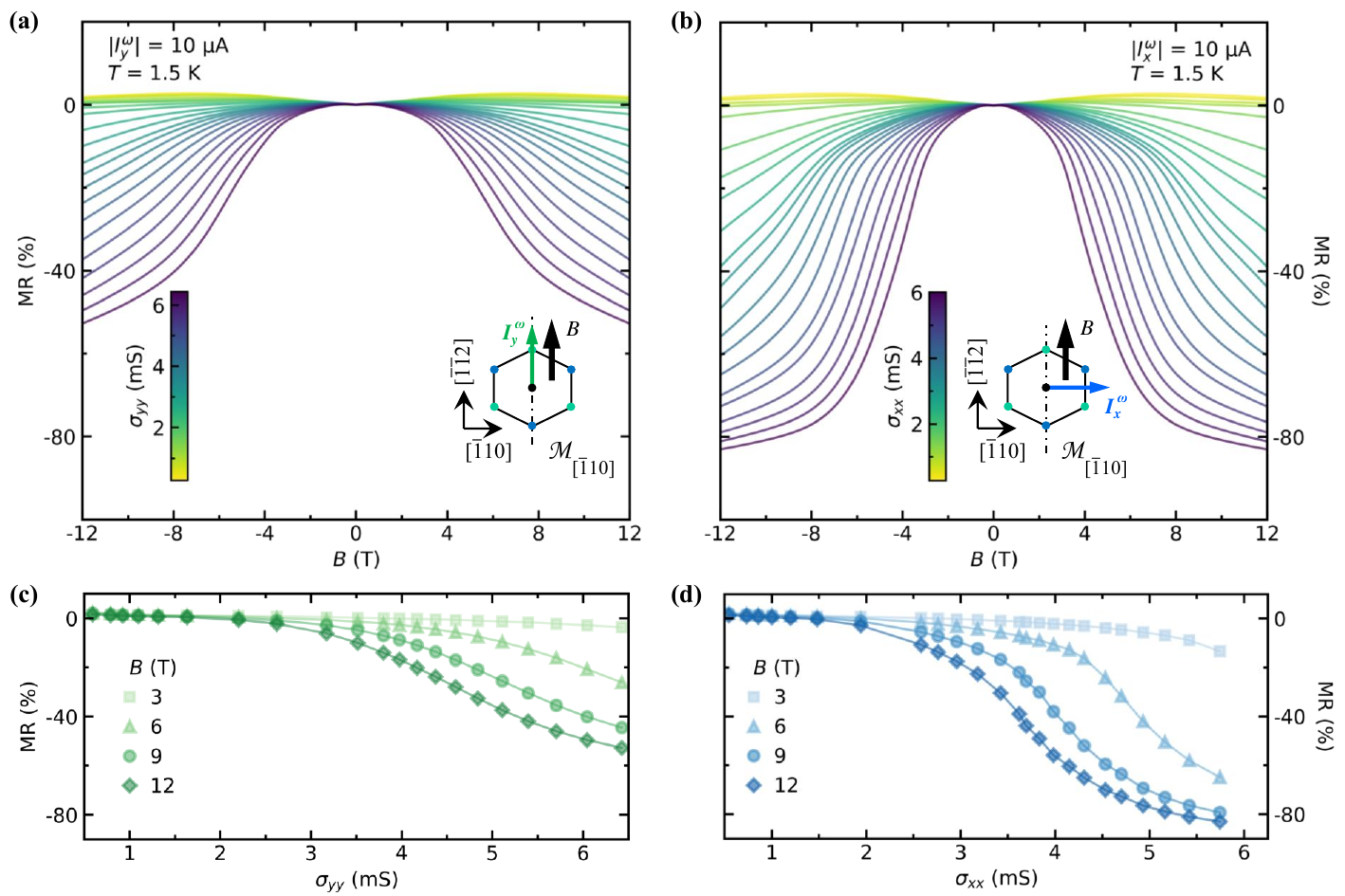}
    \caption{\textbf{Measured gate-dependent planar magnetoresistance.} Gate modulated magnetoresistance in an in-plane magnetic field for $I^\omega_{y}$ along $[\bar{1}\bar{1}2]$ parallel to $B$ \textbf{(a)}, and for $I^\omega_{x}$ along $[\bar{1}10]$ transverse to $B$ \textbf{(b)}. In both cases, the MR is seen to grow negatively above a critical planar magnetic field value. Panels \textbf{(c)} and \textbf{(d)} display the corresponding sheet conductance dependences of the MR at constant magnetic field values. For both Hall bar devices, the MR shows an onset above a given value of sheet conductance followed by a monotonic increase, and even an apparent saturation for the curve corresponding to $B=12$~T.}
\label{fig:MRmeas}
\end{figure}

We have computed the planar magnetoresistance $\textrm{MR}=\left[\rho_{xx,yy}(B) / \rho_{xx,yy}(0) - 1\right]$ considering a planar magnetic field directed along the $[\bar{1} 1 0]$ direction, thus preserving the mirror symmetry. In this case, the transverse Berry-mediated conductance $\sigma_{xy}$ vanishes. The magnetoresistance is strongly anisotropic and indeed a qualitative difference exists depending on whether the driving current is collinear or orthogonal to the magnetic field. Specifically in the former case (see Fig.~\hyperref[fig:MRcalc]{S4}) a small positive magnetoresistance starts to develop when the Fermi energy crosses the mirror symmetry-protected Dirac point (orange line). This positive magnetoresistance persists up to the Lifshitz point (red line). After the Lifshitz transition, the magnetoresistance starts decreasing and reaches a negative saturation value as also seen in experiments (see Fig.~\hyperref[fig:MRmeas]{S5}). The negative magnetoresistance can be intuitively understood by considering that after the Lifshitz transition scattering between the two energy bands is suppressed, hence enhancing the magnetoconductance.
In the configuration in which the driving electric field and the magnetic field are orthogonal, the magnetoresistance is always negative and the weak-field positive magnetoresistance does not occur [see Fig. S2(a)]. This is in agreement with the features observed at the LaVO$_3$-KTaO$_3$ interface~\cite{Wadehra2020si}. 
Note that for configurations in which the magnetic field is not mirror-preserving, a similar type of behaviour is still expected since the quantum contribution is a lower order correction to the semiclassical one.

\section{Supplementary Note III: Additional magnetotransport measurements}
\label{Snote:SIII}
\addcontentsline{toc}{subsection}{Ordinary Hall effect \& estimation of the momentum relaxation time $-$ gate dependence}
\subsection*{Ordinary Hall effect \& estimation of the momentum relaxation time $-$ gate dependence} 

\begin{figure}[h!]
\centering
\includegraphics[width=16cm,scale=1]{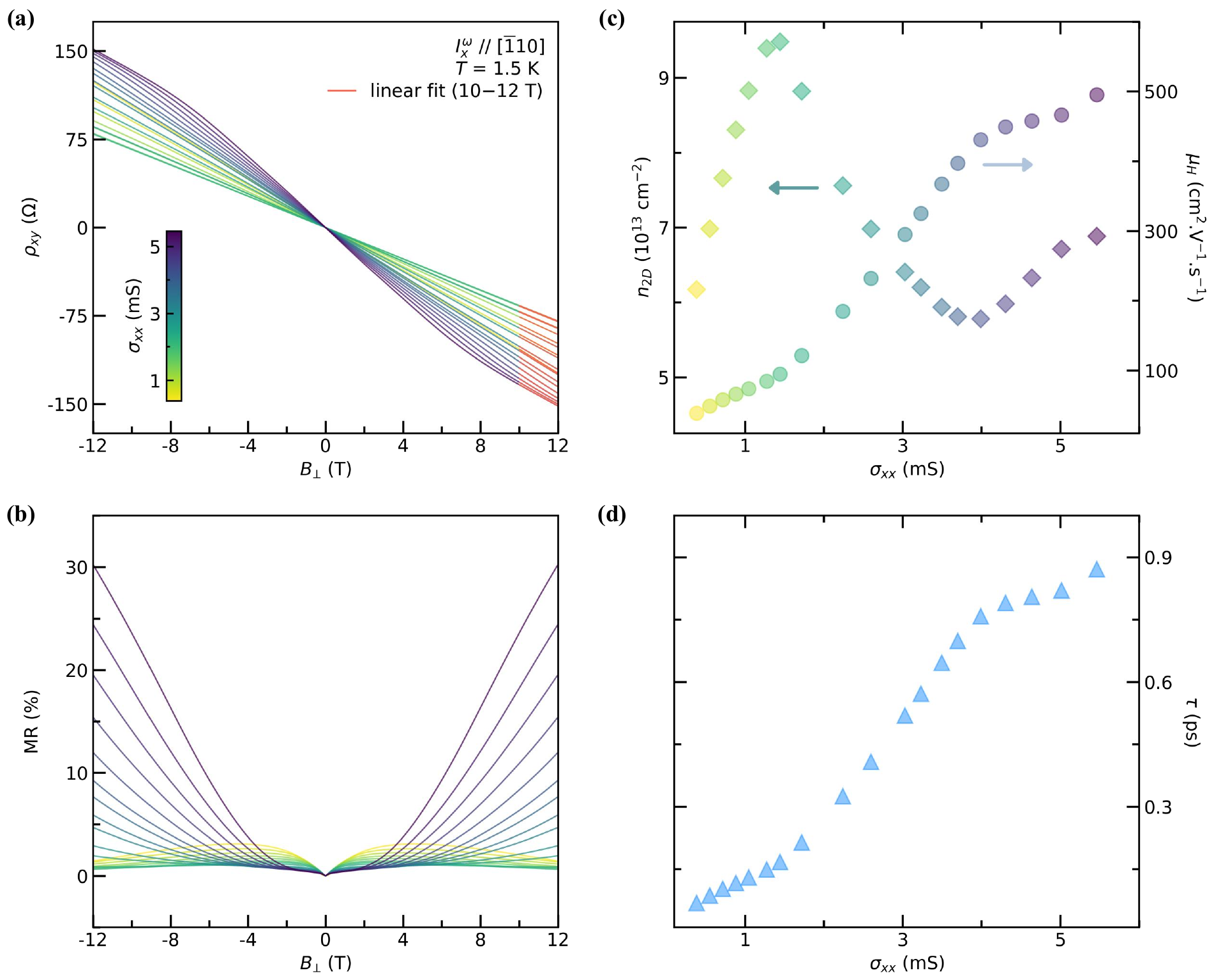}
    \caption{\textbf{Ordinary Hall effect and magnetoresistance of the 2DES.} \textbf{(a)} Gate-modulated ordinary Hall effect response of the 2DES for the current along $[\bar{1}10]$. Solid red lines are linear fits to $\rho_\textrm{xy}(B_\perp)$ performed between 10 and 12~T. \textbf{(b)} Corresponding longitudinal magnetoresistance. $B_\perp$, the out-of-plane magnetic field. \textbf{(c)} Experimentally estimated sheet carrier density $n_\textrm{2D}$ (left axis) and electronic mobility $\mu_\mathrm{H}$ (right axis) as a function of doping levels. \textbf{(d)} Momentum relaxation time $\tau$ versus sheet conductance $\sigma_\textrm{xx}$, obtained in a Drude model, following Eq.~(\ref{eq:tau})}
\label{fig:NormalHall}
\end{figure}

Figure~\hyperref[fig:NormalHall]{S6(a,b)} display the gate-modulated ordinary Hall effect, and longitudinal MR (respectively) acquired in the same device, oriented along $[\bar{1}10]$, presented throughout the manuscript; with $B_\perp$ the out-of-plane magnetic field. At low doping levels, the two-dimensional electron system (2DES) exhibits a linear Hall effect, while the low-field MR is indicative of a weak-antilocalization regime, as reported previously~\cite{Rout2017si,Monteiro2019si}.
At sheet conductance values $\sigma_\textrm{xx}$ exceeding $\approx$ 2~mS, $\rho_\textrm{xy}(B_\perp)$ is found to depart from a purely linear Hall effect. Non-linearities in the ordinary Hall effect response have been discussed extensively in the framework of a multi-carrier conduction, or multi-orbital conduction when considering SrTiO$_3$-based 2DES~\cite{Biscaras2012si,Joshua2012si,Monteiro2019si}. Here, in the case of the [111]$-$LaAlO$_3$/SrTiO$_3$ 2DES, the nonlinear Hall component has been attributed to the populating of replica sub-bands of the $t_{2g}$-manifold (derived from the Ti-3$d$ bands) in the quantum well~\cite{Monteiro2019si,Khanna2019si}.

A number of authors have further discussed the relevance of the two-band model for the determination of meaningful transport parameters values, \textit{i.e.}, carrier densities and mobilities~\cite{Joshua2012si,Biscaras2012si,Monteiro2019si}. We simply point out, that in the limit of large magnetic fields, the total carrier density in the system, which we denote $n_\textrm{2D}$, can be related to the slope of $\rho_{xy}$ vs. $B_\perp$ via:
\begin{equation}
n_\textrm{2D}=\frac{-1}{e\, R_\mathrm{H}}\,.
\label{eq:n2d}
\end{equation}
where $R_\mathrm{H}=\partial \rho_{xy}(B_\perp)/\partial B_\perp$ is the (here, high-field) Hall coefficient, whose negative sign is consistent with electron-like transport. Making use of the Drude's formula, it follows for $\mu_\textrm{H}$, the electronic mobility:
\begin{equation}
\mu_\mathrm{H} = \frac{\sigma_\mathrm{s}}{e\, n_\mathrm{2D}}\,,
\label{eq:mobility}
\end{equation}
where $\sigma_\mathrm{s}=\sigma_\mathrm{xx}$ is the sheet conductance of the considered Hall bar device (along $[\bar{1}10]$).
Figure~\hyperref[fig:NormalHall]{S6(c)} displays both the estimated areal carrier density and electron mobility of the [111]$-$LaAlO$_3$/SrTiO$_3$ 2DES across the whole accessible doping range $0.5\leq \sigma_\textrm{xx} \leq 6$~mS. While the mobility is found to increase monotonically, the apparent decrease of $n_\textrm{2D}$ versus $\sigma_\textrm{xx}$ is consistent with previous reports making use of a two-band fitting procedure up to 15~T, and which has been physically mapped by self-consistent tight-binding calculations to the redistribution of sub-bands population under the effect of electronic correlations (see Ref.~\citenum{Monteiro2019si} for details).

We estimate the momentum relaxation time $\tau_\textrm{(el)}$ [see Figure~\hyperref[fig:NormalHall]{S6(d)}], within Drude's model, which in the quasi-d.c. limit ($\omega\tau \ll 1$) is given by:
\begin{equation}
 \tau=\frac{\mu_\mathrm{H} \, m^{*}}{e}\,,
\label{eq:tau}
\end{equation}
where $m^{*}=\sqrt{ m^*_{[\bar{1}\bar{1}2]} \cdot m^*_{[\bar{1}10]}}$ is the effective mass of the multi-orbital 2DES in the SrTiO$_3$[111] quantum well, with $m^*_{[\bar{1}\bar{1}2]}=8.7m_e$ and $m^*_{[\bar{1}10]}=1.1m_e$~\cite{Rodel2014si}, $m_e$ the electron mass. The resulting calculated $m^*$ is assumed to be gate- and temperature-independent.

A linear interpolation of the measured value of the momentum relaxation time $\tau$ vs. $\sigma_\textrm{xx}$ allows to calculate the sheet conductance dependence of the BCD's magnitude $D_\textrm{x}$ (shown in Fig.~4\textbf{d}) following equation~(1) of the main manuscript. 

\addcontentsline{toc}{subsection}{Ordinary Hall effect \& estimation of the momentum relaxation time $-$ temperature dependence}
\subsection*{Ordinary Hall effect \& estimation of the momentum relaxation time $-$ temperature dependence} 

\begin{figure}[h!]
\centering
\includegraphics[width=16cm,scale=1]{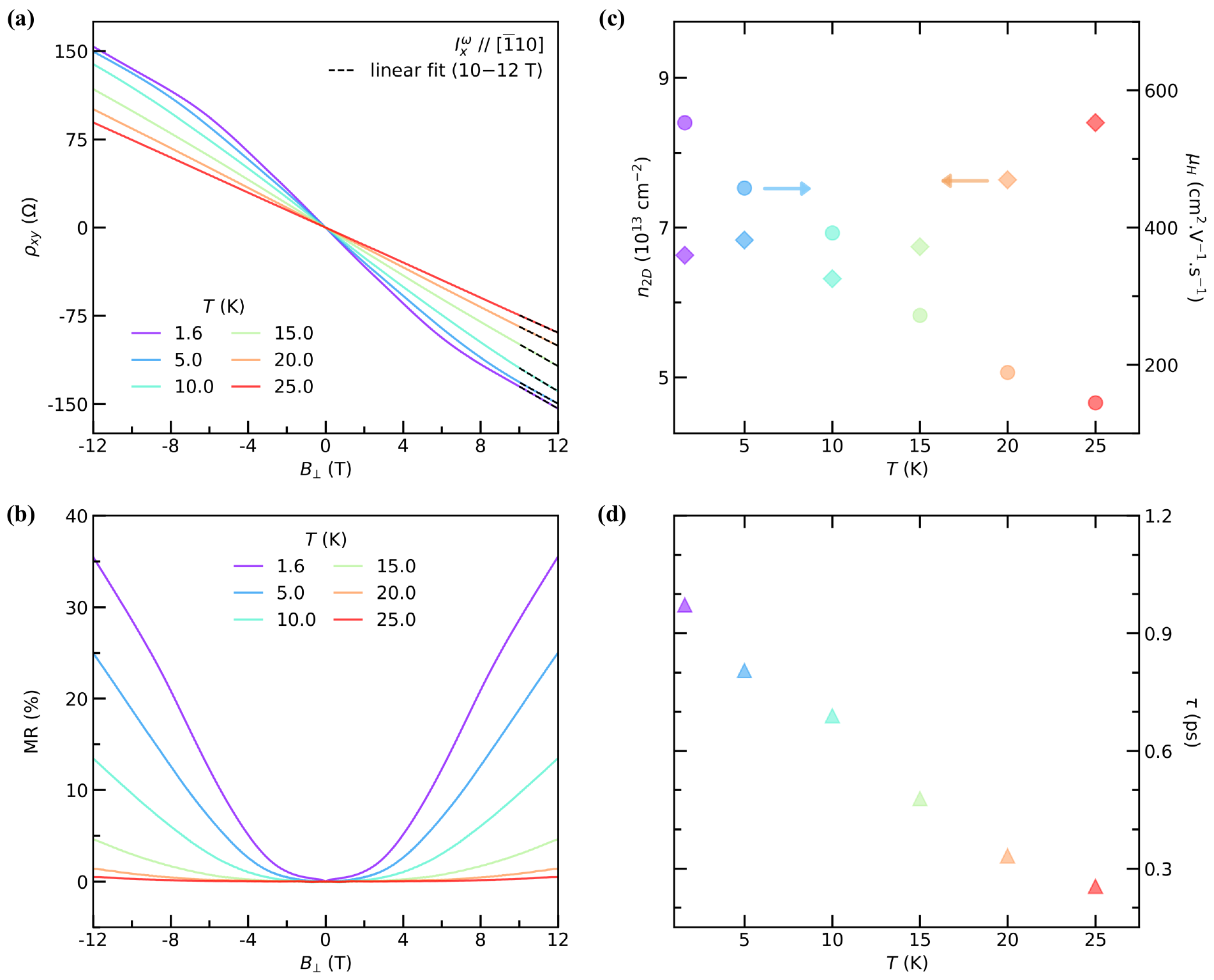}
    \caption{\textbf{Temperature dependent ordinary Hall effect and magnetoresistance of the 2DES.} \textbf{(a)} Temperature dependent Hall effect response of the 2DES for the current sourced along $[\bar{1}10]$. Dashed black lines are linear fits to $\rho_\textrm{xy}(B_\perp)$ performed between 10 and 12~T. \textbf{(b)} Corresponding longitudinal magnetoresistance. $B_\perp$, the out-of-plane magnetic field. \textbf{(c)} Experimentally estimated sheet carrier density $n_\textrm{2D}$ (left axis) and electronic mobility $\mu_\mathrm{H}$ (right axis) as a function of temperature. \textbf{(d)} Momentum relaxation time $\tau$ versus temperature $T$, obtained in a Drude model, following Eq.~(\ref{eq:tau})}
\label{fig:NormalHallvsTemp}
\end{figure}

Additionally, we also report the temperature dependence of the ordinary Hall effect and longitudinal MR of the 2DES, as shown in Fig.~\hyperref[fig:NormalHallvsTemp]{S7(a)} and \hyperref[fig:NormalHallvsTemp]{S5(b)}, respectively, for the Hall bar device oriented along [$\bar{1}10$]. Following the steps described above for the gate-dependent data set, we similarly extract the temperature dependence of the carrier density and electronic mobility [see Fig.~\hyperref[fig:NormalHallvsTemp]{S7(c)}], as well as that of the momentum scattering time [see Fig.~\hyperref[fig:NormalHallvsTemp]{S7(d)}]. Even though not shown here, we perform the same analysis for the orthogonal Hall bar device (\textit{i.e.}, oriented along [$\bar{1}\bar{1}2$]) for experimentally measured data between 1.5~K and 30~K. These combined results, together with nonlinear Hall measurements with time-reversal symmetry, allow us to calculate the temperature dependence of the nonlinear conductivity tensor elements $\chi_\textrm{yxx}$ and $\chi_\textrm{xyy}$ [see Fig.~4\textbf{e} of the main manuscript], as well as the $T$-dependence of the Berry curvature dipole $D_\textrm{x}$ [see Fig.~4\textbf{f} of the main manuscript].

\addcontentsline{toc}{subsection}{Additional doping-dependent Hall effect measurements and magnetoconductance in the weak antilocalization regime}
\subsection*{Additional doping-dependent Hall effect measurements and magnetoconductance in the weak antilocalization regime} 

\begin{figure}[h!]
\centering
\includegraphics[width=16cm,scale=1]{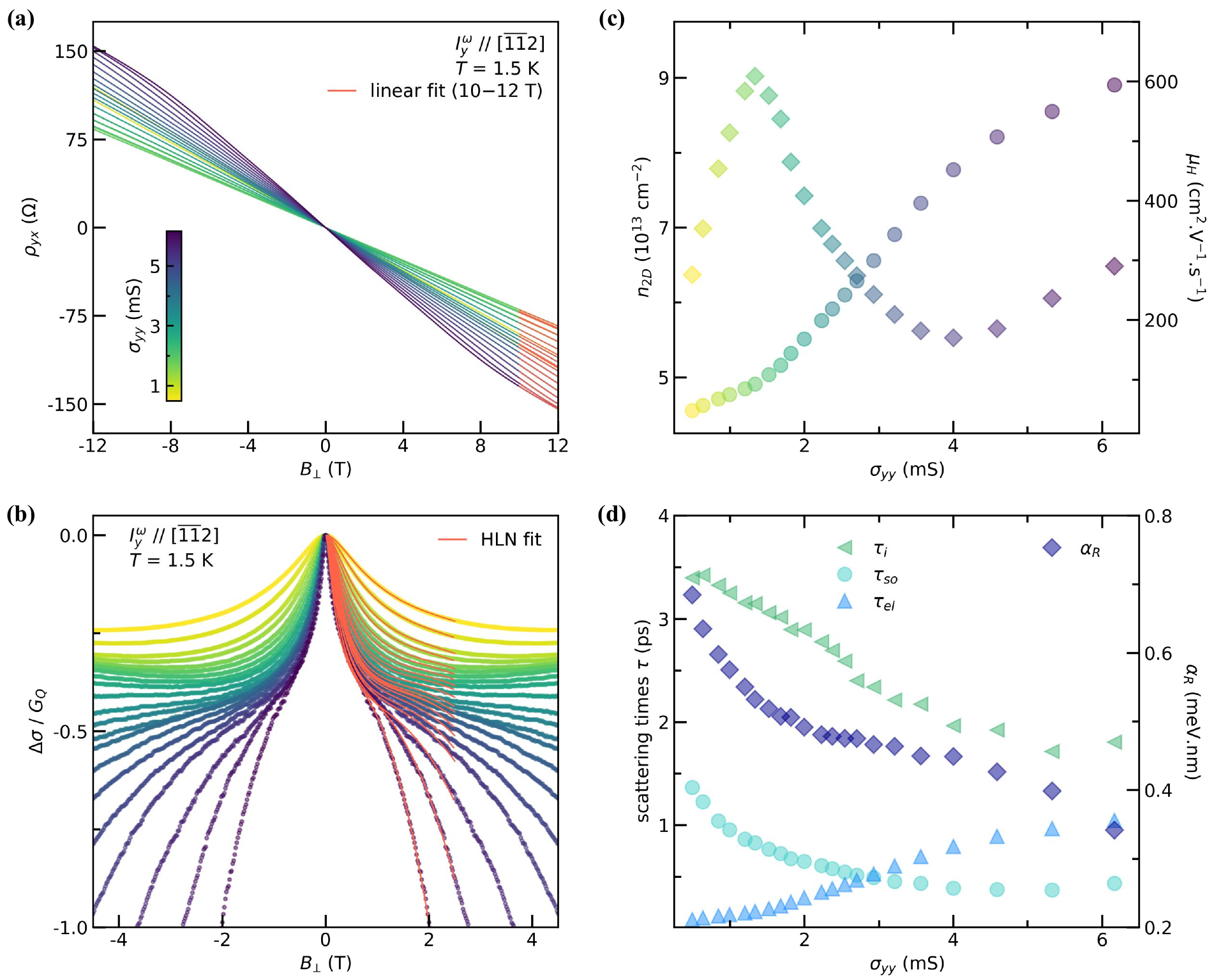}
    \caption{\textbf{Ordinary Hall effect and magnetoconductance in the WAL regime.} \textbf{(a)} Gate-modulated ordinary Hall effect response of the 2DES for the current sourced along $[\bar{1}\bar{1}2]$. Solid red lines are linear fits to $\rho_\textrm{yx}(B_\perp)$ performed between 10 and 12~T. \textbf{(b)} Corresponding magnetoconductance curves (normalized to the quantum of conductance $\textrm{G}_\textrm{Q}$) and fitted (solid red lines )using a Hikami-Larkin-Nagaoka model, following Eq.~(3) from the main manuscript. $B_\perp$, the out-of-plane magnetic field. \textbf{(c)} Experimentally estimated sheet carrier density $n_\textrm{2D}$ (left axis) and electronic mobility $\mu_\mathrm{H}$ (right axis) versus sheet conductance $\sigma_\textrm{yy}$. \textbf{(d)} Left axis: Momentum, inelastic and spin-orbit relaxation times $\tau_\textrm{el}$, $\tau_\textrm{i}$ and $\tau_\textrm{so}$, as a function of $\sigma_\textrm{yy}$. Right axis: Strength of the Rashba spin-orbit coupling $\alpha_\textrm{R}$ \textit{vs.} $\sigma_\textrm{yy}$.}
\label{fig:WAL}
\end{figure}

We present in Fig.~\hyperref[fig:WAL]{S8\textbf{(a,b)}} the gate-dependent ordinary Hall and longitudinal magnetoconductance data set collected for the Hall bar device oriented along the [$\bar{1}\bar{1}2$] direction. This allows the experimental estimation of total the carrier density ($n_\textrm{2D}$), Hall mobility ($\mu_\textrm{H}$), momentum scattering time ($\tau_\textrm{el}$), as well as inelastic and spin-orbit and scattering times ($\tau_\textrm{i}$ and $\tau_\textrm{so}$, respectively) as a function of the 2DES' sheet conductance $\sigma_\textrm{yy}$ ($I^\omega_\textrm{y} \parallel$ [$\bar{1}\bar{1}2$]).

We refer the reader to the Methods sections of the main manuscript for details regarding the Hikami-Larkin-Nagaoka (HLN) model used to fit [see Eq.~(3)] the magnetoconductance curves in the weak-antilocalization regime~\cite{HLN1980si,MF1981si}.

\addcontentsline{toc}{subsection}{Estimation of the critical magnetic field and corresponding Zeeman energy. Comparison with the Rashba spin-orbit energy}
\subsection*{Estimation of the critical magnetic field and corresponding Zeeman energy. Comparison with the Rashba spin-orbit energy} 

\begin{figure}[h!]
\centering
\includegraphics[width=15.5cm,scale=1]{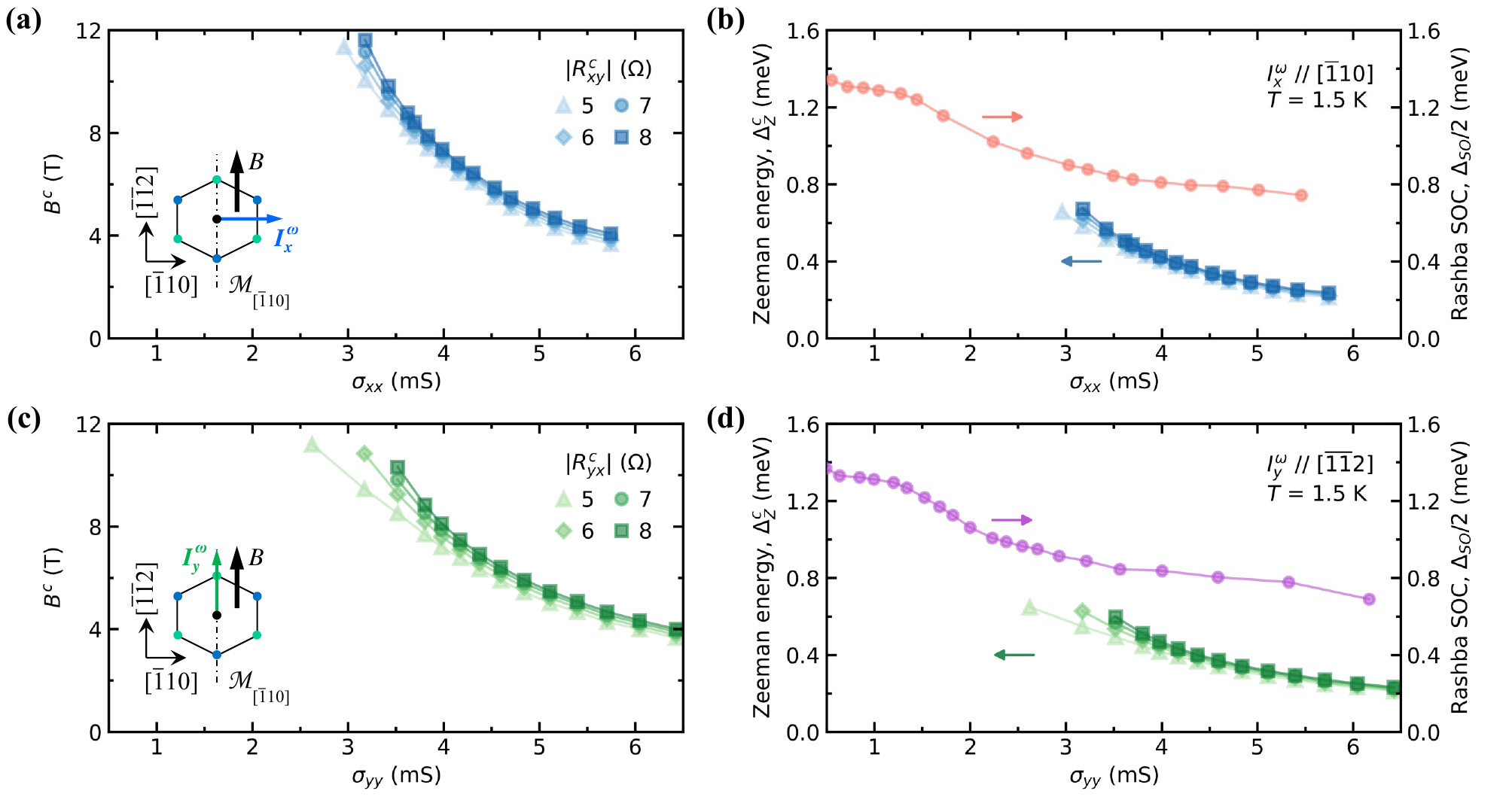}
    \caption{\textbf{Estimation of the critical magnetic field, corresponding Zeeman energy and comparison with the Rashba spin-orbit energy.} \textbf{(a)} Estimation of the critical field $B^\textrm{c}$ corresponding to the onset of the planar Hall effect, for $I_\textrm{x}^{\omega}$ along $[\bar{1}10]$. \textbf{(b)}, Corresponding effective Zeeman energy $\Delta_\textrm{Z}^\textrm{c}$ at $B=B^\textrm{c}$ (left axis), and Rashba spin-orbit energy $\Delta_\textrm{so}/2$ (right axis) versus sheet conductance. (\textbf{c-d}) \textit{Idem} for $I_\textrm{y}^{\omega}$ along $[\bar{1}\bar{1}2]$.
    }
\label{fig:Zeeman}
\end{figure}

On one hand, we establish a criterion for the determination of the `critical' magnetic field, $B^\textrm{c}$, at which the transverse magnetoresponse in a planar magnetic field exceeds a given threshold value $R_\textrm{xy}^\textrm{c}$ [shown in the legends of panels \hyperref[fig:Zeeman]{S9\textbf{(a,c)}}, respectively]. The corresponding effective Zeeman energy, at the in-plane field magnitude $B=B^\textrm{c}$, is given by:
\begin{equation}
    \Delta_\textrm{Z}^\textrm{c} = \Delta_\textrm{Z}\left(B^\textrm{c}\right) = \frac{1}{2}g\mu_\textrm{B}B^\textrm{c}\,,
\end{equation}

where $g=2$ is the electron g-factor. In the main manuscript, we keep the criterion 
$R_\textrm{xy}^\textrm{c} \equiv \left|R_\textrm{xy}\right| \geq 6$~ 
$\Omega$ 
for the determination of $B^\textrm{c}$ and $\Delta_\textrm{Z}^\textrm{c}$, as shown in Extended Data Fig.~5.

On the other hand, assuming a D'yakonov-Perel' spin relaxation mechanism~\cite{DP1971asi,DP1971bsi}, the Rashba spin-orbit energy is given by~\cite{MF1981si}:
\begin{equation}
    \Delta_\textrm{so}=2\alpha_\textrm{R}k_\textrm{F}\,,
\end{equation}
with $\alpha_\textrm{R}$ the Rashba spin-orbit coupling (SOC) determined from WAL measurements (see Methods section), and $k_\textrm{F}$ the electron wavevector at the Fermi energy.

\addcontentsline{toc}{subsection}{Out-of-plane misalignment of the planar magnetic field}
\subsection*{Out-of-plane misalignment of the planar magnetic field} 

In order to experimentally estimate the value of the out-of-plane misalignment angle, $\gamma$, for the measurements displayed in Fig.~3\textbf{c,d} of the main manuscript, we conduct a low-field analysis of both the ordinary Hall and first-harmonic planar Hall effects as a function of doping levels for the same $[\bar{1}10]-$oriented Hall bar device. 

\begin{figure}[h!]
\centering
\includegraphics[width=14.5cm,scale=1]{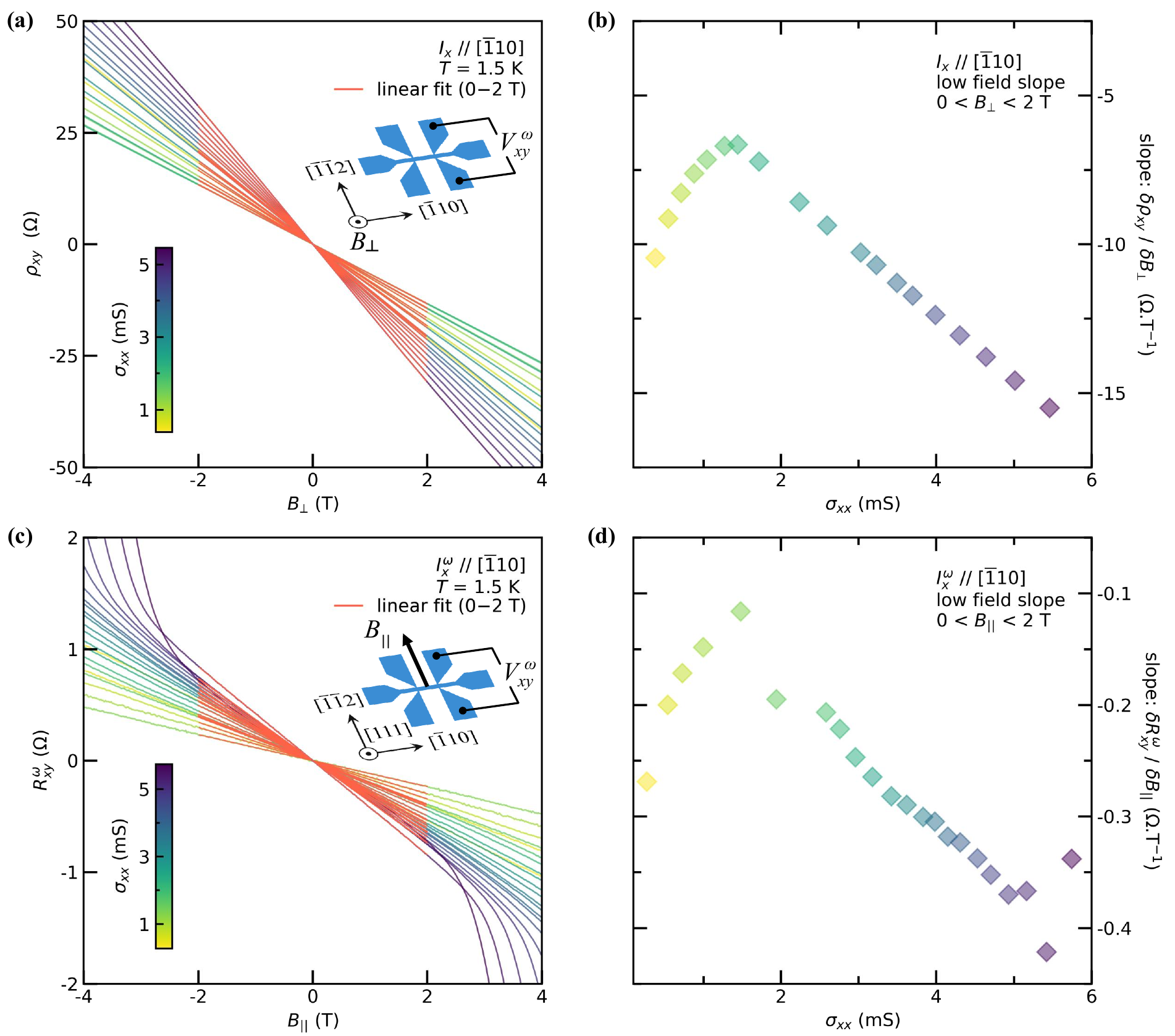}
    \caption{\textbf{Low-magnetic field linear fitting of ordinary Hall effect and first Harmonic planar Hall effect.} \textbf{(a)} Gate-modulated ordinary Hall effect response vs. $B_\perp$ of the 2DES for $I_\textrm{x}$ along $[\bar{1}10]$. \textbf{(b)} Planar Hall response versus $B_\parallel$, the out-of-plane magnetic field. Solid red lines are linear fits performed between 0 and 2~T. \textbf{(c),(d)} Corresponding low-field slopes of $\rho_\textrm{xy}(B_\perp)$, and $R^\omega_\textrm{xy}(B_\parallel)$, respectively.}
\label{fig:lowBfit}
\end{figure}

Figures~\,\hyperref[fig:lowBfit]{S10(a)} and \hyperref[fig:lowBfit]{S8(b)} are magnified low-field views of the data set displayed in Fig.~\,\hyperref[fig:NormalHall]{S6(a)} and Fig.~2\textbf{c} (see main manuscript). We perform linear fits between $\pm2$~T of the field-antisymmetrized out-of-plane $\rho_\textrm{xy}(B_\perp)$, and field-antisymmetrized in-plane $R^\omega_\textrm{xy}(B_\parallel)$ Hall magnetoresponses. The corresponding slopes are shown in Fig.~\hyperref[fig:lowBfit]{S10(c)} and \hyperref[fig:lowBfit]{S10(d)}, respectively. The common dependence of both quantities as a function of $\sigma_\textrm{xx}$ highlights their common origin. We hence attribute the linear contribution to $R_\textrm{xy}(B)$ in Fig.~2\textbf{c} [here denoted: $R^\omega_\textrm{xy}(B_\parallel)$] at low-field to a spurious contribution of the ordinary Hall component due to a small out-of-plane magnetic field component: $\Delta B_{\perp} = B_{\parallel} \sin(\gamma)$, resulting from an imperfect coplanar alignment of the field with the plane of the 2DES. We denote this out-of-plane misalignment angle $\gamma$. The contribution $\Delta R^\omega_{xy}$ to the planar Hall effect, from the conventional Hall effect due to this misalignment, is then expected to take the form: $\Delta R^\omega_{xy} = \frac{\delta \rho_\textrm{xy}}{\delta B_\perp} \cdot B_\parallel \sin(\gamma)$.

\begin{figure}[h!]
\centering
\includegraphics[width=7.5cm,scale=1]{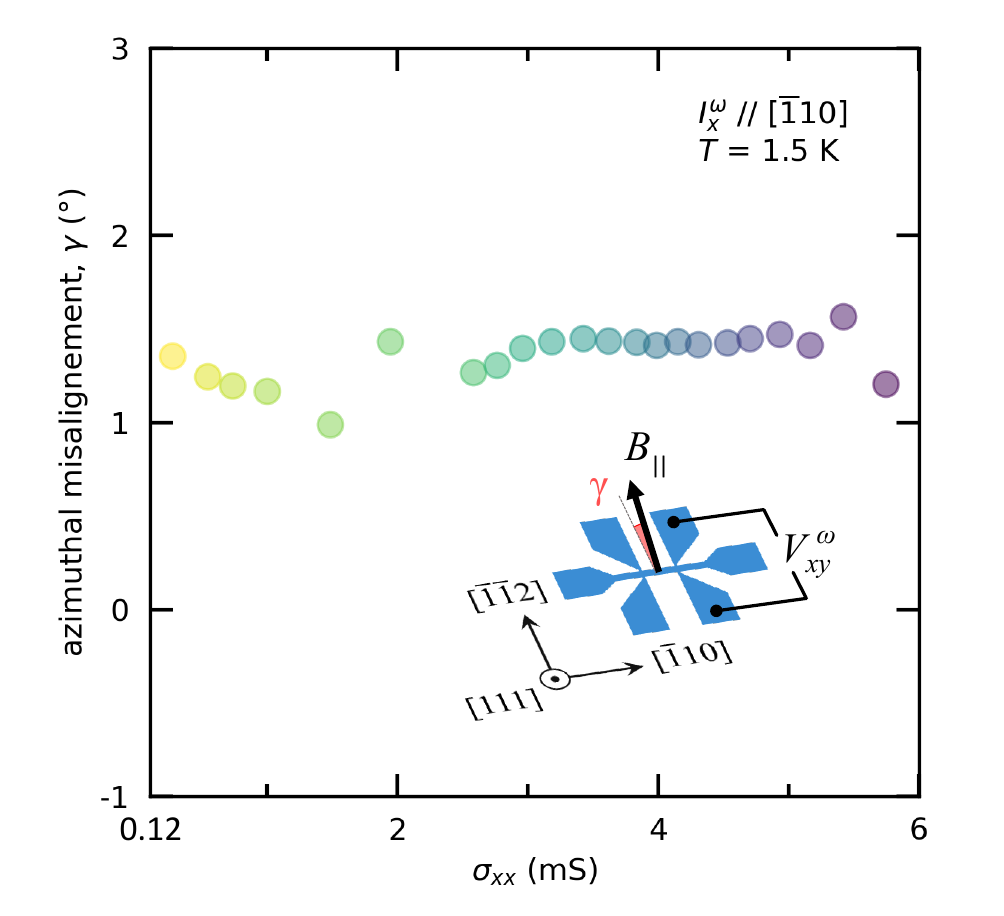}
    \caption{\textbf{Out-of-plane misalignment angle.} Experimentally estimated out-of-plane tilt value $\gamma$ of the planar magnetic field, following Eq.~(\ref{eq:gamma}), for the measurement of $R_\textrm{yx}$ displayed in Fig.~2\textbf{c} (see main manuscript).}
\label{fig:misalignment}
\end{figure}

To further support this interpretation, we display in Fig.~\hyperref[fig:misalignment]{S11} the calculated quantity $\gamma$ given by:
\begin{equation}
  \gamma = \sin^{-1} \left(\frac{\delta R^\omega_\mathrm{xy} / \delta B_\parallel}{\delta \rho_\mathrm{xy}/ \delta B_\perp}\right)
 \label{eq:gamma}
\end{equation}
where the argument is the ratio of the low-field slopes from the in-plane and out-of-plane Hall effects. We indeed find that $\gamma$ is independent of $\sigma_\textrm{xx}$, and consistently smaller than $1.5^{\circ}$. This misalignment amounts to an out-of-plane field component $\Delta B_\perp \leq 25$~mT at $B_\parallel=1$~T.

\begin{figure}[h!]
\centering
\includegraphics[width=15.5cm,scale=1]{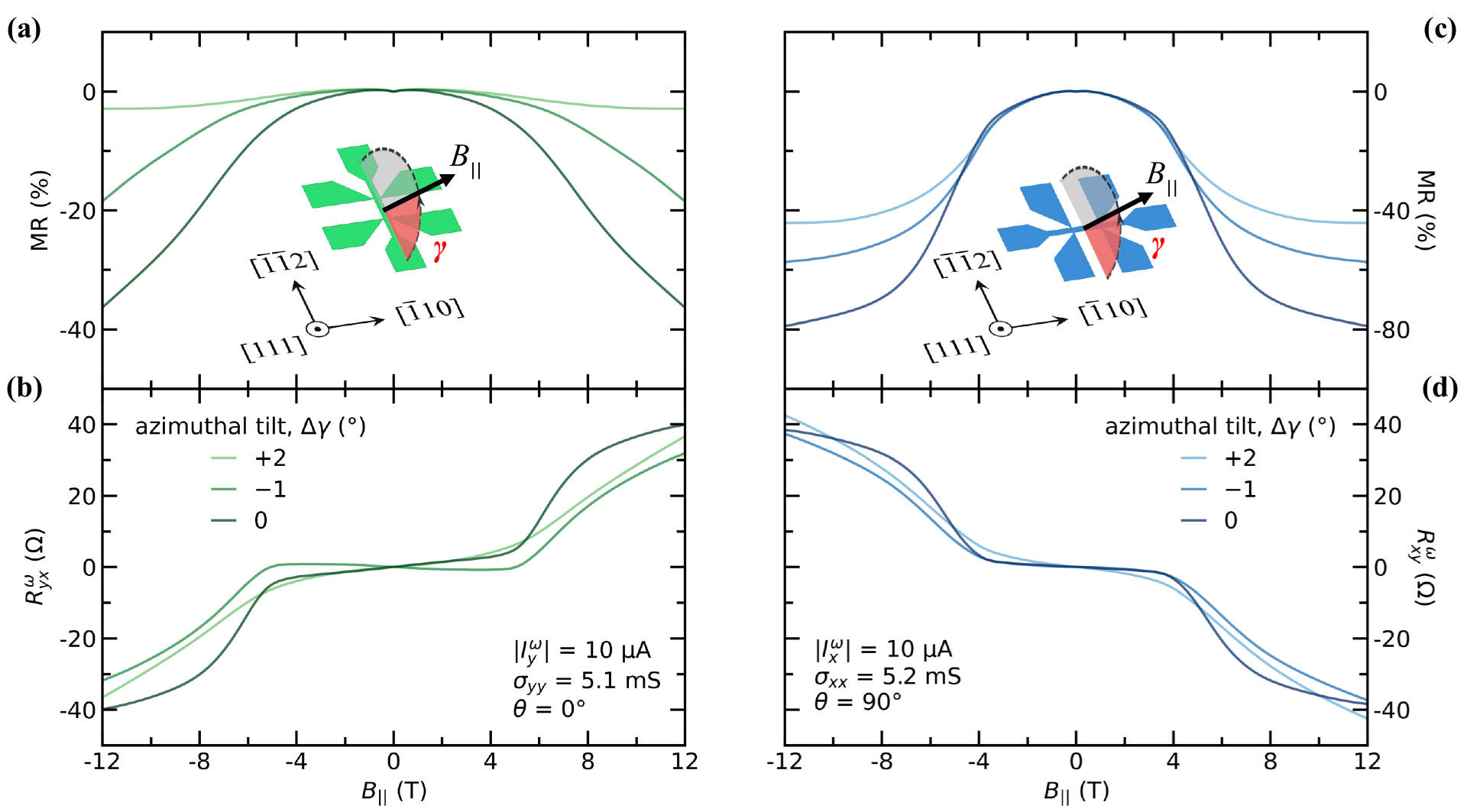}
    \caption{\textbf{Dependence of planar MR and planar Hall signals in a noncoplanar magnetic field.} Dependence of \textbf{(a)} the longitudinal MR, and \textbf{(b)} linear Hall response of the $[\bar{1}\bar{1}2]$ device ($\approx$ // $B$), for small out-of-plane deviations ($\Delta \gamma$) of the magnetic field. $\textbf{(c)}$,$\textbf{(d)}$, Idem with $B$ perpendicular to $I^\omega$ sourced along $[\bar{1}10]$ device. For both devices: while the linear planar Hall response is found to be quite insensitive to small $\Delta \gamma$ offsets, the quasi-planar MR shows a drastic change of magnitude upon small tilts of the magnetic field out-of-plane.}
\label{fig:OOPtilt}
\end{figure}

Figure~\hyperref[fig:OOPtilt]{S12} displays the change of longitudinal and transverse planar magnetoresponses, when deliberately imposing a small out-of-plane misalignment of the magnetic field (with tilt angle $\Delta \gamma$). We define $\gamma=0^\circ$ the angle at which the measurements displayed in Fig.~\textbf{2} of the manuscript were performed. Prior to any measurement campaign, we tentatively minimize $\gamma$ by finding the tilt-angle which minimizes the low-field slope of $R^\omega_\textrm{xy}(B_\parallel)$, which is typically performed at low doping levels where the BCD-induced planar Hall contribution is absent.

\begin{figure}[h!]
\centering
\includegraphics[width=17cm,scale=1]{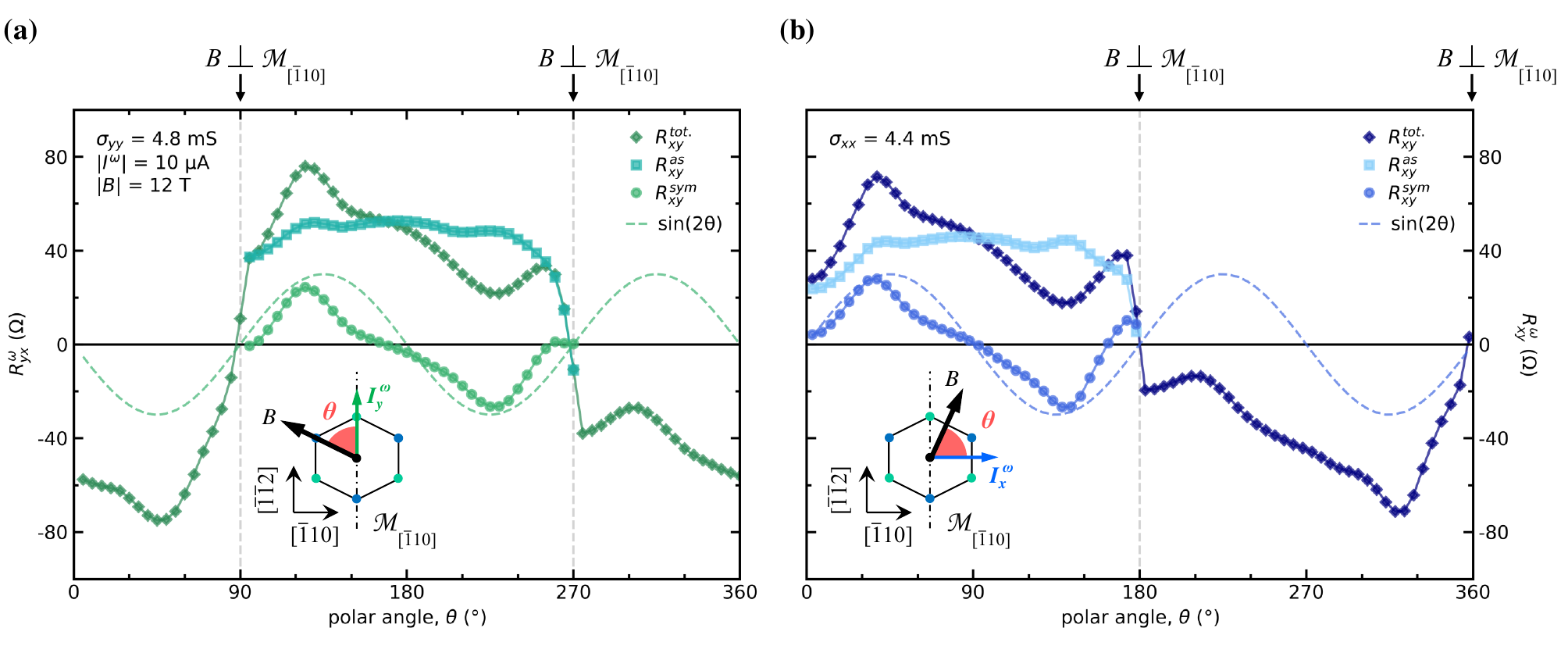}
    \caption{\textbf{Angular dependence of transverse magnetoresistance contributions.} Measured raw total signal $R_\mathrm{\alpha\beta}^\mathrm{tot.}$ at $|B|=12$~T and the corresponding field-antisymmetric $R_{\alpha\beta}^\mathrm{as}$ and field-symmetric $R_\mathrm{\alpha\beta}^\mathrm{sym}$ contributions for \textbf{(a)} $I^\omega_{y}$ along $[\bar{1}\bar{1}2]$, and \textbf{(b)} $I^\omega_{x}$ along $[\bar{1}10]$ at fixed electronic density. Remarkably, for both devices orientations, the semiclassical contribution to the planar Hall effect, \textit{i.e.}, $R_\mathrm{yx(xy)}^{sym}$,  follows the expected $\sin(2\theta)$ dependence (highlighted by the dotted lines).}
\label{fig:PHEtheta}
\end{figure}

As seen in Fig.~\hyperref[fig:OOPtilt]{S12(a,c)}, we find that the planar longitudinal MR is extremely sensitive to a very small out-of-plane misalignment, changing by a factor of two within only one degree. This explains the relative magnitude discrepancies in the planar MR between different cooldowns and measurements campaign. On the other hand, the planar Hall contribution is found to be relatively robust against small out-of-plane tilts of the magnetic field [see Fig.~\hyperref[fig:OOPtilt]{S12(b,d)}]. The low-field slope $R_{yx}$ is found to be proportional to $\Delta \gamma$, corroborating that it originates from a spurious conventional Hall component.

\addcontentsline{toc}{subsection}{Angular dependence of planar magnetoresponses}
\subsection*{Angular dependence of planar magnetoresponses} 

We acquire the full in-plane angular dependence of the longitudinal magnetoresistance (MR) at $|B|=12~$T, by sweeping the angle $\theta$ between the field and the current direction, in steps of $5^{\circ}$. Due to our mechanical rotator being limited to a 180$^{\circ}$ rotation range, we perform the rotation procedure twice, once for $B=+12~$T and a second time for $B=-12~$T, while keeping the sourced current and voltage probe contacts polarities unchanged.

\begin{figure}[h!]
\centering
\includegraphics[width=17cm,scale=1]{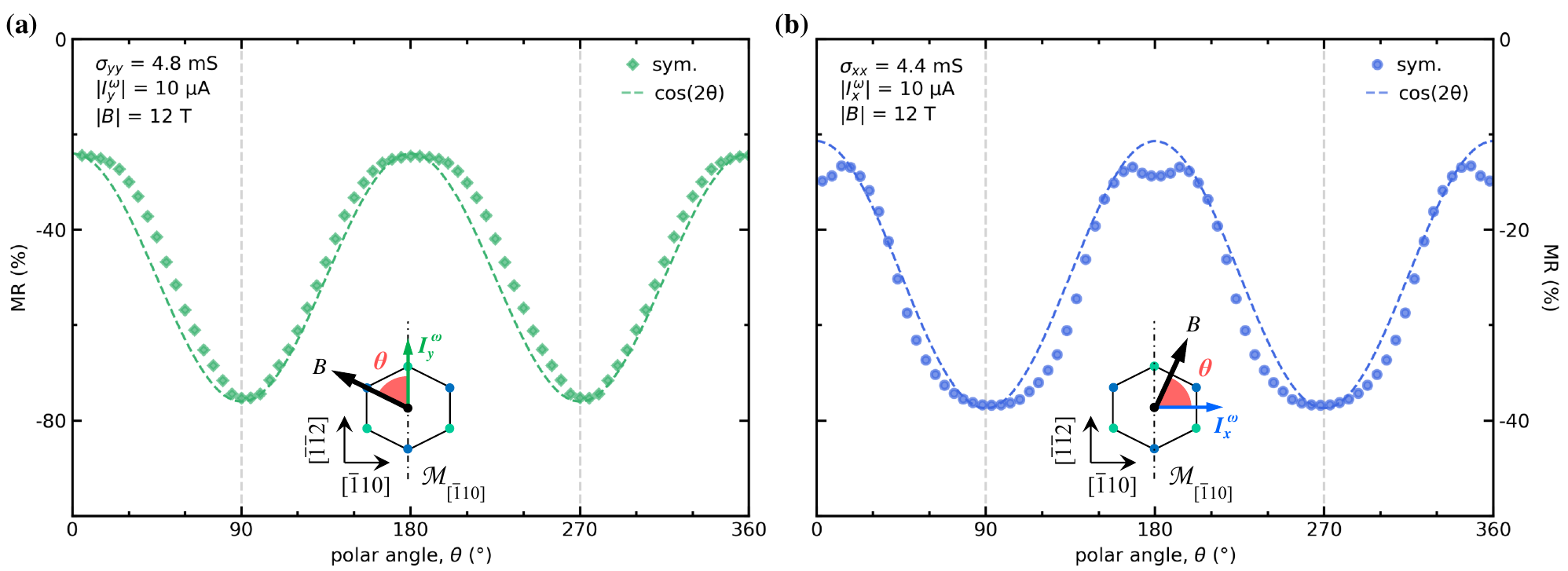}
    \caption{\textbf{Angular dependence of planar magnetoresistance.} Dependence of the field-symmetric planar magnetoresistance upon rotating a 12~T magnetic field within the sample's plane for \textbf{(a)} $I^\omega_{y}$ along $[\bar{1}\bar{1}2]$, and \textbf{(b)} $I^\omega_{x}$ along $[\bar{1}10]$; at a fixed sheet conductance value. $\theta$ is defined as the angle between the current bias direction and the external magnetic field. Dotted lines display the $\cos(2\theta)$ dependence of the MR.}
\label{fig:MRtheta}
\end{figure}

We can obtain the full field-symmetrized magnetoresistance at $|B|=12$~T, as displayed in Fig.~\hyperref[fig:MRtheta]{S13}, by virtue of Onsager's relation: $\rho_\textrm{xx,yy}(B)=\rho_\textrm{xx,yy}(-B)$. The field-symmetric longitudinal magnetoresistance is then given by:

\begin{equation}
\mathrm{MR}= \frac{\rho_{\alpha\alpha}(B)+\rho_{\alpha\alpha}(-B)}{2\rho_{\alpha\alpha}(0)}-1. 
\label{eq:eqMR}
\end{equation}

Whether the bias current is sourced along $\hat{\textbf{\textit{y}}} \parallel [\bar{1}\bar{1}2]$ or along $\hat{\textbf{\textit{x}}} \parallel [\bar{1}10]$, the planar MR follows the semiclassical $\cos(2\theta)$ dependence, where $\theta$ is the relative angle between the current direction (along a principal crystal axis) and the planar magnetic field orientation.

Concomitantly, when measuring the planar transverse magnetoresponse, we observe that the field-symmetric contributions $R^\textrm{sym}_{xy(yx)}$, of semiclassical origin (usually referred to as the 'planar Hall effect'), follows a $\sin(2\theta)$ dependence, and goes to zero at $\theta=0[\frac{\pi}{2}]$, as expected for a nonmagnetic system. 

However, we find that the total transverse resistance $R^\textrm{tot}_{\alpha\beta}$ (displayed in Fig.~2\textbf{f} of the manuscript) is dominated by the field-antisymmetric contribution (see Fig.~\hyperref[fig:PHEtheta]{S14}) , $R^\textrm{as}_{\alpha\beta}$, dubbed ``anomalous planar Hall effect"~\cite{Battilomo2021si,Cullen2021si}, which remains finite whenever the external planar magnetic field is not orthogonal to the mirror line ${\mathcal M}_{[\bar{1} 1 0]}$. Independent of whether the current is sourced along the $[\bar{1} 1 0]$ or $[\bar{1} \bar{1} 2]$ crystal axis directions, the transverse planar contribution vanishes in the linear response regime when the (${\mathcal M}_{[\bar{1} 1 0]}$-symmetry preserving) planar magnetic field is aligned with the $[\bar{1} 1 0]$ direction (\textit{i.e.} $\varphi=0[\pi]$).


%